\documentclass[a4paper,11pt]{article}
\usepackage[utf8]{inputenc}
\pdfoutput=1 

\usepackage{jheppub} 

\usepackage[T1]{fontenc} 
\usepackage{graphicx}
\usepackage{bm}
\usepackage{amssymb}
\usepackage{amsmath}
\usepackage{epsfig}
\usepackage{hyperref}
\usepackage{hhline}
\usepackage{multirow}
\usepackage[normalem]{ulem}
\usepackage{subfig}
\usepackage{tikz}
\usepackage{epstopdf}
\usetikzlibrary{snakes}
\usepackage{slashed}
\usepackage{float}
\allowdisplaybreaks

\definecolor{jd}{rgb}{0.858, 0.188, 0.478}

\def\lapp{\mathrel{\rlap{\raise.5ex\hbox{$<$}}
                    {\lower.5ex\hbox{$\sim$}}}}
\def\gapp{\mathrel{\rlap{\raise.5ex\hbox{$>$}}
                    {\lower.5ex\hbox{$\sim$}}}}

{\newcommand{\lsim}{\mbox{\raisebox{-.6ex}{~$\stackrel{<}{\sim}$~}}}
{\newcommand{\gsim}{\mbox{\raisebox{-.6ex}{~$\stackrel{>}{\sim}$~}}}

\newcommand{\bmt}{\begin{pmatrix}}
\newcommand{\emt}{\end{pmatrix}}
\newcommand{\ba}{\begin{array}{c}}
\newcommand{\ea}{\end{array}}
\newcommand{\be}{\begin{equation}}
\newcommand{\ee}{\end{equation}}
\newcommand{\bea}{\begin{eqnarray}}
\newcommand{\eea}{\end{eqnarray}}

\newcommand{\bi}{\begin{itemize}}
\newcommand{\ei}{\end{itemize}}

\newcommand{\baz}{\begin{array}{cc}}

\newcommand{\mathsym}[1]{{}}

\newcommand{\bt}{\begin{tabular}}
\newcommand{\et}{\end{tabular}}

\newcommand{\benu}{\begin{enumerate}}
\newcommand{\eenu}{\end{enumerate}}

\newcommand{\bav}{\begin{array}{cccc}}

\title{Multipartite Dark Matter with Scalars, Fermions and signatures at LHC}
\author[a]{Subhaditya Bhattacharya,}
\emailAdd{subhab@iitg.ac.in}
\author[a]{Purusottam Ghosh,}
\emailAdd{pghoshiitg@gamil.com}
\author[b]{Narendra Sahu}
\emailAdd{nsahu@iith.ac.in}
\affiliation[a]{Department of Physics, Indian Institute of Technology Guwahati, North Guwahati, Assam- 781039, India}
\affiliation[b]{Department of Physics, Indian Institute of Technology Hyderabad, Kandi, Sangareddy, Telangana- 502285, India}

\abstract{Basic idea of this analysis is to achieve a two-component dark matter (DM) framework composed of a scalar and a fermion,
with non-negligible DM-DM interaction contributing to thermal freeze out (hence relic density), but hiding them from direct detection
bounds. We therefore augment the Standard Model (SM) with a scalar singlet ($S$) and three 
vectorlike fermions: two singlets ($\chi_1,\chi_2$) and a doublet ($N$). Stability of the two DM components is achieved by a discrete
$\mathcal{Z}_2 \times {\mathcal{Z}^\prime}_2$ symmetry, under which the additional fields transform suitably. 
Fermion fields having same $\mathcal{Z}_2 \times {\mathcal{Z}^\prime}_2$ charge 
($N,\chi_1$ in the model) mix after electroweak symmetry breaking (EWSB) and the lightest component becomes one of the DM candidates, while
scalar singlet $S$ is the other DM component connected to visible sector by Higgs portal coupling. The heavy fermion ($\chi_2$) plays the 
role of mediator to connect the two DM candidates through Yukawa interaction. This opens up a large parameter 
space for the heavier DM component through DM-DM conversion. Hadronically quiet dilepton signature, arising from the fermion 
dark sector, can be observed at Large Hadron Collider (LHC) aided by the presence of a lighter scalar DM component, 
satisfying relic density and direct search bounds through DM-DM conversion.}

\begin{document}

\maketitle
\flushbottom

\setcounter{footnote}{0}
\renewcommand*{\thefootnote}{\arabic{footnote}}

\section{Introduction}
\label{sec:intro}
Observation of galactic rotation curves~\cite{Rubin:1967msa,Rubin:1970zza}, gravitational lensing and anisotropies in cosmic microwave background~\cite{Hu:2001bc} collectively
hint towards the existence of a cosmologically stable dark matter (DM) component in the the present Universe~\cite{Bertone:2004pz}. 
However, there is no such particle candidate exist within the standard model (SM), which can behave 
as DM. Hence physics beyond the SM is inevitable. Hitherto the only information known about DM is its relic abundance 
and is precisely determined by Wilkinson Microwave Anisotropy Probe (WMAP)~\cite{Hinshaw:2012aka} and PLANCK~\cite{Ade:2013zuv} to be 
$\Omega_{\rm DM}h^2=0.1161\pm 0.0028$. Apart from this, we don't have any other information about DM, such as its mass, spin, 
interaction {\it etc.} As a result, the nature of DM being a scalar, a fermion, or a vector boson or an 
admixture of them can not be avoided. In addition to gravity, if the DM is weakly interacting to visible sector, then it can thermalise in the early 
Universe at a temperature above its mass scale. As the Universe cools down due to Hubble expansion, the DM freezes-out from the 
thermal plasma at a temperature below its mass scale and gets redshifted since then. It is miraculous that the observed DM 
abundance implies to thermal freeze-out cross-section of DM: $\langle \sigma|v|\rangle\approx 10^{-36} {\rm cm}^2$, of typical 
weak interaction strength and therefore it is largely believed that the DM is a weakly interacting massive particle (WIMP)~\cite{Kolb:1990vq}.

However, the WIMP paradigm suffers from a serious threat due to the non-observation of DM in direct search experiments. In fact, in a few 
years from now the DM-nucleon cross-section measured at direct search experiments may hit the neutrino floor~\cite{Billard:2013qya}, where neutrino-nucleon 
cross-section will be a huge background for DM detection. The main problem in a WIMP paradigm is that the interactions which lead to 
the freeze-out of DM in the early Universe, also yields DM-nucleon cross-section in direct search experiments in the present epoch, 
such as LUX~\cite{Akerib:2017kat}, XENON~\cite{Aprile:2018dbl,Aprile:2015uzo}, PANDA~\cite{Cui:2017nnn} {\it etc.} The same is true for non-observation of DM in 
collider searches as well. The only difference for a WIMP of $\sim 100 ~{\rm GeV}$ is that the production of DM at collider is suppressed 
(with no electromagnetic or strong interactions with SM), so that non-observation of DM in collider provides less constraint than 
those of direct searches at terrestrial laboratories.    

Multipartite DM frameworks~\cite{Cao:2007fy,Bhattacharya:2013hva,Biswas:2013nn,Bian:2013wna,Esch:2014jpa,DuttaBanik:2016jzv,Klasen:2016qux,Bhattacharya:2016ysw,Bhattacharya:2017fid,Herrero-Garcia:2017vrl,Aoki:2017eqn,Khan:2017ygl,Ahmed:2017dbb,Herrero-Garcia:2018lga,Aoki:2018gjf,YaserAyazi:2018lrv,Poulin:2018kap,Chakraborti:2018lso} can provide a cushion to the tension of WIMP like particles to satisfy simultaneously 
relic density and direct search constraints. This is essentially due to some processes which can still contribute to the depletion of DM number density 
for thermal freeze-out, but do not contribute to direct search cross-sections. The main two contributions of such kind can arise from: 
$(i)$ Co-annihilation of DM with a heavier particle, which can not be produced in direct search for kinematic suppression~\cite{Bhattacharya:2017fid,Casas:2017jjg} or 
$(ii)$ DM-DM interactions, where the heavier DM component can annihilate to the lighter one to yield thermal relic, but do not 
contribute to direct searches of DM~\cite{Bhattacharya:2016ysw,Bhattacharya:2017fid}. 
 
Our paper investigates one of the simplest of such cases, where we assume the presence of two DM components: one scalar 
($S$) and a fermion ($N_1$). While both DMs have been studied as individual components~\cite{McDonald:1993ex,Bhattacharya:2017fid,Ghosh:2017fmr,Bhattacharya:2015qpa}, we study the interplay of DM-DM interactions 
when they are present together. In order to enhance such interactions, we insert an additional singlet fermion field ($\chi_2$), 
which works as a mediator and carries the interaction through a Yukawa term. We thereafter demonstrate that a large parameter space becomes 
available to each DM components, whichever is heavy, saved from direct search bound thanks to enhanced DM-DM interactions. 
The lighter DM component however, has the fate similar to that of a single component case, particularly when 
direct search is concerned. This shows that scalar DM can only be present in the vicinity of Higgs resonance 
($m_S\sim m_h/2$) when it is lighter than fermion DM. The presence of an additional heavy scalar ($S_H$) in the model can 
however yield a larger parameter space for the scalar DM (even when it is lighter than fermion DM). 
Efforts have already been made to accommodate scalar and fermion DM together in a single framework~\cite{Bhattacharya:2013hva,Esch:2014jpa}, 
but most often the role of DM-DM interactions has been subdued and the outcome is predictive and severely constrained. 

Collider signatures of both the DM components have also been addressed before (see for example, \cite{Bhattacharya:2017sml,Hoferichter:2017olk}). Unfortunately, 
it turns out that neither the scalar nor the fermion DM (in their single component realisation) has a possibility of producing signal excess over SM background in near 
future run of Large Hadron Collider (LHC)\footnote {Fermion DM with singlet-doublet mixing may however yield a displaced vertex signature~\cite{Bhattacharya:2017sml}.}, 
while satisfying relic density and direct search constraints. We however demonstrate here, the presence of a lighter scalar DM component helps 
in identifying hadronically quiet dilepton signal (a characteristic signature for the charged lepton components present in the fermion dark sector) at LHC, 
which was otherwise impossible due to unsurpassable SM background contribution. This is accessible due to the freedom of utilising a larger missing energy cut, 
resulting from a larger allowed mass difference between the fermion DM and its charge companions, thanks to the 
presence of a lighter DM component and non-negligible DM-DM interactions in the set up to satisfy relic density and direct search bounds. 

The paper is organised as follows. We first introduce the model framework (in Section 2). After reviewing relic density 
and direct search constraints on the individual DM components for single component frameworks (in Section 3), we discuss in details 
the case of two-component set up poised with DM-DM conversion (in Section 4). We also point out to the possibilities of having an additional heavy scalar 
in the framework (in Section 5). We then elucidate signatures of fermion dark sector at LHC accessible through two component set up (in Section 6). 
We also briefly discuss possible cosmological effect on DM particles due to early universe inflation and reheating (in Section 7).
Finally we summarise and conclude (in Section 8). Some illustrative features of fermion DM, 
Higgs invisible decay and $Z$ invisible decay constraints on the model are detailed in Appendices A, B and C respectively.

\section{The Model}
\label{sec:model}

\begin{figure}[htb!]
$$
 \includegraphics[height=7.0cm]{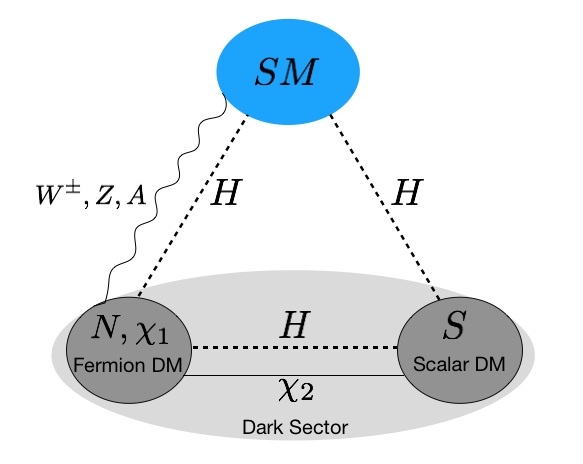}
$$
 \caption{Schematic diagram showing the interactions between scalar and fermion DM components and that to SM particles. }
 \label{fig:schmtc_diag}
\end{figure}

The model addressed here, accommodates two single component DM frameworks together: $(i)$ a 
real scalar singlet DM ($S$), connected to SM through Higgs portal~\cite{McDonald:1993ex,Feng:2014vea,Bhattacharya:2017fid,Ghosh:2017fmr} and $(ii)$ a fermion 
DM arising out of the admixture of vectorlike fermion (VF) doublet, $N=\left(\begin{matrix} N^0 &&  N^- \end{matrix}\right)^T$ 
and a vectorlike fermion singlet $\chi_1$ \cite{Bhattacharya:2015qpa,Bhattacharya:2017sml,Bhattacharya:2016rqj}, where lightest component becomes a DM. Stability of a single DM
can be ensured by an additional discrete $\mathcal{Z}_2$ symmetry, under which the DM transforms nontrivially. 
However, when two DMs are present together, the stability of both components can be ensured by enhancing the symmetry 
to $\mathcal{Z}_2 \times \mathcal{Z}^\prime_2$, where two DMs transform differently under the symmetry as we will illustrate shortly.  
Two-component DM frameworks are naturally disfavoured from direct search as each DM component acquires smaller relic density 
resulting enhanced annihilation cross-section to SM for freeze out. This enhances direct search cross-sections for both the DM components 
(resulting from same interaction vertices). This is the reason that most of the existing scalar-fermion DM scenarios have been severely 
discarded by stringent direct search limits \cite{Bhattacharya:2013hva,Esch:2014jpa}. However, DM-DM interactions may come to rescue as the freeze-out of the heavier component 
will then be additionally driven by its annihilation to lighter DM component, which do not contribute to direct search cross-section of that component. 
In order to enhance such interplay, we have introduced an additional vectorlike singlet fermion $\chi_2$, which behaves like a 
messenger between the two DM components. The interaction between the two DM components and their individual connection to the 
visible sector (SM) are shown by a schematic diagram in Fig.~\ref{fig:schmtc_diag}. 
Under the $\mathcal{Z}_2 \times \mathcal{Z}^\prime_2$ symmetry, additional dark fields transform as: 
$N$ $[-,+]$, $\chi_1$ $[-,+]$, $\chi_2$ $[+,-]$ and $S$ $[-,-]$, where all SM fields remain 
invariant: SM $[+,+]$. The quantum numbers under the SM gauge group $ SU(3)_C \times SU(2)_L \times U(1)_Y$ and 
 $\mathcal{Z}_2 \times {\mathcal{Z}^\prime}_2$ symmetry for these additional fields are shown in the Table \ref{tab:tab1}. It is remarkable 
that these additional fermions $\chi_1$, $\chi_2$ and $N$ are verctor-like and hence they don't introduce any extra anomalies. This is easy to see 
through the chiral gauge anomaly free condition coming from the one loop triple gauge boson vertex, which reads~\cite{pal2014introductory}:
\bea
\sum_{rep} Tr [\{T_L^a,T_L^b\}T_L^c]-Tr[\{T_R^a,T_R^b\}T_R^c] =0.
\eea
Here, $T$ denotes the generators for the SM gauge group and $L,R$ denotes the interactions of left or right chiral fermions with the gauge bosons. 
It is straightforward to see, that while the SM satisfies the anomaly free condition because of the presence of a quark family to each lepton family~\cite{pal2014introductory,Pisano:1993es}, the additional vector like 
fermions introduced here, have the left chiral components transforming similarly to the right chiral ones under the SM gauge symmetry. Therefore, the model is anomaly free. 

\begin{table}[htb!]
\begin{center}
\begin{tabular}{|p{3cm}|p{8cm}|}
 \hline
 \hline
  \hspace{0.2cm} {\bf Dark Fields}  &  $\hspace{0.8cm} \underbrace{SU(3)_C \times SU(2)_L \times U(1)_Y}\times \mathcal{Z}_2 \times {\mathcal{Z}^\prime}_2$  \\
 \hline
 \hline
  \hspace{0.1cm}$N=\left(\begin{matrix}
 N^0 \\  N^- 
\end{matrix}\right)$ & \hspace{1cm} 1 \hspace{1.3cm} 2 \hspace{1.1cm} -1 \hspace{0.9cm} - \hspace{0.6cm} +\\
 \hline
 \hspace{1cm}${\chi}_1$ & \hspace{1cm} 1 \hspace{1.3cm} 1 \hspace{1.15cm} 0 \hspace{0.94cm} -\hspace{0.9cm}+\\
 \hline
 \hspace{1cm}${\chi}_2$ & \hspace{1cm} 1 \hspace{1.3cm} 1 \hspace{1.15cm} 0 \hspace{0.85cm} +\hspace{0.9cm}-\\
 \hline
 \hspace{1cm}$S$ & \hspace{1cm} 1 \hspace{1.3cm} 1 \hspace{1.15cm} 0 \hspace{0.95cm} -\hspace{0.85cm} - \\
 \hline
 \hline
\end{tabular}
\end{center}
\caption{Dark sector fields and their corresponding quantum numbers under $\mathcal{G}\equiv SU(3)_C \times SU(2)_L \times U(1)_Y\times 
\mathcal{Z}_2 \times {\mathcal{Z}^\prime}_2$. }
  \label{tab:tab1}
\end{table} 
In Table~\ref{tab:tab1}, we note that $N$ and $\chi_1$ have similar $\mathcal{Z}_2 \times {\mathcal{Z}^\prime}_2$ charges. 
Hence they mix with each other after the SM Higgs acquires a vacuum expectation value (vev), while the other fermion $\chi_2$ does not. 
The lightest of such singlet-doublet admixture can be one fermion DM component of this model. 
The scalar singlet $S$ also have different charge assignment from that of all the other fermion fields, allowing it to be 
stabilized to form another DM component. The key feature of this model is the possibility of writing a Yukawa interaction between 
$\chi_1,\chi_2,~\rm{and}~S$ by the assigned $\mathcal{Z}_2 \times {\mathcal{Z}^\prime}_2$ charges, which adds to the possible DM-DM interactions as we 
explain below. This particular feature segregates this model from earlier attempts of two component scalar-fermion DM set-up like in \cite{Bhattacharya:2013hva,Esch:2014jpa}, 
where DM-DM interactions were small, so the model becomes strongly constrained by direct search or from the case where fermion DM 
doesn't have an interaction with visible sector (excepting at the loop level) to credit a large share of relic density to it and 
thus constraining the model to a particular possibility. 

Let us now describe the Lagrangian for the model, which can be segregated into three parts, constituting the vector like fermion sector, 
scalar sector and the interaction between the fermion and scalar sector as follows:

\bea
\mathcal{L}& \supset \mathcal{L}^{VF}+ \mathcal{L}^{Scalar}+\mathcal{L}^{VF+Scalar},
\eea
where,
\bea\label{lag:lagVF}
\mathcal{L}^{VF} &=& \overline{N}~[i\gamma^{\mu}(\partial_{\mu} - i g \frac{\sigma^a}{2}W_{\mu}^a - i g^{\prime}\frac{Y'}{2}B_{\mu})-m_N]~N
\nonumber \\
&&+ \overline{\chi_1}~(i\gamma^\mu \partial_{\mu}-m_{\chi_1})~\chi_1 - (Y_1\overline{N}\widetilde{H}\chi_1+h.c) \nonumber \\
&&+\overline{\chi_2}~(i\gamma^\mu \partial_{\mu}-m_{\chi_2})~\chi_2,
\eea
\bea\label{lag:scalar}
\mathcal{L}^{Scalar} && =\frac{1}{2}\partial^{\mu}S\partial_{\mu}S -\frac{1}{2} m_{S}^2 S^2 - \frac{1}{4!} \lambda_{S} S^4 - \frac{1}{2} \lambda_{SH} \Big(H^{\dagger} H -\frac{v^2}{2}\Big)S^2,
\eea 
and
\bea
\mathcal{L}^{VF+Scalar} &=&  - Y_2(\overline{\chi_1} \chi_2 S +h.c ).
\label{lag:dm-dm}
\eea

There are two Yukawa interactions present in this model. We will focus on the first in Eqn.~\ref{lag:lagVF}. Electroweak symmetry breaking (EWSB) occurs as the SM Higgs acquires a vacuum expectation value:
$
H=
\left(\begin{matrix}
 0 && \frac{1}{\sqrt{2}} (v+h)
\end{matrix}\right)^T
$ 
where $v= 246$ GeV. The Yukawa $Y_1\overline{N}\widetilde{H}\chi_1$ term in the Lagranigan (Eqn.~\ref{lag:lagVF}) mixes $N^0$ and 
$\chi_1$. Mass terms of the vector like fermions in $\mathcal{L}^{VF}$ then take the following form:

\bea
-\mathcal{L}^{VF}_{mass}&=&m_N\overline{N^0}N^0+m_NN^+N^-+m_{\chi_1} \overline{\chi_1}\chi_1+\frac{Y_1 v}{\sqrt2} \overline{N^0}\chi_1 +\frac{Y_1 v}{\sqrt2} \overline{\chi_1}N^0 \nonumber \\
&=&
\overline{\left(\begin{matrix}
\chi_1 & N^0 
\end{matrix}\right)}
{\left(\begin{matrix}
m_{\chi_1} & \frac{Y_1 v}{\sqrt2}\\
\frac{Y_1 v}{\sqrt2} & m_N
\end{matrix}\right)}
{\left(\begin{matrix}
\chi_1 \\ N^0
\end{matrix}\right)}
+m_NN^+N^-  \nonumber \\
&=&
\overline{\left(\begin{matrix}
N_1 & N_2 
\end{matrix}\right)}
{\left(\begin{matrix}
m_1 & 0\\
0 & m_2
\end{matrix}\right)}
{\left(\begin{matrix}
N_1 \\ N_2 
\end{matrix}\right)}
+m_NN^+N^-, 
\eea

where in the last step, the unphysical basis,$\left(\begin{matrix}
 \chi_1 &&  N^0 
\end{matrix}\right)^T$ is related to physical basis, $\left(\begin{matrix}
 N_1 &&  N_2 
\end{matrix}\right)^T$ through the following unitary transformation:

\bea
 \left(\begin{matrix}
 \chi_1 \\ N^0
\end{matrix}\right) 
=\mathcal U \left(\begin{matrix}
N_1 \\  N_2 
\end{matrix}\right)=\left(\begin{matrix}
 \cos\theta & -\sin\theta \\
 \sin\theta & \cos\theta
 \end{matrix}\right) 
\left(\begin{matrix}
N_1 \\  N_2 
\end{matrix}\right),
\eea

where the mixing angle
\bea\label{ref:mixang}
\tan{2\theta}= - \frac{\sqrt2 Y_1 v}{m_N-m_{\chi_1}} .
\eea

The mass eigenvalues of the physical states $N_1$ and $N_2$, for small $\sin\theta$ ($\sin\theta \rightarrow 0$) limit, can be expressed as:
\bea
m_{N_1} &\simeq& m_{\chi_1}+\frac{Y_1 v}{\sqrt 2}\sin{2\theta} \equiv  m_{\chi_1}-\frac{(Y_1 v)^2}{(m_N-m_{\chi_1})} , \nonumber \\ 
m_{N_2} &\simeq& m_{N}-\frac{Y_1 v}{\sqrt 2}\sin{2\theta} \equiv  m_{N}+\frac{(Y_1 v)^2}{(m_N-m_{\chi_1})} .
\eea

Here we have considered $Y_1 v/\sqrt2 < m _{\chi_1} < m_N$. Hence $m_{N_1} < m_{N_2}$. Therefore $N_1$ becomes the stable DM 
candidate (with a small kinematic caveat as we discuss shortly). Using Eqn.~\ref{ref:mixang}, one can find:
\bea\label{ref:reltn}
 Y_1 &=& - \frac{\Delta{m} \sin{2\theta}}{\sqrt2 v}, \nonumber \\
 m_N &=& m_{N_1}\sin^2\theta + m_{N_2} \cos^2\theta .
\eea
where $\Delta m = m_{N_2} - m_{N_1}$ is the mass difference between the two mass eigenstates and $m_N$ is the mass of electrically charged component of vectorlike fermion doublet $N^\mp$. This serves as an important parameter for the phenomenology of the model as we illustrate. Note again that due to a different $\mathcal{Z}_2 \times {\mathcal{Z}^\prime}_2$ charge, $\chi_2$ do not mix with $N$ and $\chi_1$.

Vector like fermion DM has gauge interactions to SM due to the inclusion of doublet in the model. 
Expanding the covariant derivative in $\mathcal{L}^{VF}$, one can find: 
\bea
\mathcal{L}^{VF}_{int}&=&\overline{N}i\gamma^\mu(-i g \frac{\sigma^a}{2}W_{\mu}^a + i \frac{g^{\prime}}{2}B_{\mu})N  \nonumber \\
%
&=& \Big(\frac{e_0}{2 \sin\theta_W \cos\theta_W}\Big) \overline{N^0} \gamma^{\mu} Z_{\mu} N^0 + \frac{e_0}{\sqrt2 \sin\theta_W}\overline{N^0}\gamma^{\mu}W_{\mu}^+N^-
 +\frac{e_0}{\sqrt2\sin\theta_W}N^+\gamma^{\mu}W_{\mu}^-N^0 \nonumber \\
&&- e_0 N^+\gamma^{\mu}A_{\mu}N^-  - \Big(\frac{e_0}{2 \sin\theta_W \cos\theta_W}\Big) \cos2\theta_W N^+\gamma^{\mu}Z_{\mu}N^-  
\eea
where $g=e_0/\sin\theta_W$ and $g'=e_0/\cos\theta_W$ with $e_0$ being the electromagnetic coupling constant and $\theta_W$ being the Weinberg angle. 
One can therefore express the gauge and the Yukawa interactions of $\mathcal{L}^{VF}$ in mass basis of $N_1$ and $N_2$ as:
\bea
\mathcal{L}^{VF}_{int} &=& \Big(\frac{e_0}{2 \sin\theta_W \cos\theta_W}\Big) \Big[\sin^2\theta \overline{N_1}\gamma^{\mu}Z_{\mu}N_1+\cos^2\theta \overline{N_2}\gamma^{\mu}Z_{\mu}N_2 \nonumber \\
&& ~~~~~~~~~~~~~~~~~~~~~~~~~~~~~~~~~~~~~~~~~~~~~~+\sin\theta \cos\theta(\overline{N_1}\gamma^{\mu}Z_{\mu}N_2+\overline{N_2}\gamma^{\mu}Z_{\mu}N_1)\Big] \nonumber \\
&& +\frac{e_0}{\sqrt2\sin\theta_W}\sin\theta \overline{N_1}\gamma^\mu W_\mu^+ N^- +\frac{e_0}{\sqrt2 \sin\theta_W} \cos\theta \overline{N_2}\gamma^\mu W_\mu^+ N^- \nonumber \\
&& +\frac{e_0}{\sqrt2 \sin\theta_W} \sin\theta N^+ \gamma^\mu W_\mu^- N_1 +\frac{e_0}{\sqrt2\sin\theta_W}\cos\theta N^+\gamma^\mu W_\mu^- N_2  \nonumber \\
&& - e_0 N^+\gamma^{\mu}A_{\mu}N^- - \Big(\frac{e_0}{2 \sin\theta_W\cos\theta_W}\Big) \cos2\theta_W N^+\gamma^{\mu}Z_{\mu}N^- \nonumber \\
&& -\frac{Y_1}{\sqrt2}h\Big[\sin2\theta(\overline{N_1}N_1-\overline{N_2}N_2)+\cos2\theta(\overline{N_1}N_2+\overline{N_2}N_1)\Big]
\eea

Let us now focus into the other Yukawa interaction between different DM particles as introduced in $\mathcal{L}^{VF+Scalar} $ (Eqn.~\ref{lag:dm-dm}). 
In the physical basis it reads:
\bea
\mathcal{L}^{VF+Scalar}_{int}&=& -Y_2 ( \cos\theta \overline{N_1}\chi_2 S - \sin\theta \overline{N_2}\chi_2 S + \cos\theta \overline{\chi_2} N_1 S - \sin\theta \overline{\chi_2}N_2 S). 
\label{lag:dm-dm-phy}
\eea

The scalar field $S$ do not acquire any vev and thus retains the $\mathcal{Z}_2 \times {\mathcal{Z}^\prime}_2$ symmetry intact and is eligible as a 
possible DM candidate of the model. The interaction terms involving $S$ of $\mathcal{L}^{Scalar}$ after EWSB turns out to be:
\bea
\mathcal{L}^{Scalar}_{int}&=&-\frac{\lambda_{SH} v}{2} h S^2 - \frac{\lambda_{SH}}{4} h^2 S^2 .
\eea

Following $\mathcal{L}^{Scalar}$ (Eqn.~\ref{lag:scalar}), the full scalar potential including SM Higgs can be written as: 
\bea\label{potS}
V(H,S)&=&-{\mu_H}^2 (H^\dagger H) + \lambda_H (H^\dagger H)^2 + \frac{1}{2} m_S^2 S^2 + \frac{\lambda_S}{4!} S^4 + \frac{\lambda_{SH}}{2} \Big(H^\dagger H- \frac{v^2}{2}\Big) S^2.\nonumber  \\ 
\eea

It is important now to identify the key parameters of the model which control relevant phenomenology of the model. Mainly seven independent parameters do 
the job including two DM masses, mass of the mediator, mixing angle of singlet-doublet fermion, Yukawa coupling denoting DM-DM interactions and the 
Higgs portal coupling of the scalar DM respectively:
\bea\label{parms}
\{~m_{N_1},~\Delta m,  ~m_S , ~m_{\chi_2}, ~\sin\theta , ~Y_2, ~\lambda_{SH}\}
\eea

\subsection{Constraints on the model parameters }

Before evaluating the constraints on the model parameters given in Eqn.~\ref{parms}, from DM and collider analysis, we would like to 
go through the constraints coming from stability of the potential, perturbativity of the parameters and invisible decay widths of $Z$ and $H$ to determine 
the broad parameter space available for our numerical scan.
\begin{itemize}
 \item {\bf Stability of potential}: For the tree-level vacuum stability of the scalar potential as in Eqn.~(\ref{potS}), one requires to 
satisfy the following co-positivity conditions ~\cite{Kannike:2016fmd}:  
 \bea
  ~~~~~~ \lambda_H \geq 0 , ~~ \lambda_S \geq 0 , ~~~~ \rm and ~~~~\lambda_{SH}~+\sqrt{\frac{2}{3}\lambda_H~ \lambda_S} ~~~\geq ~0 ~.
  \eea
This essentially means, we satisfy the constraints if we assume $\lambda_H, \lambda_S, \lambda_{SH} \ge$ 0 throughout the scan. 

 \item {\bf Perturbativity}: The upper limit of perturbativity bound on quartic and Yukawa couplings of the model are given by, 
 \begin{align}
  |\lambda_S| , ~|\lambda_{SH} | ~ < 4 \pi,\nonumber \\
  \rm ~and~ ~~ |Y_1| , ~|Y_2| < \sqrt{4 \pi} ~.
 \end{align}
 
 \item {\bf Relic density of DM}: The total relic density of DM is limited by the combined WMAP~\cite{Hinshaw:2012aka} 
 and PLANCK~\cite{Ade:2013zuv} data as:
\bea 
\Omega_{\rm DM}h^2=0.1161\pm 0.0028.  
\label{eq:relicdata}
\eea
 \item {\bf Invisible decay width of Higgs }: Invisible Higgs decay width puts strong constraints on light DM having masses $<m_h/2$ if they are connected through 
 Higgs portal, as we have in the model. Current bound from LHC on Higgs invisible branching fraction is given by~\cite{Tanabashi:2018oca}:
 \begin{align}
  \rm~Br~(\rm ~Higgs~ \rightarrow inv.) <~0.24.
 \end{align}
 Details have been furnished in Appendix B.
\item {\bf Invisible decay width of Z }: Z boson can decay to DM with $m_{DM}<M_Z/2$, whenever the DM has a weak charge as is the case for the fermion DM component of the model. 
Invisible decay of $Z$ is strongly constrained from observable data. The upper limit of invisible Z decay width is \cite{Tanabashi:2018oca}:
\begin{align}
 \Gamma( Z \rightarrow \rm inv.) \leq 499 \pm 1.5 ~~\rm MeV.
 \end{align}
 One may find the details about this constraint applied to our case in Appendix C.
\end{itemize}

\subsection{Possible multipartite DM scenarios}
\label{diff-scenarios}

\begin{figure}[htb!]
$$
 \includegraphics[height=7.0cm]{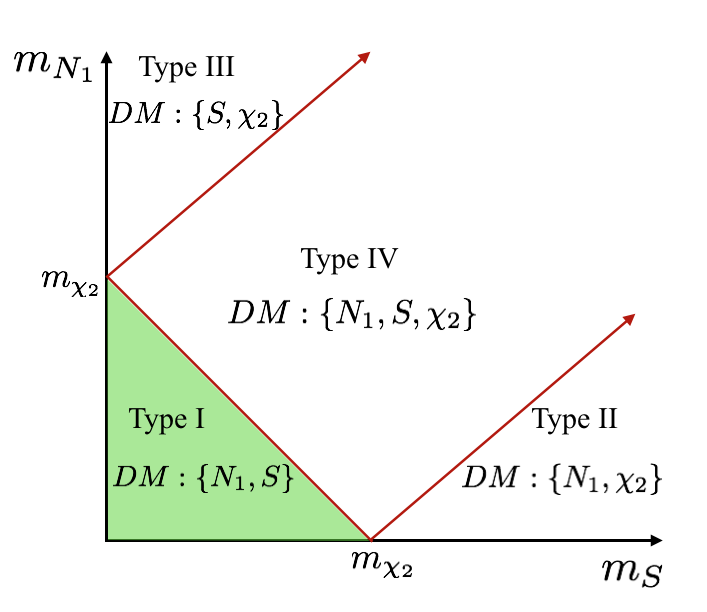}
 $$
 \caption{Different types of multicomponent DM scenarios that can be realised in the model depicted in $m_{N_1}-m_S$ plane, given that a hierarchy 
 among $m_{N_1}, m_{S}, m_{\chi_2}$. Type-I scenario (coloured in green) is analysed in this paper.}
 \label{fig:diff_multipaticle_DM_scn}
\end{figure}

We have four electromagnetic charge neutral particles in the model:  $N^0, ~\chi_1 , ~\chi_2$ and $S$. Given the same charge of $N^0$ and $\chi_1$ under 
$\mathcal{Z}_2 \times {\mathcal{Z}_2}^\prime$, they mix and the lighter eigenstate $N_1$ (with $m_{N_1}<m_{N_2}$) 
can not decay to SM, while $N_2$ decays to $N_1$. 
Then, we are left with three possible DM candidates, i.e. $N_1 , ~\chi_2$ and $S$. However, the absolute stability will be dictated by 
other Yukawa coupling present in dark sector $Y_2\overline{\chi_1} \chi_2 S \to Y_2\overline{N_1} \chi_2 S$ (as in Eqn.~\ref{lag:dm-dm-phy}). 
Evidently, if one of the physical states is heavier than the other two, then it can decay to the other two lighter particles and become unstable. 
As a result, the two lighter physical states will be the viable DM candidates. Therefore, depending on the mass hierarchy, 
the model offers four different types of multipartite DM scenarios as illustrated in Fig.~\ref{fig:diff_multipaticle_DM_scn}. 

\begin{itemize}
 \item  Type-I :~~~$ m_{\chi_2} > m_{N_1}+m_S$ :~ $N_1$ and $S$ are the stable DM components.
 \item  Type-II:~~$m_{S} > m_{N_1}+m_{\chi_2}$ : ~$N_1$ and $\chi_2$ are stable DM components.
 \item  Type-III:~~$m_{N_1} > m_{S}+m_{\chi_2}$ : ~$S$ and $\chi_2$ are stable DM components.
 \item  Type-IV:~~If $ m_{\chi_2} < m_{N_1}+m_S$, $m_{S} < m_{N_1}+m_{\chi_2}$ and $m_{N_1} < m_{S}+m_{\chi_2}$, then all three particles $N_1,~\chi_2$ and $S$ are stable and will yield a three-component DM scenario.
\end{itemize}
In this paper, we focus mostly on Type-I scenario (green region in Fig.~\ref{fig:diff_multipaticle_DM_scn}). 
This gives us an opportunity to compare with the single component cases of the corresponding DM components 
($N_1$ and $S$), which are very well studied, and indicate the effects of DM-DM conversion employed in this set-up.

\section{Review of single component DM frameworks with $N_1$ and $S$}
\label{revw:singleComp}
Before we discuss the two component DM set up (of Type-I) as advocated above, we 
need to know the fate of the individual DMs in single component frameworks. We 
review relic density and direct search allowed parameter space for both vectorlike 
fermion DM ($N_1$) and singlet scalar DM ($S$) in the next two consecutive subsections. 
 
\subsection{Single component fermion DM ($N_1$)}

The presence of vector-like fermion singlet $(\chi_1)$ and a doublet $(N)$ can give rise to a 
fermion DM~\cite{Bhattacharya:2015qpa}, where both transform under a $\mathcal{Z}_2$ symmetry. The relevant Lagrangian is still given by 
$\mathcal{L}^{VF}$ as in Eq.~\ref{lag:lagVF}. As described above, the singlet and the
neutral component of the doublet mix after EWSB, and the lightest component of 
the neutral physical states $N_1$ becomes a stable single component DM. 

We note here that the freeze-out abundance of $N_1$ DM is controlled by the annihilation and co-annihilation channels as detailed in Appendix~\ref{apnd:SVF} (Fig.~\ref{fd:an-coan}, 
\ref{co-ann-2} and \ref{co-ann-3}). Therefore, the important parameters which decide the relic abundance of $N_1$ are
\bea
\nonumber
\{m_{N_1},~\Delta m,~\sin \theta\}.
\eea
Due to singlet-doublet mixing, the DM in direct search experiments can scatter off the target nucleus via both $Z$ and Higgs mediated processes (shown in top pannel of  Fig.~\ref{fig:fd-direct}). 

 \begin{figure}[htb!]
$$
 \includegraphics[height=6.5cm]{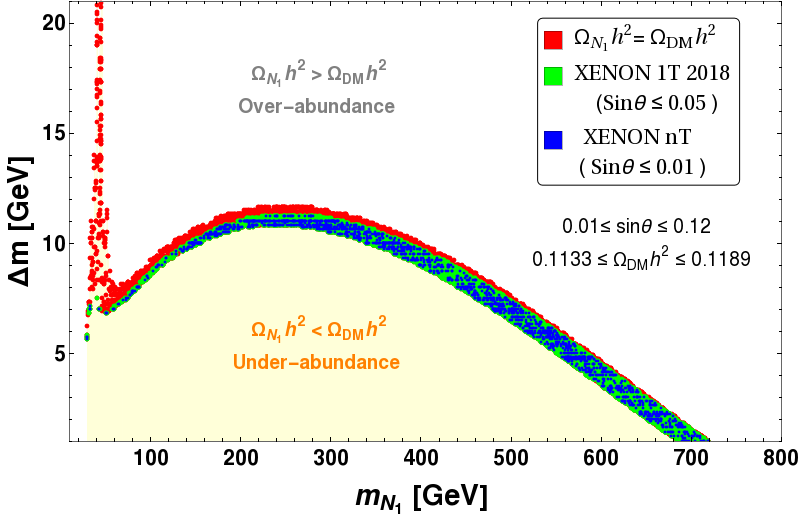} 
$$
 \caption{Relic density allowed (red points) and direct detection \{XENON 1T (green points), XENON nT (blue points)\} allowed parameter space for single component fermion DM ($N_1$) 
 is shown in $m_{N_1}-\Delta m$ plane. Under abundance $(\Omega_{N_1}h^2 < \Omega_{DM}h^2)$ (yellow region below the red points) and over-abundance 
 $(\Omega_{N_1}h^2 > \Omega_{DM}h^2)$ (the region above the red points) are also indicated. $\Omega_{DM}h^2$ range is mentioned in the figure inset and also in Eqn.~\ref{eq:relicdata}.}
 \label{fig:vf-relic-dd}
\end{figure}

The relic density and direct search allowed parameter space for $N_1$ DM is shown in Fig.~\ref{fig:vf-relic-dd}. This is shown in $m_{N_1}-\Delta m$ plane for 
small values of $\sin\theta$. It has already been noted~\cite{Bhattacharya:2015qpa} that due to $Z$ mediation, $\sin\theta$ is limited to very small 
values $\lsim 0.1$ by the non-observation of DM in direct search experiments. We therefore choose only such small mixing regions for 
illustration in Fig.~\ref{fig:vf-relic-dd}. Essentially, the whole relic density allowed plane is also allowed by direct search constraints 
(XENON1T~\cite{Aprile:2018dbl}, XENON nT~\cite{Aprile:2015uzo} as shown in the Fig.~\ref{fig:vf-relic-dd}). The under and over abundant regions are also indicated, 
which will be more useful for discussing the two component framework. The important 
message from this graph is that for small $\sin\theta$, $\Delta m$ has to be small ($\lsim 12$ GeV) to satisfy relic density, 
except for those low DM mass resonance regions ($\sim m_Z/2,~m_h/2$). For under abundance, the DM has to obtain even higher 
annihilation cross-section. For small $\sin\theta$ to satisfy direct search, the only way to probe under abundant regions is to 
have even smaller $\Delta m$ to enhance co-annihilation effects. 
Therefore, when we embed the fermion DM in a non-interacting two component DM framework, the under abundant regions 
(as indicated in Fig.~\ref{fig:vf-relic-dd}) are going to be allowed. However, the situation alters in presence of an interacting 
two-component framework as we will demonstrate in Section \ref{two_component_DM}.

\subsection{Single component scalar DM ($S$)} 

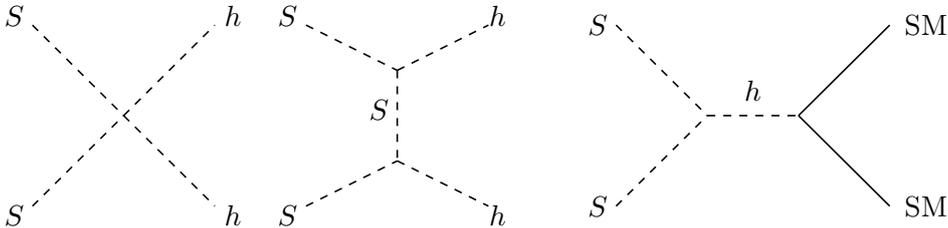
\begin{figure}[htb!]
 \begin{center}
    \begin{tikzpicture}[line width=0.6 pt, scale=1.2]
        \draw[dashed] (-5,1)--(-4,0);
	\draw[dashed] (-5,-1)--(-4,0);
	\draw[dashed] (-4,0)--(-3,1);
	\draw[dashed] (-4,0)--(-3,-1);
	\node at (-5.2,1.1) {$S$};
	\node at (-5.2,-1.1) {$S$};
	\node at (-2.8,1.1) {$h$};
	\node at (-2.8,-1.1) {$h$};
	\draw[dashed] (-2.0,1.0)--(-1.0,0.5);
	\draw[dashed] (-2.0,-1.0)--(-1.0,-0.5);
	\draw[dashed] (-1.0,0.5)--(-1.0,-0.5);
	\draw[dashed] (-1.0,0.5)--(0.0,1.0);
	\draw[dashed] (-1.0,-0.5)--(0.0,-1.0);
	\node at (-2.2,1.1) {$S$};
	\node at (-2.2,-1.1) {$S$};
	\node at (-1.2,0.07) {$S$};
	\node at (0.1,1.1) {$h$};
	\node at (0.1,-1.1) {$h$};
	%
        \draw[dashed] (1.4,1)--(2.4,0);
	\draw[dashed] (1.4,-1)--(2.4,0);
	\draw[dashed] (2.4,0)--(3.4,0);
	\draw[solid] (3.4,0)--(4.4,1);
	\draw[solid] (3.4,0)--(4.4,-1);
	\node  at (1.2,-1) {$S$};
	\node at (1.2,1) {$S$};
	\node [above] at (2.9,0.05) {$h$};
	\node at (4.8,1.0){SM};
	\node at (4.8,-1.0) {SM};
     \end{tikzpicture}
 \end{center}
\caption{Feynman diagrams for Scalar DM $S$ annihilating to SM particles i.e $S ~S \rightarrow $ SM ~SM. }
\label{ann_diag_scalar}
 \end{figure}

The Lagrangian $\mathcal{L}^{Scalar}$ in Eqn.~\ref{lag:scalar}, describes the case of single component scalar DM $S$. The relevant 
parameters describing the scalar DM interaction with SM is given by 
\bea
\nonumber
\{m_S, \lambda_{SH}\}.
\eea
The annihilation process which controls the freeze-out of $S$ are shown in Fig.~\ref{ann_diag_scalar}.  

\begin{figure}[htb!]
$$
 \includegraphics[height=7.0cm]{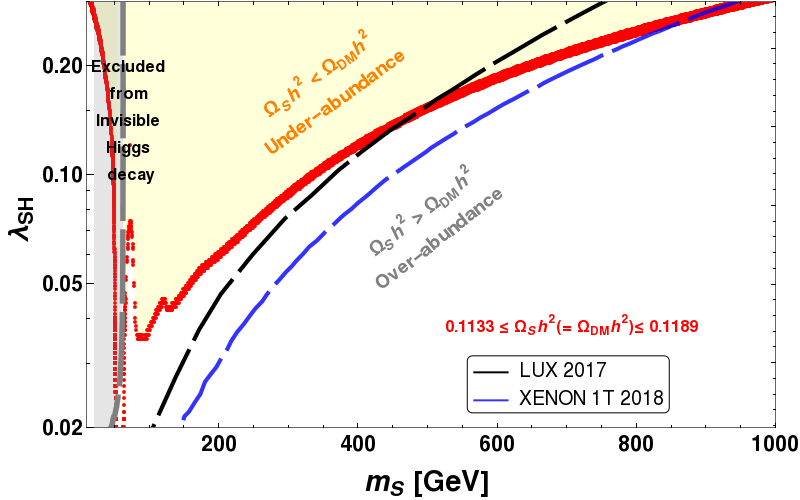} 
$$
 \caption{Relic density allowed $(\Omega_{S}h^2 = \Omega_{DM}h^2)$ (red region) parameter space for scalar DM ($S$) is shown in $m_{S}-\lambda_{SH}$ plane. LUX~\cite{Akerib:2017kat} (black dashed) and 
 XENON 1T~\cite{Aprile:2018dbl} (blue dashed) exclusion limits are also shown. The region `above' the red patch (in yellow) corresponds to under abundance $(\Omega_{S}h^2 < \Omega_{DM}h^2)$ and 
 the one below corresponds to over abundance $(\Omega_{S}h^2 > \Omega_{DM}h^2)$. Exclusion limit from invisible Higgs decays is shown by grey region. 
 $\Omega_{DM}h^2$ range is mentioned in the figure inset and also in Eqn.~\ref{eq:relicdata}.}
 \label{fig:scalar-relic-dd}
\end{figure}

The relic density allowed parameter space for the scalar DM is well studied~\cite{Bhattacharya:2016ysw,Bhattacharya:2017fid,McDonald:1993ex,Ghosh:2017fmr,Feng:2014vea} and is summarised in Fig.~\ref{fig:scalar-relic-dd}, in terms of DM mass ($m_{S}$) and Higgs portal 
coupling ($\lambda_{SH}$). Direct search sensitivities of LUX~\cite{Akerib:2017kat} and XENON 1T~\cite{Aprile:2018dbl} from null detection are also shown in the same graph for $S$ which only has a $t$-channel 
Higgs mediation with nucleus (shown in bottom pannel of  Fig.~\ref{fig:fd-direct}). This essentially shows that if $S$ contributes to the full DM relic density, it lives either in resonance region ($m_S \sim m_h/2$) or in 
high DM mass regions ($m_S \gsim 900$ GeV) to satisfy null observations from direct search experiments. Under abundance for $S$ can only be achieved with 
larger annihilation cross-section, that can only occur with larger Higgs portal coupling ($\lambda_{SH}$) and that is even more constrained from direct search data. 
If the scalar DM is embedded in an non-interacting multi-component DM framework, it is further restricted by direct search, discarding $m_S$ upto TeV or more. We will show 
in Section \ref{two_component_DM}, that the situation alters in presence of an additional DM component, with which the scalar DM has non-negligible interactions. 
We also point out that the presence of a heavy scalar $S_H$ (also a SM singlet) having same $\mathcal{Z}_2$ charge as of $S$, can change the conclusion significantly allowing a larger 
parameter space through co-annihilation (in Section~\ref{two_component_DM_SH}).

\section{Two Component DM with $N_1$ and $S$}
\label{two_component_DM}

\begin{figure}[htb!]
\begin{center}
   \begin{tikzpicture}[line width=0.6 pt, scale=1.2]
        \draw[dashed] (-5.8,1.0)--(-4.8,0.0);
	\draw[dashed] (-5.8,-1.0)--(-4.8,0.0);
	\draw[dashed] (-4.8,0.0)--(-3.8,0.0);
	\draw[solid] (-3.8,0.0)--(-2.8,1.0);
	\draw[solid] (-3.8,0.0)--(-2.8,-1.0);
	\node at (-6.0,-1.1) {$S$};
	\node at (-6.0,1.1) {$S$};
	\node at (-4.3,0.2) {$h$};
	\node at (-2.6,1.1) {$\overline{N_i}$};;
	\node at (-2.6,-1.1) {$N_j$};
        \draw[dashed] (-1.4,1.0)--(-0.4,0.5);
	\draw[dashed] (-1.4,-1.0)--(-0.4,-0.5);
	\draw[solid] (-0.4,0.5)--(-0.4,-0.5);
	\draw[solid] (-0.4,0.5)--(0.4,1);
	\draw[solid] (-0.4,-0.5)--(0.4,-1.0);
	\node  at (-1.6,-1.1) {$S$};
	\node at (-1.6,1.1) {$S$};
	\node [above] at (-0.7,-0.2) {$\chi_2$};
	\node at (0.8,1.1) {$\overline{N_i}$};;
	\node at (0.8,-1.1) {$N_j$};
     \end{tikzpicture}
 \end{center}
 \caption{Diagrams contributing to DM-DM conversion $(i=1,2)$ between fermion ($N_i$) and scalar DM ($S$) components.}
\label{fig:conversion}
 \end{figure}
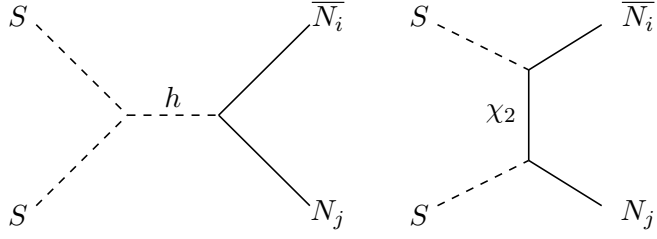

As already discussed in Section~\ref{diff-scenarios}, we choose Type-I case for illustrating a two-component interacting DM model with $m_{\chi_2} > m_{N_1} + m_S$, 
where $N_1$ froms a vectorlike fermion DM component and $S$ forms a scalar DM component. The heaviest field $\chi_2$ in the dark sector, which can decay to $N_1$ and $S$, 
act as a mediator between the two DM components through the Yukawa interaction: $Y_2 \overline{\chi_2} \chi_1 S $. These DMs can also interact with each other through Higgs portal  
couplings: $Y_1\overline{N}\widetilde{H}\chi_1$ and $\lambda_{SH} ( H^\dagger H) S^2$.  
 The DM-DM interactions of this model is shown by the Feynman diagrams in Fig.~\ref{fig:conversion}.


DM-DM conversion diagrams will dominantly help the heavier DM component to annihilate into the lighter one and therefore contribute to its 
thermal freeze-out and relic density. Apart from DM masses and mediator mass ($m_{\chi_2}$), the DM-DM conversion is a function of the following couplings 
\bea
\nonumber
\{Y_1,Y_2,\lambda_{SH}\}.
\eea
However, the Higgs portal couplings $Y_1$~(as a function of $\Delta m$ and $\sin\theta$, see in Eqn.~\ref{ref:reltn}) ~and $~\lambda_{SH}$ are strongly constrained from direct detection bound (already discussed in section~\ref{revw:singleComp}). Therefore, DM-DM interaction through Higgs mediation will be 
negligible in relic density and direct search allowed parameter space of the two component model and can be identified with $Y_2 = 0$ situation. 
We will show that in such a case, the two DMs are almost decoupled and behave like single component cases to occupy the under abundant regions of their corresponding 
DM parameter space. Here lies the importance of assuming the presence of a heavy mediator $\chi_2$ in this model to carry out DM-DM interactions through Yukawa coupling $Y_2$.
 
\subsection{Coupled Boltzmann Equations}

The thermal freeze-out of two component DM framework is described by a coupled Boltzmann equations (BEQs) and can be written as a function of 
reduced $x$, where $x=\mu/T$, with $\frac{1}{\mu}=\frac{1}{m_{N_1}}+\frac{1}{m_{N_2}}+\frac{1}{m_S}$ \cite{Bhattacharya:2016ysw,Ahmed:2017dbb,Aoki:2012ub}. The one here reads:
   
\bea
\frac{dY_{N_i}}{dx}&=& -0.264  M_{Pl} \sqrt g_*\frac{\mu}{x^2} \bigg[  \sum_j \bigg\{\langle \sigma v _{\overline{N_i} N_j \rightarrow SM}\rangle \Big(Y_{N_i} Y_{N_j}-{Y_{N_i}^{EQ}} {Y_{N_j}^{EQ}}\Big) \nonumber \\
 && + \langle \sigma v_{\overline{N_i} N_j \rightarrow S S}\rangle \Big(  Y_{N_i} Y_{N_j}-\frac{Y_{N_i}^{EQ}Y_{N_j}^{EQ}}{{Y_{S}^{EQ}}^2} Y_S^2 \Big) \Theta(m_{N_i}+m_{N_j}-2 m_S) \nonumber \\
&& - \langle \sigma v_{S S \rightarrow \overline{N_i} N_j}\rangle \Big( Y_{S}^2-\frac{{Y_{S}^{EQ}}^2}{Y_{N_i}^{EQ}Y_{N_j}^{EQ}} Y_{N_i} Y_{N_j} \Big) \Theta(2 m_S-m_{N_i}-m_{N_j}) \bigg\} \nonumber \\
&& + \langle \sigma v_{\overline{N_i} N^{\pm} \rightarrow SM}\rangle \Big(  Y_{N_i} Y_{N^\pm} -{Y_{N_i}^{EQ}}{Y_{N^\pm}^{EQ}} \Big)\bigg], \nonumber \\ 
\frac{dY_{S}}{dx} &=& -0.264  M_{Pl} \sqrt g_*\frac{\mu}{x^2} \bigg[\langle \sigma v_{ S S \rightarrow SM}\rangle \Big({Y_{S}}^2 -{Y_{S}^{EQ}}^2 \Big) \nonumber \\
 &&+ \sum_{i,j} \bigg\{- \langle \sigma v_{\overline{N_i} N_j \rightarrow S S}\rangle \Big(  Y_{N_i} Y_{N_j}-\frac{Y_{N_i}^{EQ}Y_{N_j}^{EQ}}{{Y_{S}^{EQ}}^2} Y_S^2 \Big) \Theta(m_{N_i}+m_{N_j}-2 m_S) \nonumber \\
&& + \langle \sigma v_{S S \rightarrow \overline{N_i} N_j}\rangle \Big(  Y_{S}^2-\frac{{Y_{S}^{EQ}}^2}{Y_{N_i}^{EQ}Y_{N_j}^{EQ}} Y_{N_i} Y_{N_j} \Big) \Theta(2 m_S-m_{N_i}-m_{N_j})\bigg], \nonumber \\ 
\label{eq:BEQx}
\eea
where the subscripts $i,j=1,~2$ describes the fermion DM and the heavy neutral fermion component of the model respectively. 
In the above equations, we note that the annihilation contribution of $N_i$ to $S$ or otherwise depending on the mass hierarchy is included. The equilibrium distributions now recast in terms of $\mu$ takes the form: 
\bea
Y_{N_i}^{EQ}(x)=0.145 \frac{g}{g_*}{x}^{\frac{3}{2}}\Big(\frac{m_{i}}{\mu}\Big)^{\frac{3}{2}} e^{-x\Big(\frac{m_{i}}{\mu}\Big)} \nonumber \\
Y_S^{EQ}(x)=0.145\frac{g}{g_*}{x}^{\frac{3}{2}}\Big(\frac{m_S}{\mu}\Big)^{\frac{3}{2}} e^{-x\Big(\frac{m_{S}}{\mu}\Big)}
\label{eq:Yeqx}
\eea
The relic density allowed parameter space of the two-component framework is then given by the solution of the above Boltzmann equations, that 
determine the freeze-out of the individual components depending on annihilations plus co-annihilations and DM-DM interactions.
Obviously, total DM relic density for the two component case will be the sum of individual relic density as: 
\bea
\Omega_{T}h^2 = \Omega_{N_1}h^2 + \Omega_{S}h^2,
\eea
which should satisfy combined WMAP and PLANCK limit $0.1133 \le \Omega_{T}h^2~(=\Omega_{DM}h^2)\le 0.1189 $~\cite{Ade:2013zuv}.
Individual relic density in interacting multipartite DM case can be found out by numerical solution to the coupled Boltzmann equations or 
approximate analytical solution of coupled BEQ \cite{Bhattacharya:2016ysw} and that of the $i$-th DM candidate is given by: 
\bea
\Omega_i h^2 =\frac{854.45 \times 10^{-13}}{\sqrt{g_*}}\frac{x^i_f}{{\langle \sigma v\rangle}^T_i},
\eea
where ${\langle \sigma v\rangle}^T_i$ is the total effective annihilation cross-section and $x^i_f$ corresponds to freeze-out temperature of the $i$th DM component. Note however, for the ease of the analysis, we are not using the approximate solution here; relic density and direct search cross-sections for both the DM components are obtained numerically by inserting the model in {\tt MicrOmegas} package~\cite{Belanger:2014vza}.

If fermion DM is heavier than scalar DM $(m_{N_1} > m_S)$, then heavier DM component ($N_1$) can annihilate to lighter component ($S$) following processes as in Fig.~\ref{fig:conversion}. Such DM-DM conversion affects the freeze out of heavier DM component and hence its relic density \cite{Bhattacharya:2016ysw}.  The lighter DM candidate on the other hand, have no new channel to deplete its number density and behave almost like single component DM. Then ${\langle \sigma v\rangle}^{T}_{N_1}$ for fermionic DM assuming $m_{N_1} > m_S$ will be given by:
\bea
{\langle \sigma v\rangle}^{T}_{N_1} &\simeq&{\langle \sigma v\rangle}^{eff}_{N_1} + {\langle \sigma v\rangle}_{\bar{N_1} N_1 \rightarrow S S}+{2\langle \sigma v\rangle}_{\bar{N_1} N_2 \rightarrow S S}(1+\frac{\Delta m}{m_{N_1}})^{3/2}e^{\frac{-\Delta m}{T}}, 
\label{eq:ann-NS1}
\eea
where ${\langle \sigma v\rangle}^{eff}_{N_1}$ is the annihilation plus co-annihilation cross-section of fermion DM to SM given by Eq.~\ref{eq:vf-ann}. The last term in the above equation represents co-annihilation to scalar DM component and is therefore aided by the Boltzmann factor along with  a symmetry factor of 2 (assuming $m_{N_1}\simeq m_{N_2}$). In this limit of $m_{N_1} > m_S$, the annihilation cross-section for scalar DM ($S$) only captures the annihilations to SM as in a single component framework:
\bea
{\langle \sigma v\rangle}^T_{S} &=&{\langle \sigma v\rangle}_{ S S \rightarrow SM ~SM}~. 
\label{eq:ann-NS2}
\eea

Evidently, for the opposite hierarchy, $m_S > m_{N_1}$:
\bea
{\langle \sigma v\rangle}^{T}_{N_1} &=&{\langle \sigma v\rangle}^{eff}_{N_1}, \nonumber\\
{\langle \sigma v\rangle}^T_{S} &=&{\langle \sigma v\rangle}_{ S S \rightarrow SM SM}+{\langle \sigma v\rangle}_{ S S \rightarrow \overline{N_i}N_j}.
\label{eq:ann-SN}
\eea
We note here that the relic density of the lighter DM may also get affected by DM-DM conversion when the production from the 
heavier component becomes significantly large and comparable to its annihilation to SM. 
This is to remind again that the parameter space scan performed in the subsequent analysis do not use approximate solutions to the coupled BEQs derived above
but uses the numerical results obtained from the code {\tt MicrOmegas} which nicely captures all the features of individual DM relic density affected by DM-DM conversion.

\subsection{Relic density and direct search outcome}

\begin{figure}[htb!]
$$
 \includegraphics[height=5.0cm]{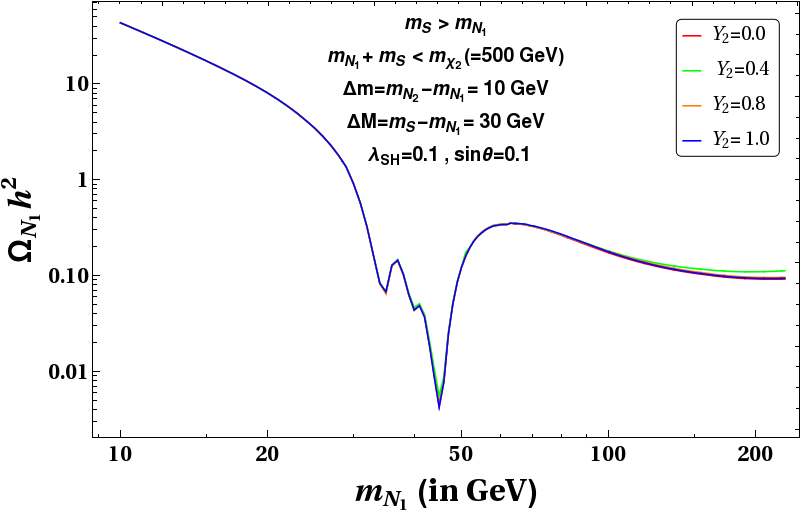}
 \includegraphics[height=5.0cm]{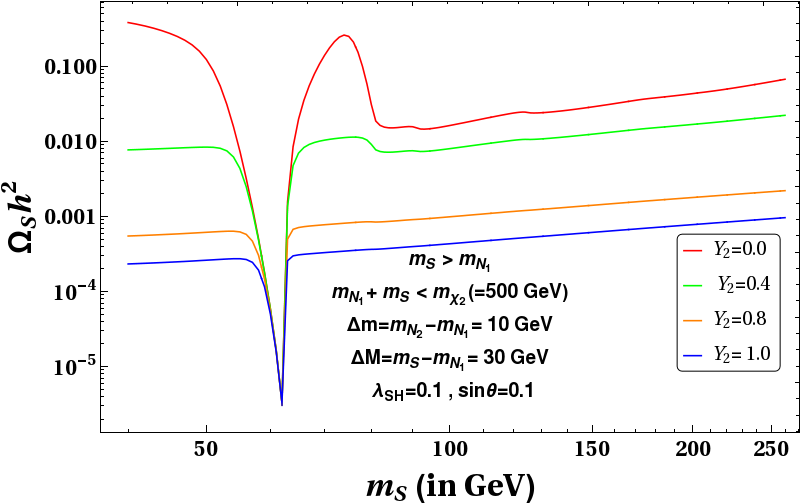}
$$
$$
\includegraphics[height=5.0cm]{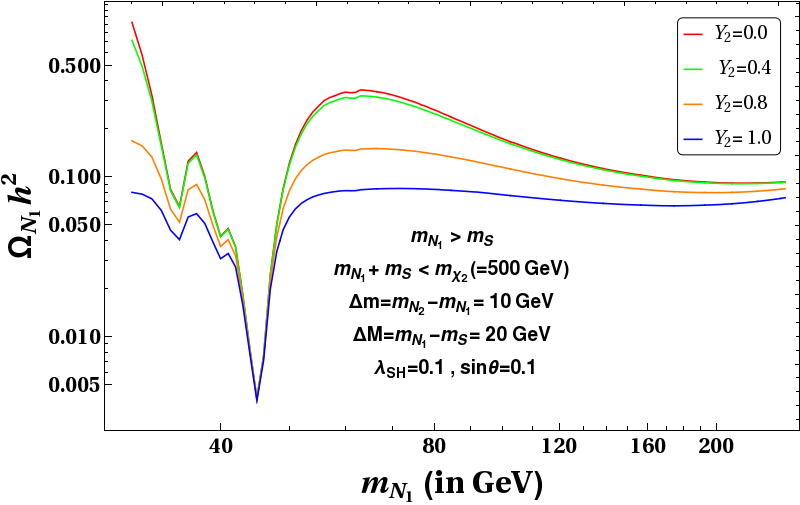}
\includegraphics[height=5.0cm]{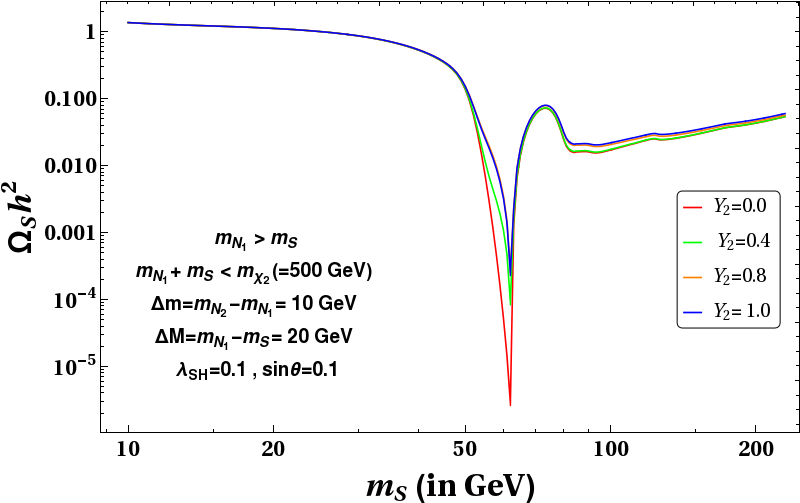} 
$$
 \caption{Relic densities $(\Omega_{N_1}h^2 \rm ~and~ \Omega_{S}h^2 )$ of the individual DM components as a function of respective DM masses $(m_{N_1} \rm ~and~ m_s)$ shown in left and right panel respectively. 
 Two possible mass hierarchies are shown: $m_S > m_{N_1}$ (top panel) and $m_{N_1} > m_{S}$ (bottom panel). Different values of $Y_2=0.0 \rm ~(red)~,0.4 \rm ~(green)~,0.8 \rm ~(orange)~,1.0 \rm ~(blue)~$ 
 are chosen keeping other parameters fixed (as mentioned in the plots) to decipher DM-DM interactions. }
 \label{fig:relic-2-comp}
\end{figure}

 \begin{figure}[htb!]
$$
 \includegraphics[height=5.0cm]{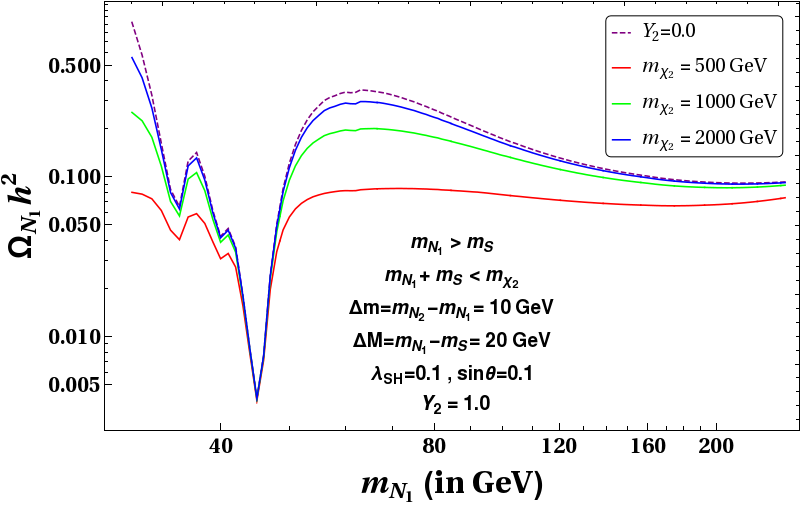}
 \includegraphics[height=5.0cm]{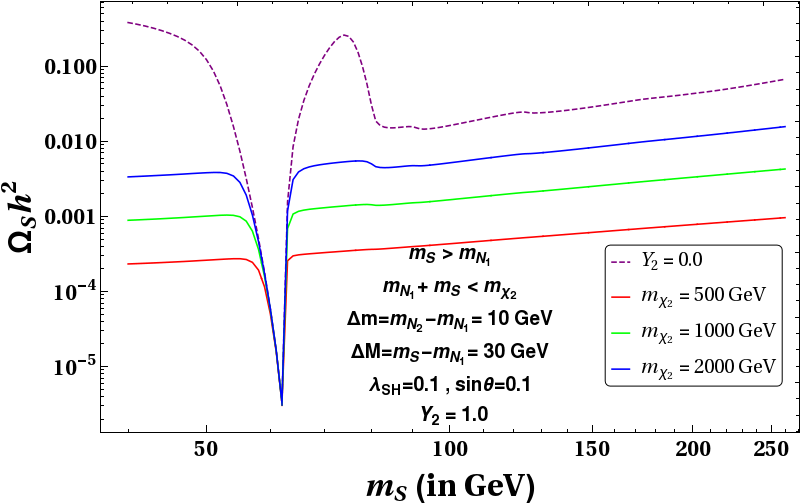}
$$
 \caption{Sensitivity of mediator mass ($m_{\chi_2}$) to DM-DM conversion and that to relic density of the heavier component is demonstrated. 
 [Left] $\Omega_{N_1}h^2$ as a function of $m_{N_1}$ for different values of $m_{\chi_2}= 500 \rm ~(red)~,1000 \rm ~(green)~,2000 \rm ~GeV~ \rm ~(blue)~$ assuming $m_{N_1}>m_S$. 
 [Right]  $\Omega_{S}h^2$ as a function of $m_{S}$ for $m_S>m_{N_1}$. Other parameters kept fixed at different values are mentioned in the plot along with $Y_2=1.0$ and $\lambda_{SH}=0.1$. 
 Purple dotted line in both graphs correspond to $Y_2 = 0$ case, shown for comparison. }
 \label{fig:relic-2-comp-diff-mx2}
\end{figure}

We first study the variation of individual relic densities with corresponding DM masses as shown in Fig.~\ref{fig:relic-2-comp}. 
Two possible mass hierarchies are shown; in top we choose $m_S>m_{N_1}$ and in the bottom panel we have $m_S<m_{N_1}$. 
Relic density of fermion DM ($N_1$) is shown in the left panel and that of the scalar ($S$) is shown in the right panel. 
We see that for $m_S>m_{N_1}$ (top left panel of Fig.~\ref{fig:relic-2-comp}), 
$\Omega_{N_1} h^2$ do not change with different choices of Yukawa coupling $Y_2$. However with same hierarchy ($m_S>m_{N_1}$) for $S$, 
relic density is steadily reduced with larger choice of $Y_2$ (top right panel). It is exactly the other way round, when we have $m_S<m_{N_1}$ 
(bottom panel of Fig.~\ref{fig:relic-2-comp}). In such a case, relic density for $N_1$ decreases with larger $Y_2$, while it remains unaltered for $S$. 
This follows from the analytic solution of the effective annihilation cross sections as mentioned in Eqs.~\ref{eq:ann-NS1}, \ref{eq:ann-NS2}, 
\ref{eq:ann-SN} showing the importance of DM-DM conversion. In this plot we have kept other parameters fixed as mentioned in the plot, 
particularly with a moderate value of the mediator mass fixed at $m_{\chi_2}=500$ GeV.

The sensitivity of individual relic densities to DM-DM conversion as a function of mediator mass ($m_{\chi_2}$) is shown in 
Fig.~\ref{fig:relic-2-comp-diff-mx2}. Evidently, we demonstrate it for the heavier component ($N_1$ on the left and $S$ on the right) 
with different choices of mediator masses: $m_{\chi_2}$: 500 (red), 1000 (green) and 2000 (blue) GeV, keeping $Y_2=1.0$ and $\lambda_{SH}=0.1$ fixed. 
In the same Fig.~\ref{fig:relic-2-comp-diff-mx2}, we have also demonstrated the case of $Y_2 = 0$ (purple dotted line), 
when $\chi_2$ does not take part in the DM-DM conversions. It is evident that with large $m_{\chi_2}$, 
DM-DM conversion becomes feeble and closely resembles $Y_2 = 0$ (purple dotted line) case.  
Therefore, large Yukawa $Y_2$ can play an important role in relic density, but with 
not-so-heavy mediator mass ($m_{\chi_2}$). The lighter DM component is again mostly unaffected by DM-DM conversion as has already 
been discussed. One important point to note is the difference between $Y_2=1.0$ and $\lambda_{SH}=0.1$ chosen for illustration. 
This is because $Y_2$ remains unconstrained (excepting for large perturbative limit $\le \sqrt{4\pi}$), while $\lambda_{SH}$ is highly restricted by 
direct search (recall Fig.~\ref{fig:scalar-relic-dd}). We can also see that $m_{\chi_2}=2$ TeV closely mimic $Y_2=0$ case for fermion DM, 
while it does not completely do so for $\Omega_S$. This is because of very small annihilation cross-section of the scalar DM to SM compared 
to DM-DM conversion due to the choice of small $\lambda_{SH}$. 

\begin{figure}[htb!]
$$
 \includegraphics[height=3.5cm]{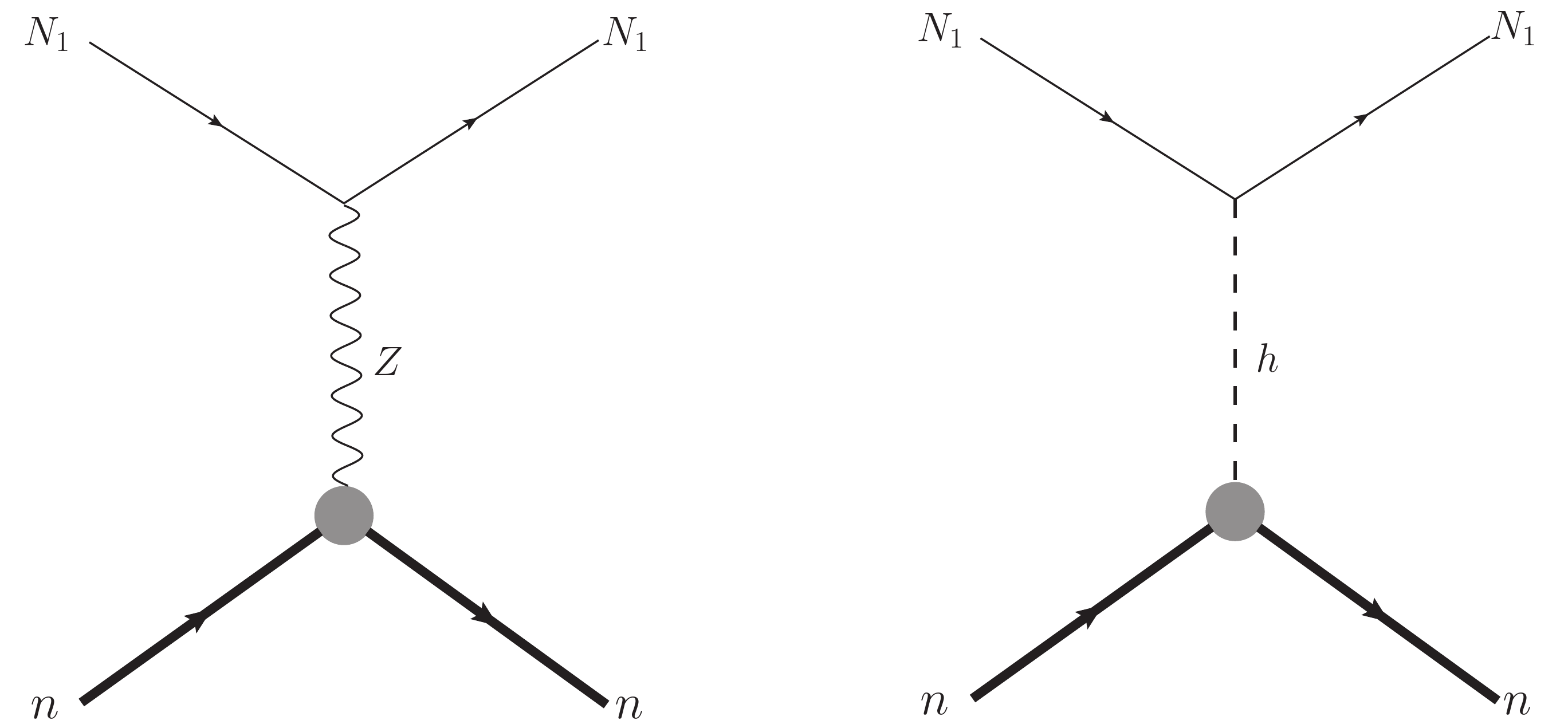}
 $$
 $$
 \includegraphics[height=3.5cm]{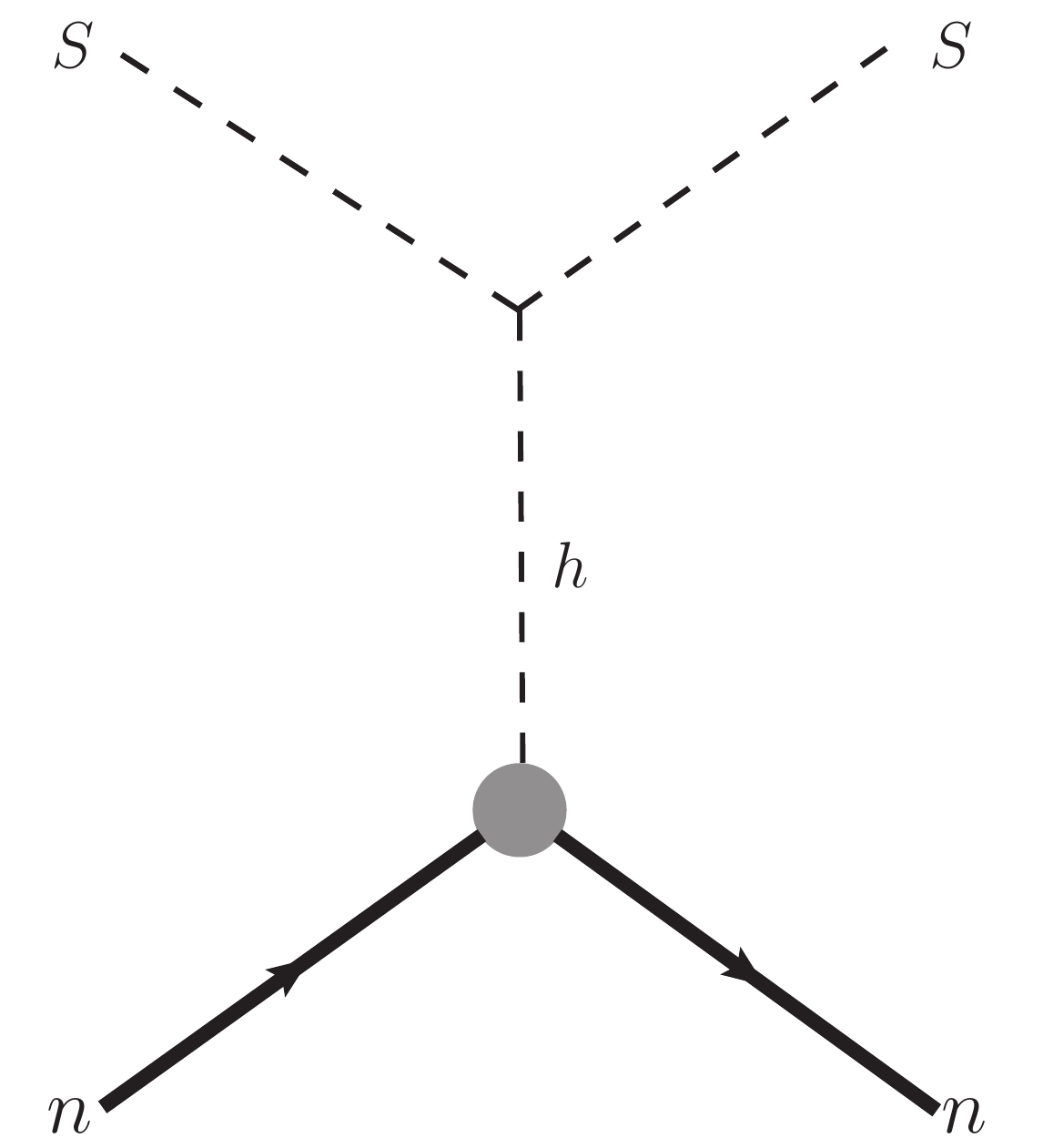}
$$
 \caption{Feynman diagrams of spin independent (SI) direct detection of fermion DM (top panel) and scalar DM (bottom panel).}
 \label{fig:fd-direct}
\end{figure}

Let us now turn to direct search constraints of this two component DM set up. Feynman graphs for direct search contribution of the DM components are shown in Fig.~\ref{fig:fd-direct}. Fermion DM ($N_1$) has both $Z$ and Higgs mediated interaction, while the scalar ($S$) interacts only through Higgs mediation. Direct search cross-sections for individual components are well known; however in two-component set up, the direct search cross-section for each component is folded by their fraction of relic density as\footnote{The actual limit from direct search 
on multipartite DM scenarios need to account for mass sensitivity on the nuclear recoil, the details can be found here ~\cite{Bhattacharya:2016ysw,Ahmed:2017dbb}.}:
\begin{eqnarray}
 \sigma_{eff}^{SI}(S)= \Big(\frac{\Omega_S h^2}{\Omega_T h^2}\Big)\sigma_{S}^{SI}, \label{eq:dd1}
 \end{eqnarray} 
and 
\begin{eqnarray}
 \sigma_{eff}^{SI}(N_1)= \Big(\frac{\Omega_{N_1} h^2}{\Omega_T h^2}\Big)\sigma_{N_1}^{SI}.
 \label{eq:dd2}
 \end{eqnarray} 
 
Spin independent direct search cross-sections for both DM components ($\sigma_{S}^{SI}$ and $\sigma_{N_1}^{SI}$) are obtained from 
inserting the model into the code {\tt MicrOmegas} ~\cite{Belanger:2014vza}. No signal for DM in direct search experiments like LUX~\cite{Akerib:2017kat}, XENON 1T~\cite{Aprile:2018dbl} so far put a strong constraint on the 
WIMP-like DM scenarios as we have here. Recall that scalar DM lives only in the high mass region ($\ge 900$ GeV) except for resonance ($\sim m_h/2$) and fermion DM 
lives in $\sin\theta \le 0.1$ region with a small $\Delta m$ in their single component set up. The question is how much the above conclusions get relaxed 
in a two component set up with large DM-DM conversion as adopted here.

\subsection*{Case I: Feeble DM-DM interactions with $Y_2=0$}

\begin{figure}[htb!]
$$
 \includegraphics[height=5.0cm]{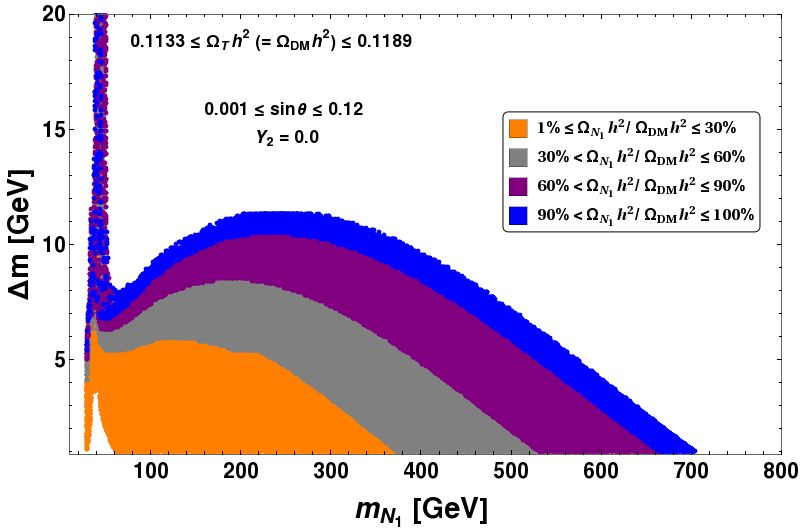}
 \includegraphics[height=5.0cm]{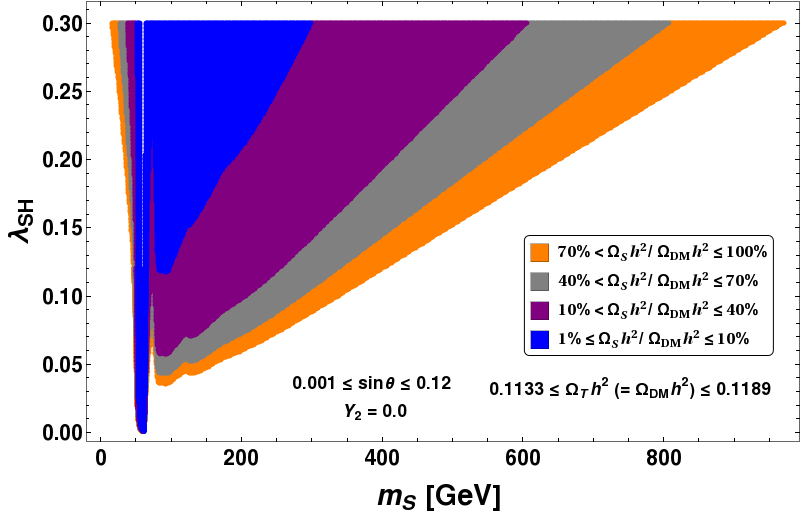}
$$
$$
\includegraphics[height=5.0cm]{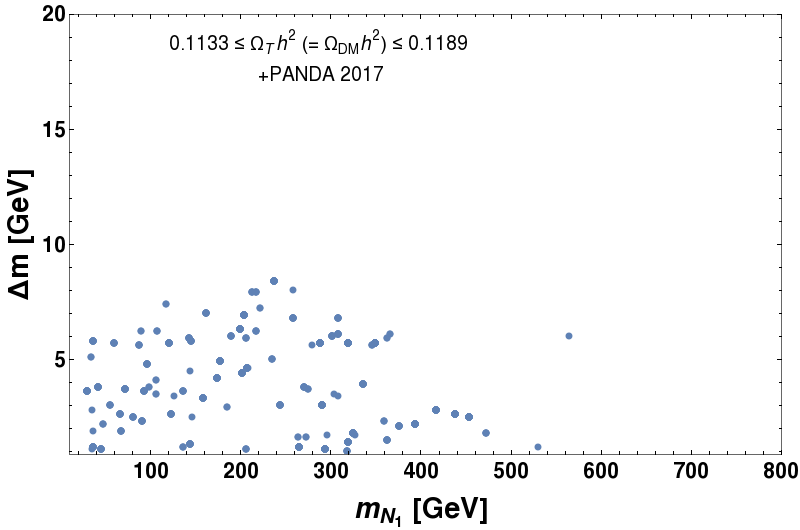}
\includegraphics[height=5.0cm]{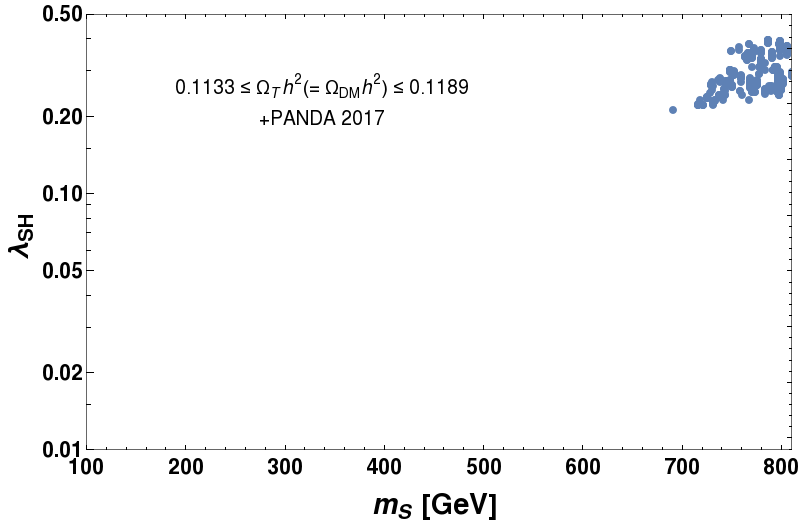} 
$$
 \caption{[Top Panel] Relic density allowed regions of two component DM scenario $\{N_1,S\}$ in $m_{N_1}-\Delta m$ (top left) and $m_S-\lambda_{SH}$ (top right) for $Y_2=0$. 
 Different colour codes indicate the fraction of individual relic density $\frac{\Omega_i}{\Omega_T}$ varied in different ranges as mentioned in inset. 
 [Bottom Panel] Relic density and direct detection (PANDA 2017~\cite{Cui:2017nnn}) allowed parameter space in $m_{N_1}-\Delta m$ (bottom left) and $m_S-\lambda_{SH}$ (bottom right) planes.}
 \label{fig:relic-2comp-y2zero}
\end{figure}

Let us now turn to relic density ($0.1133 \leq \Omega_T h^2\leq 0.1189$) and SI direct search allowed parameter space 
of this two component model. We will first study the case for negligible DM-DM interactions with $Y_2=0$. The results are summarised in 
Fig.~\ref{fig:relic-2comp-y2zero}. We show the relic density allowed parameter space in upper panel, in the left for $N_1$ and  in the right for 
$S$. With $Y_2=0$, the two DM-components behave as if they are decoupled and the allowed parameter space only opens up in the 
under-abundant regions of those individual DMs (compare the single component cases as demonstrated before in Fig.~\ref{fig:vf-relic-dd} and Fig.~\ref{fig:scalar-relic-dd}). 
Different colour codes indicate the percentage of the individual DM density as indicated in the figure inset. It is understood that given a certain percentage 
of one DM, rest of DM relic density is composed of the other component. So any combination is essentially possible by relic density constraint. 
In the bottom panel of Fig.~\ref{fig:relic-2comp-y2zero}, we show the allowed parameter space after direct search constraints from PANDA where both DMs simaltaneously satisfy direct seach bound from PANDA~\cite{Cui:2017nnn}. Note here that there are no parameter space where effective DD cross-section~(in Eqs.~\ref{eq:dd1},~\ref{eq:dd2}) of both $N_1$ and $S$ DM simaltaneously goes beyond recent XENON-1T limit~\cite{Aprile:2015uzo}. For fermion DM, 
direct search allowed parameter space spans the whole of under-abundant parameter space as it doesn't constrain the small $\Delta m$ region further 
with small singlet-doublet mixing $\sin\theta \le 0.05$, as we have chosen for the scan. 
We have already explained that for fermion DM, direct search crucially controls $\sin\theta$ only, which is well below the 
required cut-off. On the other hand, scalar DM is severely constrained by direct search constraint in $m_S-\lambda_{SH}$ plane, 
which leaves Higgs resonance (not shown in the plot) and heavy Scalar DM mass region ($m_S \ge 900$ GeV) only. 
In the heavy scalar mass region, the relic density is $\gsim 80 \%$, therefore allowing only a tiny fraction of fermion DM. 
The whole analysis at $Y_2=0$ also shows that the presence of $s$-channel Higgs mediated DM-DM interactions to be very feeble to 
alter the freeze-out of any of the DM component as mentioned earlier.     

\subsection*{Case II: The case of DM-DM interactions with $Y_2 \ne 0$}

\begin{figure}[htb!]
$$
 \includegraphics[height=4.8cm]{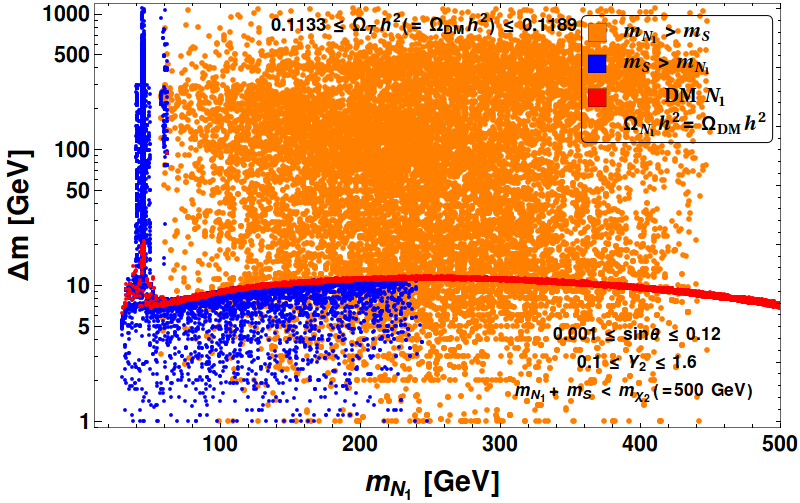}
 \includegraphics[height=4.8cm]{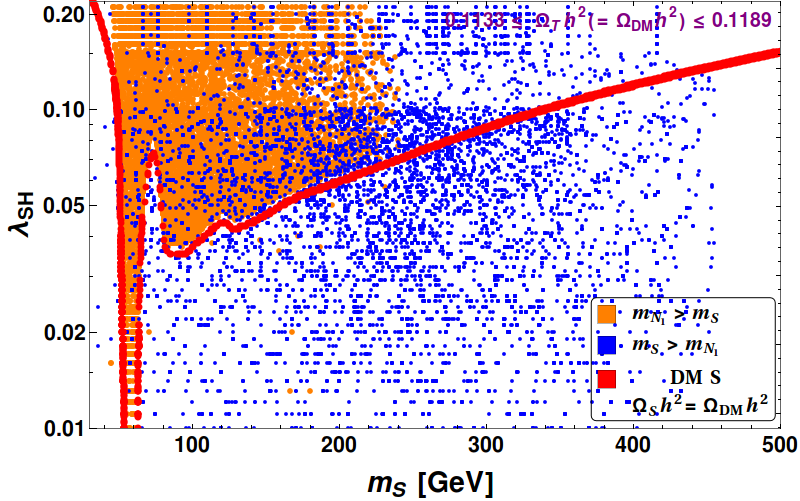}
$$
$$
 \includegraphics[height=4.7cm]{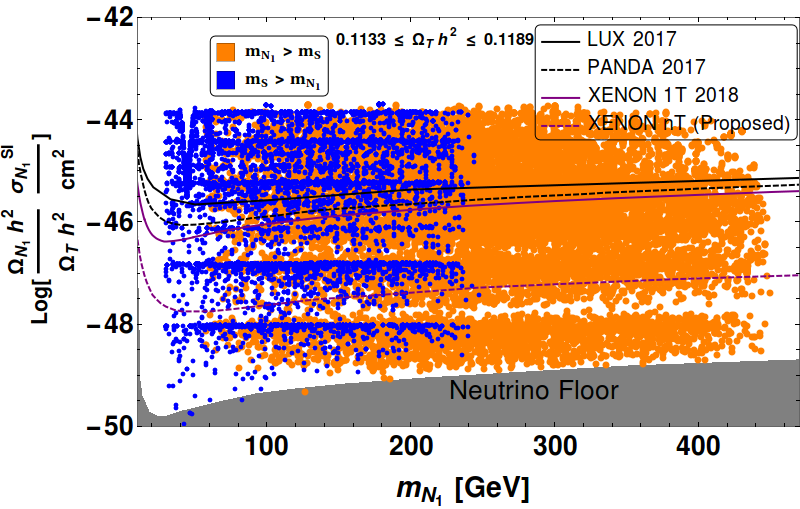}
 \includegraphics[height=4.7cm]{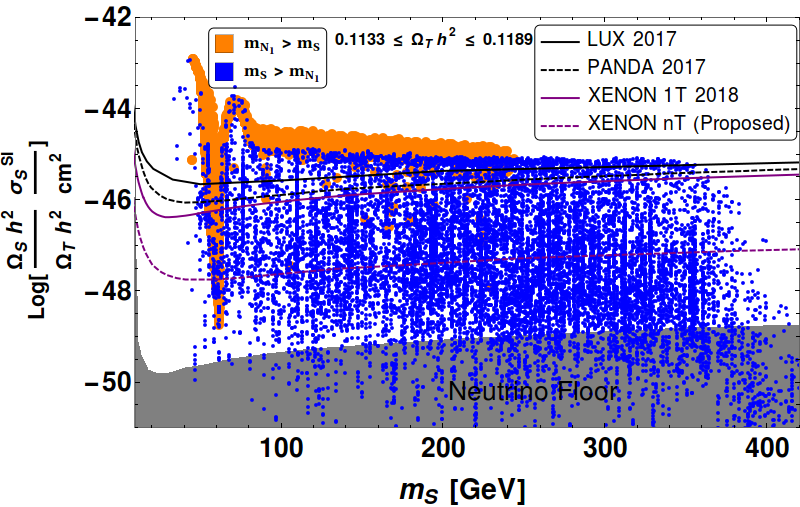}
$$
 \caption{[Top Panel] Relic density allowed parameter space for two component DM model in $m_{N_1}-\Delta m$ (top left) and 
 $m_S-\lambda_{SH}$ (top right) planes. Two mass hierarchies : $m_{N_1} > m_S$ (orange points) and $m_S > m_{N_1}$ (blue points) are shown in both plots. 
 Red points depict the case of single component DM scenarios, for $N_1$ on the left and for $S$ on the right panel.  
 [Bottom Panel] Relic density allowed points are shown in DM mass vs 
 effective SI DM-nucleon cross-section planes; $\Big(\frac{\Omega_{N_1}h^2}{\Omega_{T}h^2}\Big) \sigma_{N_1}^{SI}-m_{N_1}$ in bottom left and 
 $\Big(\frac{\Omega_{S}h^2}{\Omega_{T}h^2}\Big)\sigma_{S}^{SI}-m_S$ in bottom right. Limits from different DD experiments, 
 LUX~\cite{Akerib:2017kat}(black solid line), recent PANDA~\cite{Cui:2017nnn} (black dashed), XENON 1T~\cite{Aprile:2018dbl} (purple solid line) and predicted XENON nT~\cite{Aprile:2015uzo} (purple dotted line) are also indicated in the figures. Shadded region correspond to Neutrino floor where DM signal cannot be distinguish from neutrino background.}
 \label{fig:relic-2comp-y2}
\end{figure}

In Fig.~\ref{fig:relic-2comp-y2}, we show the relic density and direct search allowed parameter space of the model with a non-zero Yukawa coupling ($Y_2 \ne 0$, $0.1\leq Y_2 \leq 1.6$). 
Relic density allowed parameter space is shown in the upper panel for $N_1$ (in $m_{N_1}-\Delta m$ plane) on left and for $S$ (in $m_S-\lambda_{SH}$ plane) on right. 
Both possible mass hierarchies are studied and depicted; (i) $m_{N_1} > m_S$ by orange and (ii) $m_S > m_{N_1}$ by blue points. 
We see that when $m_S>m_{N_1}$, the whole $m_S-\lambda_{SH}$ parameter space is allowed (blue points in top right plot), 
where smaller $\lambda_{SH}$ is substituted by larger $Y_2$ appropriately. On the other hand,  ${N_1}$ DM has the fate of single 
component DM with under abundance adjusted to the other component when $m_S > m_{N_1}$ (blue points in top left plot). This is exactly the other way round, when we choose 
$m_S<m_{N_1}$; the whole $m_{N_1}-\Delta m$ plane becomes allowed (orange points in top left plot) and $S$ has the fate of single component DM filling the under abundance 
region (orange points in top right plot). This is possible because of DM-DM conversion that we introduced in this model through the heavy mediator $\chi_2$ with $Y_2$ Yukawa interaction. 
With $m_{N_1}>m_S$, the effective annihilation required for fermion DM to acquire required relic density ($\Omega_{N_1}h^2<\Omega_{DM}h^2$) in small $\sin\theta$ region no longer 
depends on small $\Delta m$ through co-annihilation because of additional annihilation channel to scalar DM. We will focus on this particular case for collider signatures of this model at the LHC. 
In the bottom panel of Fig.~\ref{fig:relic-2comp-y2}, we show the effective SI direct search cross-section for both DM components at relic density allowed points 
($\Omega_ih^2<\Omega_{DM}h^2$) for both the mass hierarchies. The limits from LUX~\cite{Akerib:2017kat}, PANDA~\cite{Cui:2017nnn}, XENON 1T~\cite{Aprile:2018dbl} and XENON nT~\cite{Aprile:2015uzo} are shown. The plots in the bottom panel point out 
to a larger available parameter space for the heavier DM component. This is simply due to freeze-out of the heavier component being governed by DM-DM conversion, 
not affecting direct search significantly. The scans in Fig.~\ref{fig:relic-2comp-y2} are limited to DM mass within $\lsim$ 500 GeV as it has been done for a mediator mass 
$m_{\chi_2}=500$ GeV to satisfy $m_{N_1}+m_S<m_{\chi_2}$. 

\begin{figure}[htb!]
$$
\includegraphics[height=5.0cm]{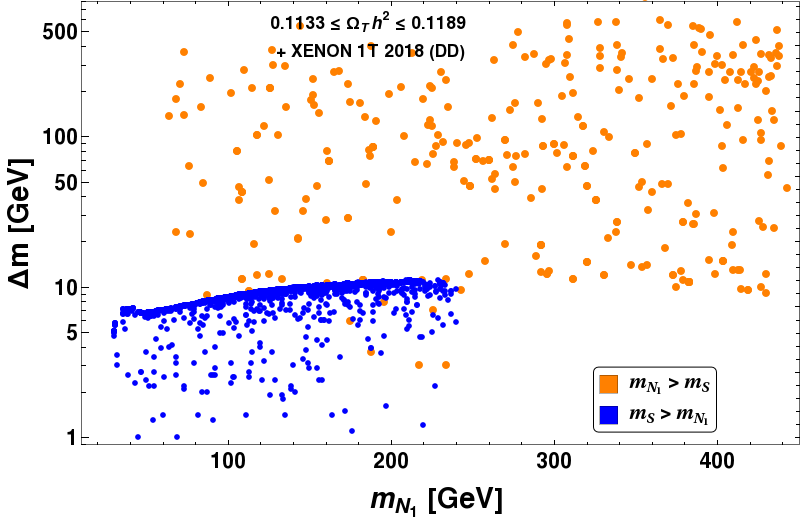}
\includegraphics[height=5.0cm]{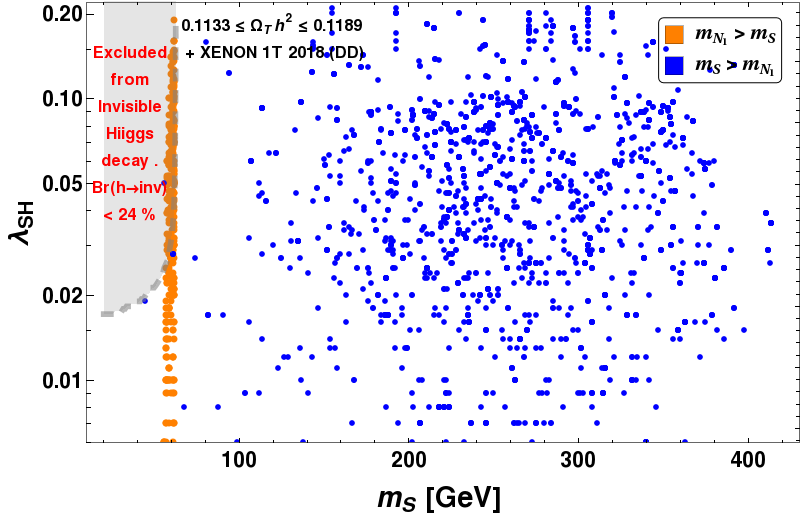} 
$$
 \caption{Relic density and direct search (XENON 1T data~\cite{Aprile:2018dbl}) allowed parameter space is shown for both $N_1$ and $S$ components in $m_{N_1}-\Delta m$ (left) and $m_S-\lambda_{SH}$ (right) plane. Two possible mass hierarchies: $m_{N_1} > m_S$ (orange points) and $m_S > m_{N_1}$ (blue points) are indicated in both planes. Invisible Higgs branching constraint is also shown in the right panel which discards a significant part of $m_S<m_h/2$ region.}
 \label{fig:DD-2comp-XENON1T-y2}
\end{figure}

The outcome of relic density and direct search (XENON 1T limits~\cite{Aprile:2018dbl} from the bottom panel of Fig.~\ref{fig:relic-2comp-y2}) constraints put together yield 
Fig.~\ref{fig:DD-2comp-XENON1T-y2}. The constraints on fermion DM in $m_{N_1}-\Delta m$ plane (left) is obviously less restrictive as we choose small $\sin\theta \lsim 0.05$ 
for the scan, thus allowing the whole parameter space with upto $\Delta m \gsim 500$ or more for $m_{N_1}>m_S$ (orange points in left plot) thanks to conversion to the scalar 
DM. This feature serves as the most interesting phenomenological outcome of this model, as we discuss in collider section. For $m_{N_1}>m_S$, the scalar DM however is 
allowed only in the resonance region ($m_S\sim m_h/2$) as can be seen by orange points in the right plot of Fig.~\ref{fig:DD-2comp-XENON1T-y2}. This is already expected as direct search tames down the relic 
density allowed scalar DM parameter space absent DM-DM conversion. For the reverse hierarchy $m_{N_1}<m_S$ (blue points), fermion DM is allowed only in the 
under-abundant regions of its single component manifestation, whereas it allows a larger mass range of scalar DM, thanks again to the possible DM-DM conversion with a lighter $N_1$. 
Invisible Higgs branching ratio $Br(h \to \rm{inv})<0.24$ ~\cite{Tanabashi:2018oca}, puts a significant constraint for the scalar DM with $m_S<m_h/2$. But for fermion DM, this doesn't discard any 
parameter space given the small values of $\sin\theta$ chosen for the scan. 

\begin{figure}[htb!]
$$
\includegraphics[height=5.0cm]{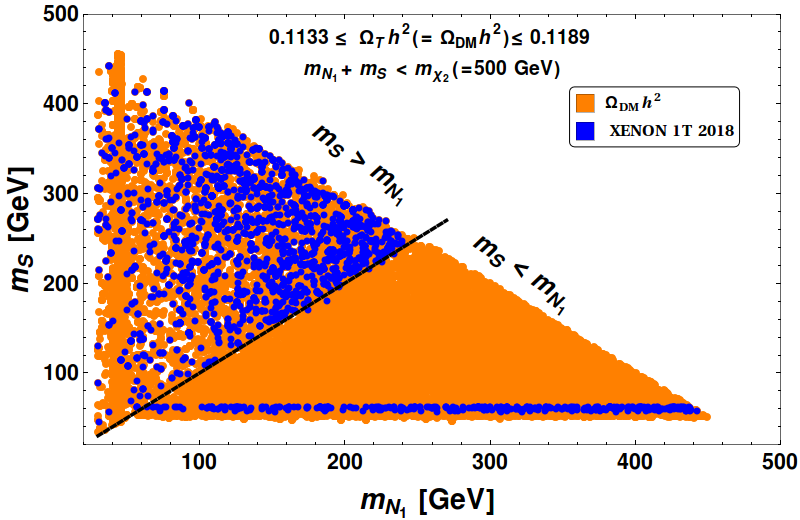}
\includegraphics[height=5.0cm]{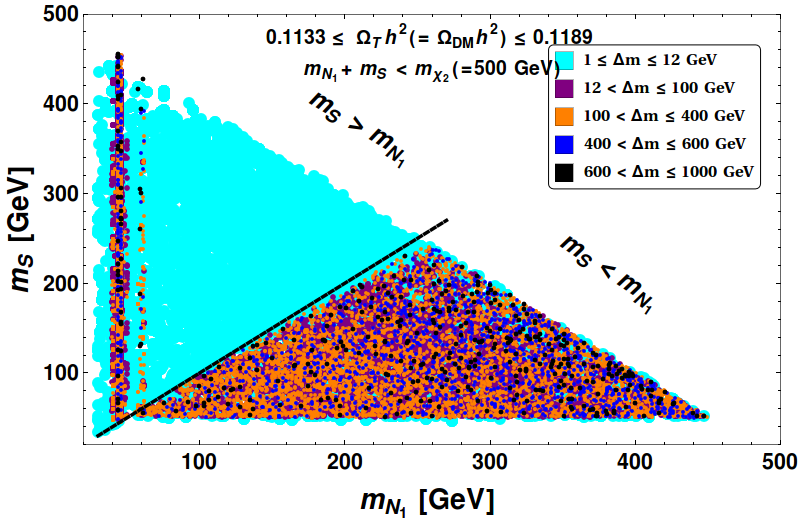} 
$$
 \caption{Mass correlation of the two DM components in $m_{N_1}- m_S$ plane. [Left Panel] Relic density allowed parameter space is shown by orange points and direct search constraint 
 from XENON 1T~\cite{Aprile:2018dbl} on both $N_1$ and $S$ is shown by blue points. Black solid line corresponding to $m_{N_1} ~=~m_S$ segregates the two possible hierarchies: the one above corresponds to 
 $m_S ~>~m_{N_1}$ and the region below has $m_S ~ < ~ m_{N_1}$. [Right Panel] Relic density allowed points for different ranges of $\Delta m$ shown with different colour codes.}
 \label{fig:DD-2comp-mN1-mS-y2}
\end{figure}

A possible mass correlation of these two DM components is studied next and depicted in Fig.~\ref{fig:DD-2comp-mN1-mS-y2} in $m_{N_1}- m_S$ plane for satisfying relic density 
and direct search constraints. On the left panel, we show that the whole triangle designated by the kinematic limit $m_{N_1}+m_S<m_{\chi_2}$, 
with $m_{\chi_2}=500$ GeV chosen for the scan is allowed by relic density constraint. However direct search (XENON1T data~\cite{Aprile:2018dbl}) restricts it significantly for $m_{N_1}>m_S$, 
allowing only scalar DM to lie in resonance $m_h/2$, while it is not that restrictive for the other hierarchy $m_{N_1}<m_S$, as shown by the spread of blue points filling 
almost entirely the upper part of the triangle. The thick black line depicting $m_{N_1} ~=~m_S$ separates these two heirarchies. 
The plot on the right panel shows the allowed points in $m_{N_1}- m_S$ plane to satisfy relic density for different ranges of 
$\Delta m$. It shows that small $1\leq\Delta m\leq12$ GeV is allowed throughout the parameter space while large $\Delta m$ is restricted to $m_{N_1}>m_S$ as we already discussed. 
For $m_{N_1}<m_S$, one can have larger $\Delta m$ allowed only in the resonance region $m_{N_1} \sim m_h/2$ and $\sim m_Z/2$ .

\begin{figure}[htb!]
$$
 \includegraphics[height=5.0cm]{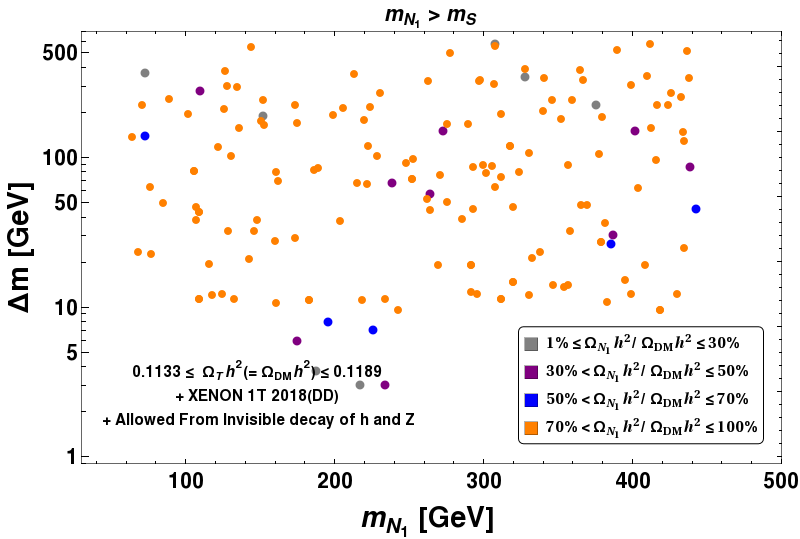}
 \includegraphics[height=5.0cm]{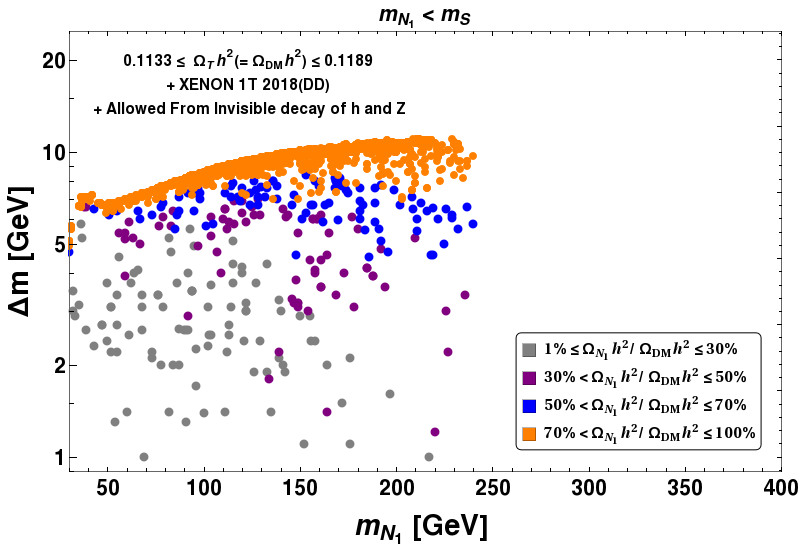}
$$
\caption{Relic density, direct search (XENON 1T~\cite{Aprile:2018dbl}) and invisible decay constrain of Higgs and $Z$ boson~\cite{Tanabashi:2018oca} allowed parameter space of the two component model (with $Y_2 \ne 0$) in $m_{N_1}-\Delta m$ plane. Percentage of fermion DM component 
in total relic density within different ranges are shown by different coloured points as detailed in the figure inset. Two different hierarchies $m_S ~<~m_{N_1}$ and 
$m_S ~ > ~ m_{N_1}$ are shown separately in left and right panel respectively. }
 \label{fig:relic-percent}
\end{figure}

Another important question is to know the percentage of fermion or scalar DM component present in the allowed parameter space of this two component model. We show the outcome of this 
exercise in Fig.~\ref{fig:relic-percent}, for fermion DM in $m_{N_1}-\Delta m$ plane. The other component ($S$) just fills the rest of it and can be gauged from this figure itself. 
Two possible mass hierarchies $m_S ~<~m_{N_1}$ and $m_S ~ > ~ m_{N_1}$ are shown separately in left and right panel respectively. Fermion DM content in total relic density 
(for different ranges in percentage) is shown by different colour codes mentioned in the figure inset. All the points also additionally satisfy direct search constraint from XENON1T data~\cite{Aprile:2018dbl} and invisible decay constraint of Higgs and $Z$~\cite{Tanabashi:2018oca}. 
The bottom line is that for $m_{N_1}>m_S$, the larger share of DM density is carried by fermion DM with $\Delta m \gg 12$ GeV as it becomes enough to bring the annihilation 
in the right ballpark through conversion to the scalar DM component (with small $\sin\theta$), while the scalar DM anyway has a large annihilation cross section 
(and therefore smaller relic density) as it requires to be in the Higgs resonance region ($m_S \sim m_h/2$) to address direct search bound. 
For the other hierarchy $m_{N_1}<m_S$, under abundant regions of the single component fermion DM is filled up with different percentage as the scalar DM has the freedom to 
adjust its relic density through its annihilation to SM plus fermion DM.

 \begin{figure}[htb!]
$$
 \includegraphics[height=5.0cm]{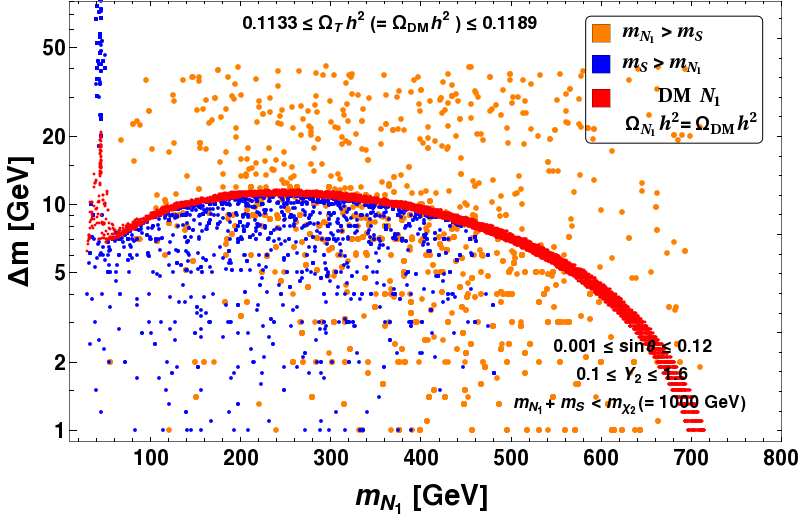}
 \includegraphics[height=5.0cm]{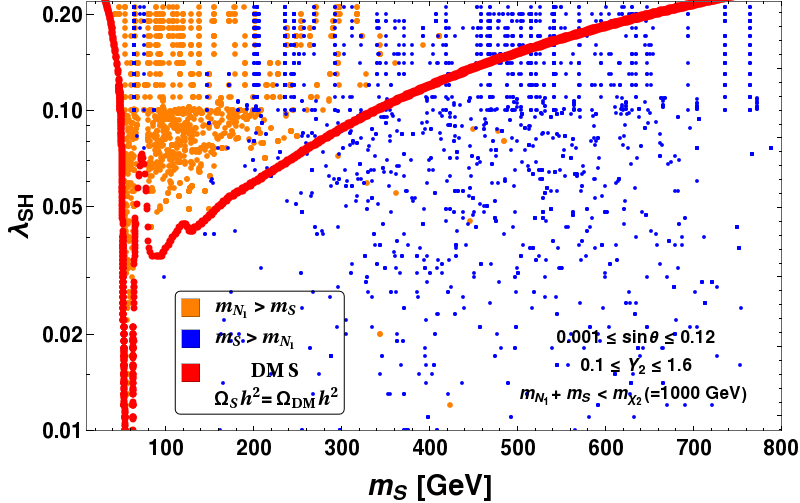}
$$
\caption{Total relic density allowed regions of two component DM model in $m_{N_1}-\Delta m $ (left panel) and $m_S-\lambda_{SH}$ plane (right panel) with $Y_2 \neq 0$ for mediator 
mass $m_{\chi_2} = 1000 $ GeV. Two mass hierarchies are shown in different colour codes: $m_{N_1} > m_S$ (orange points) and $m_S > m_{N_1}$ (blue points). Red points correspond to the 
case of single component DM scenarios for $N_1$ on left and for $S$ on right panel. }
 \label{fig:relic-2comp-y2_mx2_1000}
\end{figure}

\begin{table}[htb!]
\begin{tabular}{|p{1.0cm}|p{4.4cm}|p{0.8cm}|p{1.0cm}|p{1.0cm}|p{2.1 cm}|p{2.1 cm}| }
 \hline
 \hline
  BPs & {$\{~m_{N_1},m_S,\lambda_{SH},Y_2, \sin\theta~\}$}& $\Delta m $ &${\Omega_{N_1}} h^2  $ & ${\Omega_{S}} h^2$ &$\Big(\frac{\Omega_{N_1}h^2}{\Omega_{DM}h^2}\Big) \sigma_{N_1}^{SI}$ (in $cm^2$)  &  $\Big(\frac{\Omega_{S}h^2}{\Omega_{DM}h^2}\Big) \sigma_{S}^{SI}$ (in $cm^2$) \\
 \hline
 \hline
 BPA1&$\{~79,256,0.029,0.2,0.02~\}$ & ~6.1~ & 0.0546 & 0.0641 & $6.8 \times 10^{-48}$  & $5.9 \times 10^{-47}$ \\
 \hline
 BPA2&$\{~276,58,0.010,0.9,0.02~\}$ & ~50~ & 0.1092 & 0.0054 & $1.5 \times 10^{-48}$  & $1.5 \times 10^{-47}$ \\
 \hline
 BPA3&$\{~131,61,0.026,0.9,0.01~\}$ & ~101~ & 0.1171 & 0.0012 & $1.0 \times 10^{-48}$  & $1.5 \times 10^{-47}$ \\
 \hline
 BPA4&$\{~102,62,0.010,0.9,0.02~\}$ & ~193~ & 0.1144 & 0.0010 & $1.8 \times 10^{-47}$  & $2.0 \times 10^{-48}$ \\
 \hline
 BPA5&$\{~135,58,0.004,1.0,0.02~\}$ & ~295~ & 0.0840 & 0.0313 & $1.5 \times 10^{-47}$  & $1.0 \times 10^{-47}$ \\
 \hline
 BPA6&$\{~127,62,0.020,0.9,0.01~\}$ & ~377~ & 0.1136 & 0.0004 & $1.4 \times 10^{-48}$  & $3.0 \times 10^{-48}$ \\
 \hline
 BPA7&$\{~144,62,0.032,0.9,0.02~\}$ & ~541~ & 0.1152 & 0.0002 & $2.7 \times 10^{-47}$  & $4.2 \times 10^{-48}$ \\
 \hline
 \hline
\end{tabular}
\caption{Some benchmark points allowed by relic density, direct search and invisible Higgs and $Z$ decay limit for mediator mass, $m_{\chi_2}=500$ GeV. 
DM masses, couplings, relic density of individual components and effective SI direct search cross-sections are mentioned. 
All the masses are in GeVs. We mainly focus on $m_{N_1}>m_S$ excepting for BPA1.}
\label{tab:BP1}
\end{table}

So far we have discussed the allowed DM parameters space for the model with a moderate choice of mediator mass, $m_{\chi_2}=500$ GeV. 
Now we choose a higher value of $\chi_2$ mass, $m_{\chi_2}=1000$ GeV to depict relic density allowed limit in $m_{N_1}-\Delta m$ plane (left) and $m_S-\lambda_{SH}$ plane (right) of 
Fig.~\ref{fig:relic-2comp-y2_mx2_1000}. Allowed parameter space in $m_{N_1}-\Delta m$ plane becomes more restrictive ($\Delta m$ spanning roughly upto 
$\sim 50$ GeV compared to 500 GeV with $m_{\chi_2}=500$ GeV) even with $m_{N_1} > m_{S}$ due to suppressed t-channel DM-DM conversion 
$\overline{N_i} N_j \rightarrow S S$ due to the heavy mediator ($m_{\chi_2}$). Comparatively, larger parameter space is available for scalar DM $S$ as shown in right 
panel of Fig.~\ref{fig:relic-2comp-y2_mx2_1000} in $m_S-\lambda_{SH}$ plane. This is possible as $S S \rightarrow \overline{N_i} N_j$ with $(Y_2 \neq 0.0)$ still 
dominate over scalar DM annihilation to SM (controlled by portal coupling $\lambda_{SH}$) even with a heavy mediator mass. 
This feature has already been pointed out while discussing the outcome of DM-DM conversion cross-sections in Fig.~\ref{fig:relic-2-comp-diff-mx2}. 

Finally, to summarise the main outcome of the DM analysis is to see that heavier DM component enjoys annihilation to lighter DM for thermal freeze out, 
relaxing its interaction to visible sector and thus reducing the constraints from direct search cross-sections. Specifically for the two-component case, 
when scalar DM is heavier than the fermion DM, the Higgs portal coupling can be reduced significantly allowing the scalar DM to be allowed through 
the entire DM mass plane. On the other hand, when the fermion DM is heavier than the scalar DM, it relaxes the mass difference with the charge companion, 
allowing larger $\Delta m$. No relaxation is possible however for $\sin\theta$ as larger values of mixing is still discarded by $Z$ mediated direct search. 
The relaxation of $\Delta m$ plays a crucial role in achieving collider signatures of fermion DM as we illustrate next. 
We choose a set of benchmark points allowed by relic density and direct search in Table \ref{tab:BP1} for performing collider analysis, where above features are apparent.

\section{Two Component DM in presence of additional heavy scalar}
\label{two_component_DM_SH}

In the two component DM set up, lighter DM component behaves almost as a single component candidate, 
due to the absence of additional channels for annihilation, thus occupying only under abundant regions accessible from relic density. 
For $m_{N_1} > m_{S}$, a large mass splitting $\Delta m$ can be achieved  for a moderated value of mediation mass $m_{\chi_2} \sim 500$ GeV. 
But at the same time, the scalar DM can only be accommodated at the resonance region, $m_S \approx \frac{m_H}{2}$ (see Fig.~\ref{fig:DD-2comp-XENON1T-y2} and Table~\ref{tab:BP1}). 
This is predictive and restrictive at the same time. This situation however alters significantly if the scalar sector is enlarged with a heavy real scalar 
$S_H$ which has same charge like $S$ under $\mathcal{Z}_2 \times \mathcal{Z}_2^\prime$  as: $S ~ [- , -] ~\rm and~ S_H ~[-,-]$~\cite{Casas:2017jjg,Bhattacharya:2017fid}. 
We briefly discuss such a possibility here. The relevant interacting scalar potential is given by: 

\bea\label{potSSH}
V(S,S_H)& \supset & \frac{1}{2} m_S^2 S^2 + \frac{1}{2} m_{S_H}^2 S_H^2 + \frac{\lambda_{SH}}{2} \Big(H^\dagger H- \frac{v^2}{2}\Big) (S^2 + S_H^2 ) + \lambda_{CH} \Big(H^\dagger H- \frac{v^2}{2}\Big) S  S_H , \nonumber \\
\eea

where $m_{S_H}$ is the heavy scalar mass and $\lambda_{CH}$ is additional (co-annihilation type) Higgs portal coupling. Due to the presence of this interaction, 
$\lambda_{CH} \Big(H^\dagger H \Big) S  S_H$, a new co-annihilation channel, $S~ S_H \rightarrow SM ~SM$ opens up. 
$S_H$ having same charge as of $S$, is not stable and therefore is not a DM. But the possibility of co-annihilation provides additional channel for scalar DM to 
freeze out, while it does not contribute to direct search. This is similar to the co-annihilation processes already present in the fermion DM sector. With this, 
even for $m_{N_1} > m_S$, the scalar DM can be allowed in a large parameter space beyond resonance. Presence of this heavy scalar, also augments 
dark sector Yukawa interaction providing additional contribution to DM-DM conversion: 
\bea
\mathcal{L_{DM}}^{Yuk} &=&  - Y_2(\overline{\chi_1} \chi_2 S + h.c) -Y^{'}_2(\overline{\chi_1} \chi_2 S_H +h.c ).
\eea
In our numerical analysis, we assume $Y_2=Y^{'}_2$ for simplicity.

\begin{figure}[htb!]
$$
 \includegraphics[height=4.8cm]{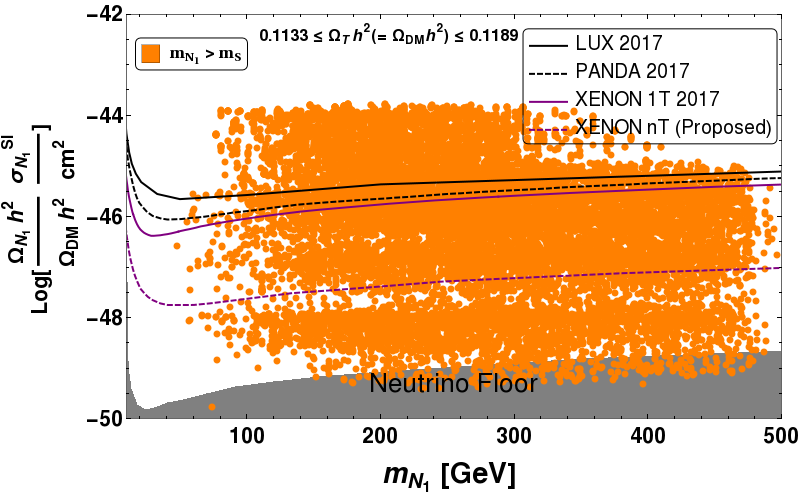}
 \includegraphics[height=4.8cm]{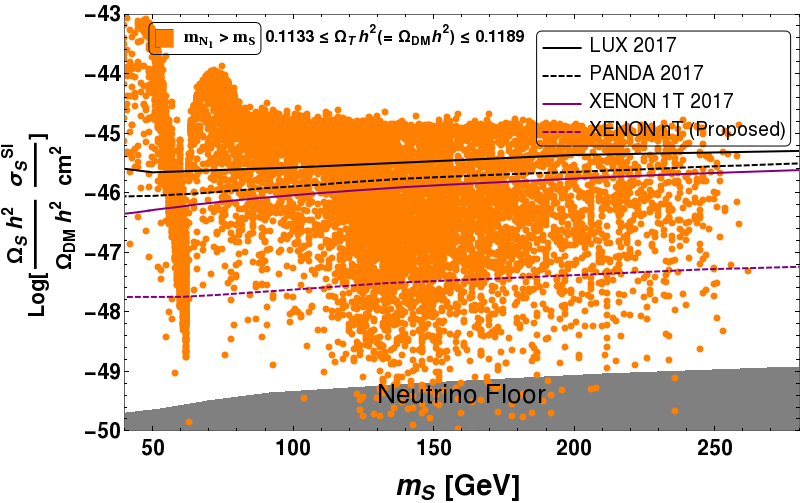}
$$
 \caption{Relic density allowed points plotted in DM mass vs effective SI DM-nucleon cross-section plane: 
 $\Big(\frac{\Omega_{N_1}h^2}{\Omega_{DM}h^2}\Big) \sigma_{N_1}^{SI}-m_{N_1}$ plane (left) and 
 $\Big(\frac{\Omega_{S}h^2}{\Omega_{DM}h^2}\Big)\sigma_{S}^{SI}-m_S$ plane (right) in presence of heavy scalar $S_H$. Upper bounds on SI DM-nucleon 
 cross-section from LUX~\cite{Akerib:2017kat} (black solid line), recent PANDA~\cite{Cui:2017nnn} (black dashed), XENON 1T~\cite{Aprile:2018dbl} (purple solid line) and predicted XENON nT~\cite{Aprile:2015uzo} (purple dotted line) are also indicated in the figures. We have chosen the mass hierarchy: $m_{N_1} > m_S$ and the mediator mass $m_{\chi_2} = 500$ GeV.}
 \label{fig:relic-2comp-SH-y2}
\end{figure}

The first outcome of this extended two component framework is to show a large parameter space available to the scalar DM through relic density and direct search bounds with 
the hierarchy $m_{N_1} > m_S$. This is illustrated in Fig.~\ref{fig:relic-2comp-SH-y2}. The direct search cross-section for fermion DM in relic density allowed points is shown on the 
left plot, while that for the scalar is shown in the right panel. We see in the right plot that orange points now span allover the plane with a huge number 
below the direct search limit unlike being only available in resonance region with the previous case (compare bottom right plot of Fig.~\ref{fig:relic-2comp-y2}).

\begin{figure}[htb!]
$$
\includegraphics[height=5.0cm]{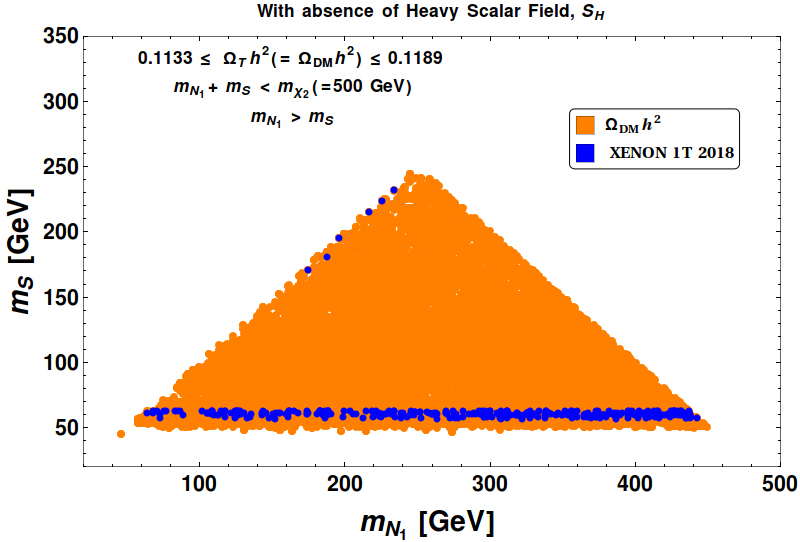}~
\includegraphics[height=5.0cm]{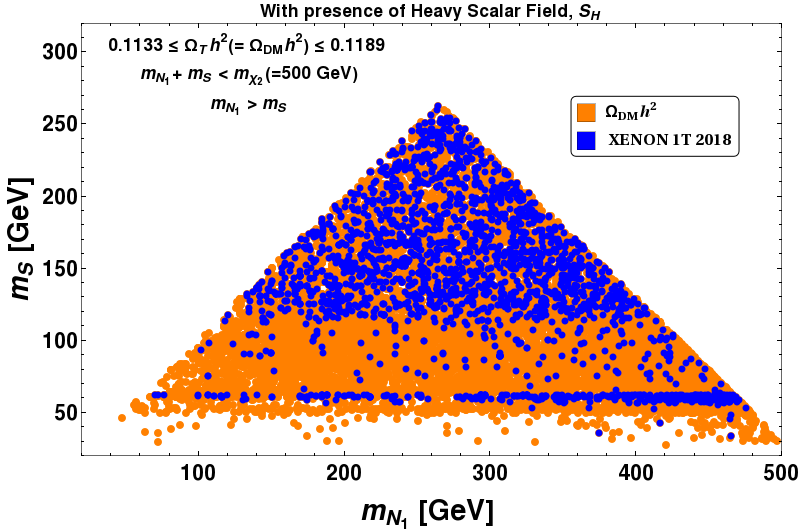} 
$$
\caption{Allowed region of parameters space in $m_{N_1}-m_S$ plane, which satisfy relic density (Orange points) and direct search constraints for both $N_1$ and $S$ by 
XENON1T data (blue points) for  $m_{N_1} > m_S$. In left panel, we show the original two component scenario in absence of heavy scalar ($S_H$) and in the right panel, 
we show it in presence of additional heavy scalar $S_H$. }
\label{fig:DD-2comp-mass}
\end{figure}

\begin{figure}[htb!]
$$
 \includegraphics[height=5.0cm]{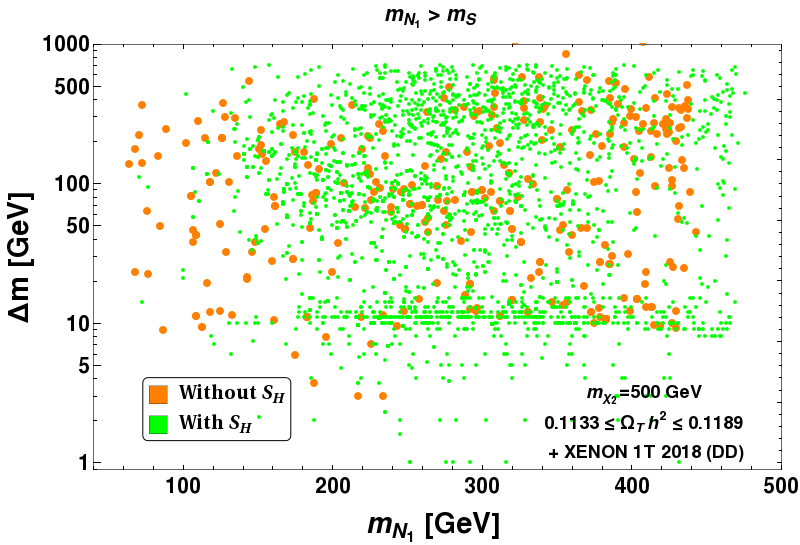}~
 \includegraphics[height=5.0cm]{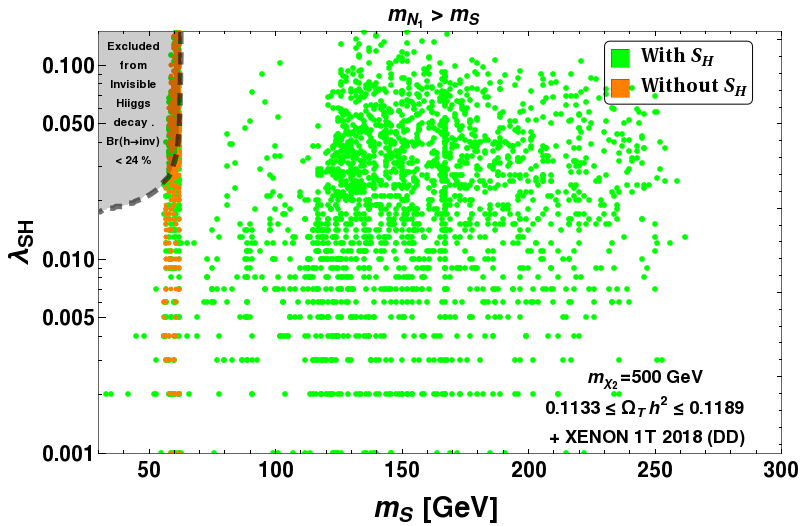}
$$
\caption{Relic density and direct search (XENON 1T~\cite{Aprile:2018dbl}) allowed parameter space of the two component model compared between two cases: 
(i) original model, in absence of the heavy scalar ($S_H$) (orange points) and (ii) in presence of a heavy scalar ($S_H$) (green points) in 
$m_{N_1}-\Delta m$ plane (left) and $m_S-\lambda_{SH}$ plane (right).}
\label{fig:compare-SH-2}
\end{figure} 

A mass correlation for two DM components is shown in Fig.~\ref{fig:DD-2comp-mass} and compared between the original framework (left) to that in presence of an 
additional heavy scalar (right). We show that with $m_{N_1} > m_S$, presence of co-annihilation in the scalar sector allows the scalar DM to be present in a 
larger parameter space after satisfying direct search constraints (plot on the right panel).

Finally, we compare relic density and direct search (XENON 1T~\cite{Aprile:2018dbl}) allowed parameter space of the two component model in presence of 
$S_H$ (green points) to that in absence of $S_H$ (orange points) for both fermion DM and scalar DM components in Fig.\ref{fig:compare-SH-2}. 
As expected, we see that for fermion DM, in $m_{N_1}-\Delta m$ plane (on left panel), there is no difference between these two cases, 
while for the scalar DM $S$, the presence of the heavy scalar $S_H$ allows almost all of the plotted parameter space (green points on the right panel) due to coannihilation. 
A few benchmark points are indicated in Table~\ref{tab:BP2} to show the effect of relaxing the case for scalar DM in presence of $S_H$ for $m_{N_1} > m_S$. 
They should be contrasted with those in Table~\ref{tab:BP1}. 


\begin{table}[htb!]
\begin{tabular}{|p{0.9cm}|p{5.8cm}|p{0.7cm}|p{1.0cm}|p{1.0cm}|p{1.96 cm}|p{1.96 cm}| }
 \hline
 \hline
  BPs~~ & {$\{m_{N_1},m_S,m_{S_H},\lambda_{SH},\lambda_{CH},Y_2, \sin\theta \}$}& $\Delta m $ &${\Omega_{N_1}} h^2  $ & ${\Omega_{S}} h^2$ &$\Big(\frac{\Omega_{N_1}h^2}{\Omega_{DM}h^2}\Big) \sigma_{N_1}^{SI}$ (in $cm^2$)  &  $\Big(\frac{\Omega_{S}h^2}{\Omega_{DM}h^2}\Big) \sigma_{S}^{SI}$ (in $cm^2$) \\
  \hline
  \hline
 BPB1&$\{200,~150,~205,0.006,~0.8,~0.7,0.01\}$ & ~50~ & 0.1144 & 0.0030 & $9.7 \times 10^{-49}$  & $3.3 \times 10^{-49}$ \\
   \hline
 BPB2&$\{202,~118,~131,0.002,~0.3,~0.9,0.04\}$ & ~101~ & 0.0303 & 0.0838 & $7.2 \times 10^{-47}$  & $1.5 \times 10^{-48}$ \\
   \hline
 BPB3&$\{183,~113,~135,0.009,~0.8,~0.8,0.04\}$ & ~201~ & 0.0462 & 0.0680 & $1.2 \times 10^{-46}$  & $3.0 \times 10^{-47}$ \\
   \hline
 BPB4&$\{310,~153,~203,0.052,~0.6,~0.7,0.02\}$ & ~300~ & 0.1112 & 0.0100 & $2.0 \times 10^{-47}$  & $5.8 \times 10^{-47}$ \\
   \hline
 BPB5&$\{424,~91,~109,0.004,~0.6,~1.1,0.03\}$ & ~503~ & 0.0238 & 0.0945 & $2.7 \times 10^{-47}$  & $1.1 \times 10^{-47}$ \\
 \hline
 \hline
\end{tabular}
\caption{Benchmark points allowed by relic density, direct search and invisible Higgs decay limit in presence of a heavy scalar $S_H$. Input parameters 
(masses and couplings), relic densities of individual components and direct search cross-sections are mentioned. All the masses are in GeVs.}
  \label{tab:BP2}
\end{table}  

\section{Collider searches at LHC}

\begin{figure}[htb!]
$$
 \includegraphics[height=4.0cm]{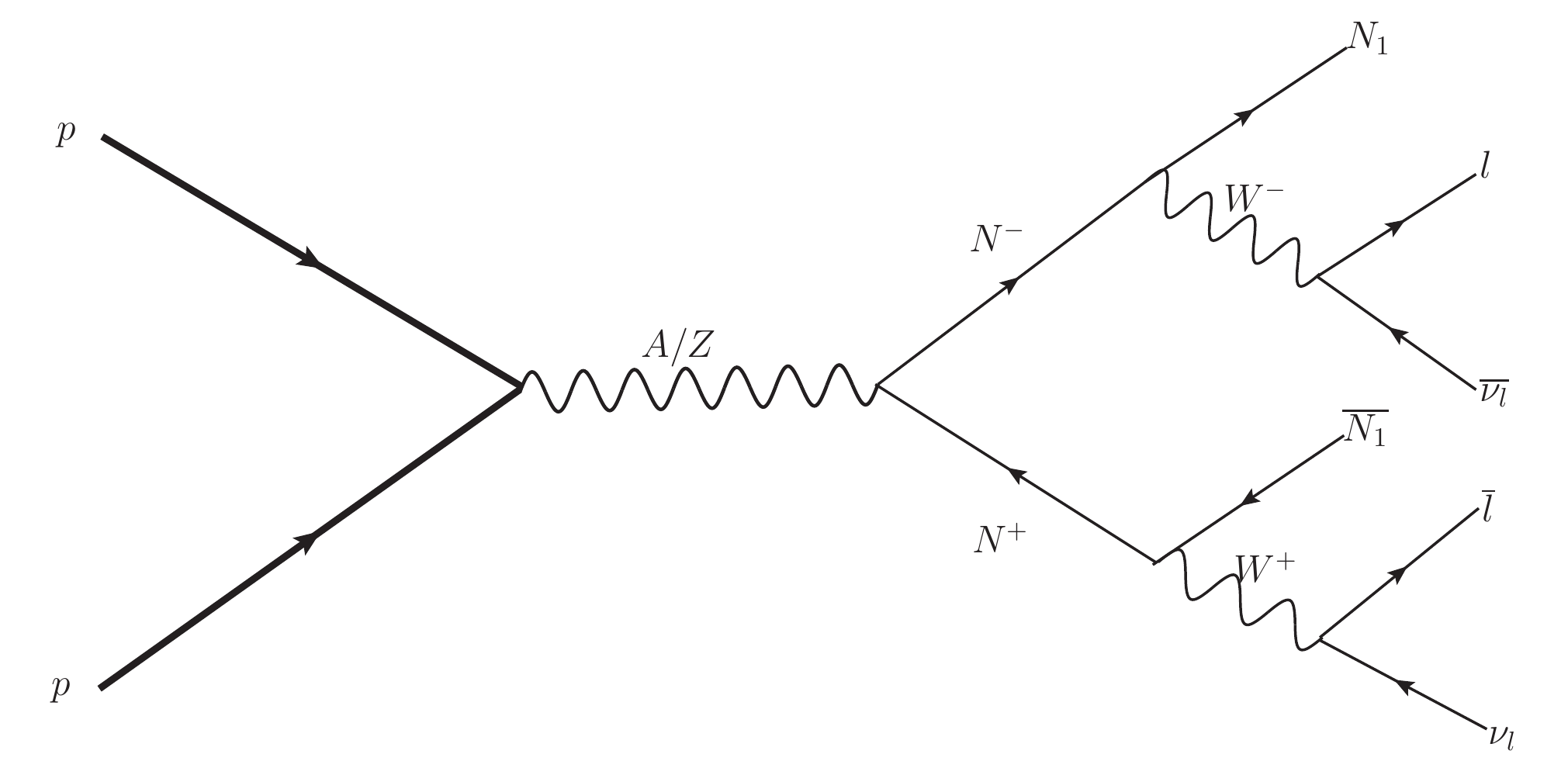}
 \includegraphics[height=5.0cm]{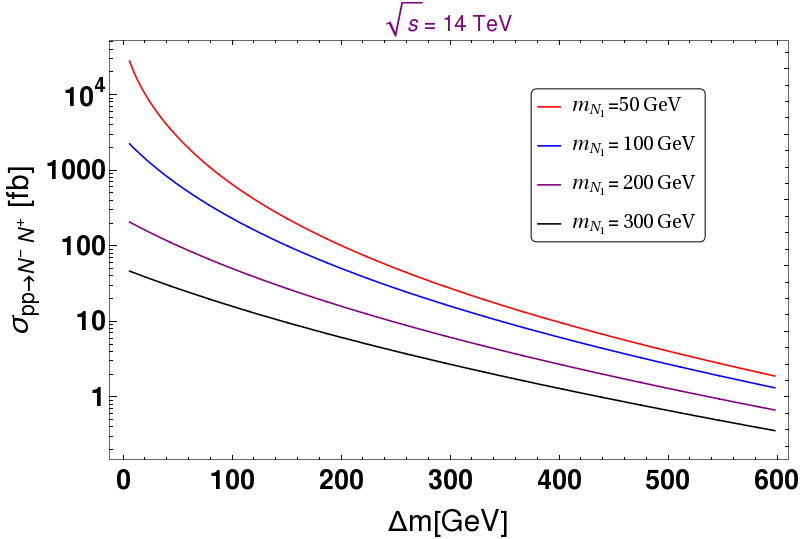}                             
$$
 \caption{[Left] Feynman diagram for signal process $pp \to N^+N^-$, resulting in hadronically quiet opposite sign dilepton plus missing energy ($\ell^+\ell^{-} +({\slashed E}_T)$) events. [Right] Variation in production cross section, $\sigma_{p p \rightarrow N^+N^-}$ with $\Delta m ~(= m_{N^\pm} - m_{N_1})$ for different values of DM mass $m_{N_1}$ [mentioned at figure inset] for centre-of-mass energy $\sqrt{s} = 14 $~ TeV at LHC. }
 \label{fig:LHC-production}
\end{figure}

Collider signature of this model includes searches for scalar and fermion DM. The scalar DM sector doesn't give any novel 
signature being comprised only of a singlet. Only possible signature can be the production of $S$ through Higgs portal coupling associated with initial state 
radiation (ISR), yielding mono jet/mono-X signal (higher jet multiplicity can occur suppressed by further jet radiation) plus missing energy~\cite{Han:2016gyy}. Given the limit on the 
Higgs portal coupling ($\lambda_{SH}$) and DM mass set by the relic density and direct search bound of the model,  even in the two component set up, 
the signal cross-section is very weak to probe anything at near future run of LHC given a huge SM background for such final states\footnote {Even though 
the presence of a heavy scalar $S_H$ adds to the freedom of choosing a larger span of scalar DM mass, the strength of the cross-section still is determined by the 
small $\lambda_{SH}$.}. On the other hand, fermion DM consisting of an admixture of vector-like singlet and doublet leptons, has better 
prospect of getting unravelled at LHC. This is of particular interest due to the possibility of producing the charged companions of fermion doublet ($N^+N^-$) at LHC. 
They eventually decay to DM with off/on- shell W mediation to leptonic final states to yield opposite sign 
dilepton plus missing energy as pointed out in the left side of Fig.~\ref{fig:LHC-production}. Therefore our interest lies in :
\bea
\nonumber
\rm{Signal :}~ \ell^+\ell^{-} +({\slashed E}_T),
\eea
where $\ell$ includes electrons and muons\footnote{Tau detection is harder due to hadronic decay modes.}. However, the detectability of such a signal depends 
on the effective reduction of corresponding SM background contribution. We will discuss below how the presence of a second (lighter) DM component as considered 
in this model framework, enhance the possibility of detecting such signals at LHC. Similar signal events appear for different other models, see for example~\cite{Bahrami:2016has,Barman:2018esi}.

Signal strength is mainly dictated by the production cross-section for $pp \to N^+N^-$ at LHC. This cross-section is essentially a function of $m_{N^\pm}$ and is 
independent of mixing angle $\sin\theta$. Therefore, one can recast the cross-section as a function of $\Delta m$ for a fixed DM mass (Given $m_{N^\pm}=m_N+\Delta m$). 
This is shown in the right panel of Fig.~\ref{fig:LHC-production} for some different fixed DM masses (mentioned in the figure inset) with centre-of-mass-energy $\sqrt s=$ 14 TeV. 
Essentially, this is to show that production cross-section is a falling function of charged fermion masses, but, as $\Delta m$ plays a crucial role in further decay of 
the produced charged fermions, we have chosen such parametrisation. We already elaborated that even in the two component set up, direct search crucially tames  
$\sin\theta \le 0.1$, it is important to choose a process which is not suppressed by small mixing angle. Therefore, this is the only process of interest. 
However, also note that, we do not consider the production of heavy neutral fermion $N_2$ in this analysis (although some of the processes like $N^{\pm}N_2$ 
are not suppressed by small $\sin\theta$), which decays through neutral current ($Z$ mediation) interaction to DM $N_1$ with 100 percent branching ratio. But such 
signals will be completely washed out by the invariant mass-cut of the leptons not to lie within $Z$-mass window, that we must apply to suppress SM background 
(as explained shortly). There are two kinematic constraints that we obey for characteristic collider signal that we discuss here: (i) $ m_{\chi_2} > m_{N_1}+m_S$ and 
(ii) $m_{N_1}>m_S$. The second constraint allows us to choose a large $\Delta m$ as explained earlier and plays an important role in separating the signal from 
SM background. Most of the benchmark points in Table~\ref{tab:BP1} and all in Table~\ref{tab:BP2}, follow the characteristics mentioned above. 
We will analyse signal strength for some such benchmark points. Although, we use benchmark points from Table~\ref{tab:BP1} here, they can also be thought as similar 
points (with same $\Delta m$) from Table~\ref{tab:BP2}, where we have further relaxation on scalar DM mass (which do not play a role in the collider signature for fermion DM). 

%

Before getting into the collider analysis, let us briefly explain the experimental environment of LHC, which mainly involves identification of leptons, jets and unclustered objects.  
Some important variables are also used in the analysis such as missing energy, invariant mass of the dilepton in the final state and scalar sum of the 
transverse momentum of all the visible objects in the final state. They are identified as follows:

\begin{itemize}
 \item {\it Lepton ($l=e,\mu$):} Leptons are identified with a minimum transverse momentum $p_T>20$ GeV and pseudorapidity $|\eta|<2.5$. Although the present sensitivity 
 of the detector allows further soft leptons to be identified, we find that such a $p_T$ cut also helps to tame SM background. Leptons require to be isolated if their 
 mutual distance in the $\eta-\phi$ plane is $\Delta R=\sqrt{\left(\Delta\eta\right)^2+\left(\Delta\phi\right)^2}\ge 0.2$, while the separation with a jet requires $\Delta R\ge 0.4$.
 
 \item {\it Jets ($j$):} Jets are formed for simulated signal and background events using cone algorithm {\tt PYCELL} inbuilt in {\tt Pythia} event generator. All the 
 partons within $\Delta R=0.4$ from the jet initiator cell are included to form the jets. We require $p_T>20$ GeV for a clustered object to be identified as jets in 
 hadron calorimeter (HCAL). Jets are isolated from unclustered objects with $\Delta R>0.4$. Note here, that although jets are not present in the final state, we 
 require a specific jet identification criteria to demand the final state has zero jets.   
 
 \item {\it Unclustered Objects:}  All the final state objects with low $p_T$, which are neither clustered to form jets, nor passes through the identification criteria to become isolated leptons, belong to such category. Hence all particles with 
 $0.5<p_T<20$ GeV and $2.5<|\eta|<5$, are considered as unclustered objects. They only contribute to missing energy. 
 
 \item {\it Missing Energy ($\slashed{E_T}$):} The transverse momentum of all those electromagnetic charge neutral particles not registered in the detector, can be estimated form the momentum imbalance in the transverse direction associated to the visible particles. 
 Thus missing energy (MET) is defined as:
 
 \bea
 \slashed{E_T} = -\sqrt{(\sum_{\ell,j,unc.} p_x)^2+(\sum_{\ell,j,unc.} p_y)^2},
 \eea
 where the sum runs over all visible objects that include the leptons and jets, and the unclustered components. Missing energy is the most significant variable to identify DM at collider.
 
 \item {\it Effective Mass ($H_T$):} Effective mass of an event is identified here with the scalar sum of the transverse momentum of detectable objects in an event, namely lepton and jets as follows:
 \bea
 {H_T} = \sum_{\ell,j} \Big({p_T}\Big)_{\ell,j}~.
 \eea
Effective mass usually also includes missing energy as a component added in the scalar sum. However, here we use $H_T$ without including $\slashed{E_T}$, as we will use $\slashed{E_T}$ as a separate variable in combination of $H_T$ cut anyway to segregate signal from SM background.
 
 \item {\it Invariant mass ($m_{\ell\ell}$):} Invariant mass of opposite sign dilepton is an important variable to segregate SM background from the signal, as it hints to the parent particle mass from which the leptons have been produced. This is defined as:
  \bea
 {m_{\ell\ell}} = \sqrt{ (\sum_{\ell} p_x)^2+ (\sum_{\ell} p_y)^2+(\sum_{\ell} p_z)^2}.
 \eea
\end{itemize}
\begin{figure}[htb!]
$$
 \includegraphics[height=5.0cm]{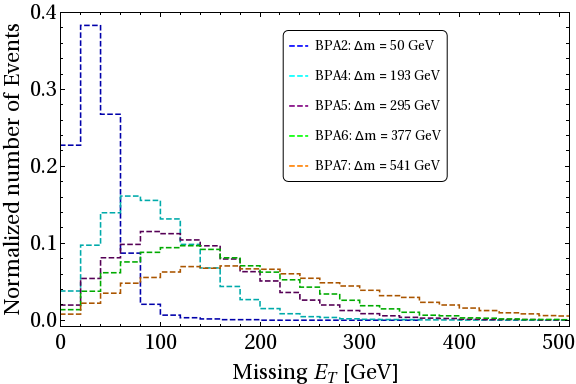}
 \includegraphics[height=5.0cm]{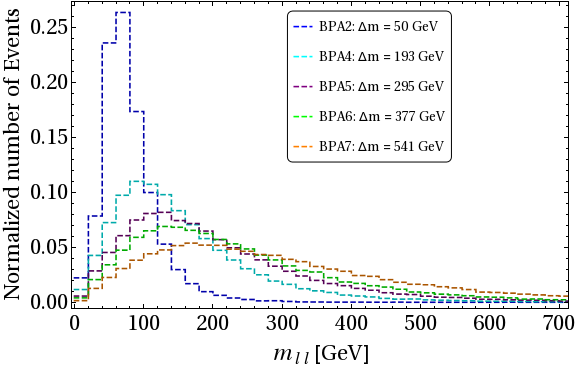}                             
$$
$$
\includegraphics[height=5.0cm]{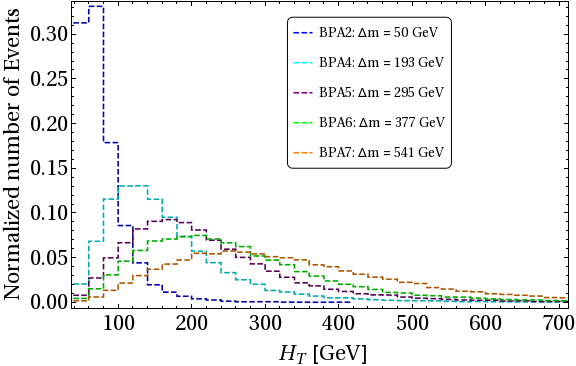} 
$$
 \caption{Missing energy ($\slashed{E}_T$), invariant mass of dilepton ($m_{ \ell \ell}$) and effective mass ($H_T$) distributions of $\ell^+\ell^{-} +({\slashed E}_T)$ events from signal at LHC are shown at $\sqrt{s} = 14 \rm ~TeV$. We have chosen different values of $\Delta m$ corresponding to different benchmark points as indicated in Table~\ref{tab:BP1}.}
 \label{fig:signal-dist}
\end{figure}

\begin{figure}[htb!]
$$
 \includegraphics[height=5.0cm]{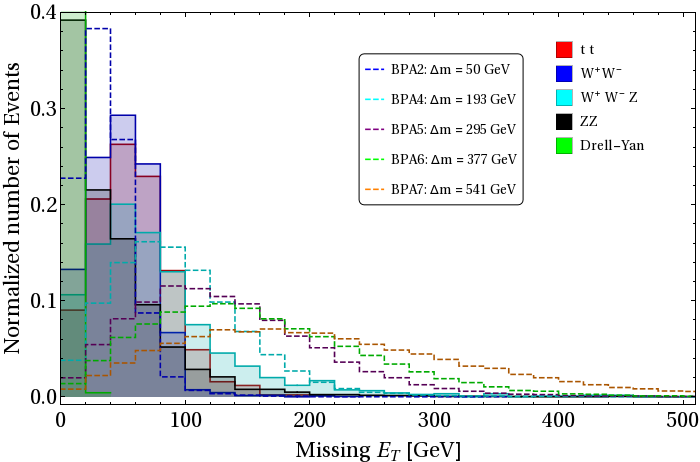}
 \includegraphics[height=5.0cm]{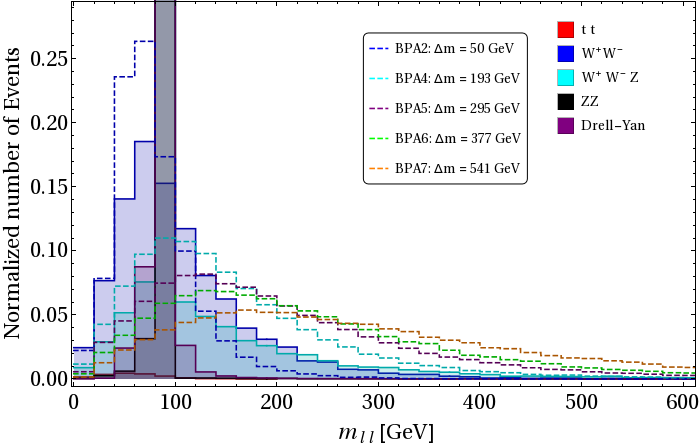}                             
$$
$$
\includegraphics[height=5.0cm]{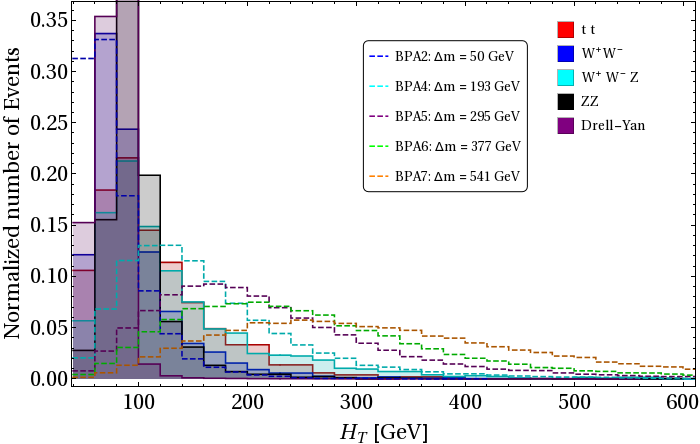} 
$$
 \caption{Missing energy ($\slashed{E}_T$), invariant mass of dilepton ($m_{ \ell \ell}$) and effective mass ($H_T$) distributions of $\ell^+\ell^{-} +({\slashed E}_T)$ events from signal (Benchmark points as in Table~\ref{tab:BP1}) and dominant SM background events at LHC with $\sqrt{s} = 14 \rm ~TeV~$.}
 \label{fig:signal+backg-dist}
\end{figure}

We inserted the model in {\tt Feynrules} ~\cite{Alloul:2013bka} and passed to {\tt Madgraph}~\cite{Alwall:2011uj} to generate signal events, which were further analysed in {\tt Pythia}~\cite{Sjostrand:2006za} to reconstruct leptons, 
jets and other variables discussed above. The dominant SM backgrounds have been generated in {\tt Madgraph}~\cite{Alwall:2011uj} and then showered through {\tt Pythia}~\cite{Sjostrand:2006za}.  
We have identified dominant SM backgrounds for hadronically quiet opposite sign dilepton events as the production of: $t\bar{t}$, $W^{+}W^{-}$, $W^{\pm}Z$, $ZZ$, 
$W^{+}W^{-}Z$ and $Drell-Yan$. We have also used appropriate $K$-factors to incorporate the Next-to-Leading order (NLO) cross section for the backgrounds.
The $K$-factors chosen are as~\cite{Alwall:2014hca} for $t\bar{t}:~K=1.47$, $WW:~K=1.38$, $WZ:~K=1.61$, $ZZj:~K=1.33$, $Drell$-$Yan: ~K=1.2$. We have used {\tt CTEQ 6L}~\cite{Placakyte:2011az} 
parton distribution function and subprocess centre-of-mass-energy ($\sqrt{\hat{s}}$) as jet energy scale for the analysis.  

\begin{table}[H]
\begin{center}
\begin{tabular}{|c|c|c|c|c|c|c|c|c|}
\hline
BPs & $\Delta m$ (GeV) & $\sigma_{pp \to N^+N^-}$ (fb) & $\slashed{E}_T$ (GeV)  & $H_T$ (GeV) & $\sigma^{\text{OSD}}$(fb) & $N^{\text{OSD}}_{\text{eff}}$ \\
\hline\hline 

%
%
%
 
 & & & & $>$100  & 0.002 &$<~1$\\
 
 BPA2 & 50 & 1.73 & $>$100  & $>$200 & 0.001 &$<~1$\\
 
 & & &        & $>$300 & 0.00 & 0 \\
 
 \cline{4-7}
 \hline 

& & & $>$100 & $>$100  & 0.155 &  15 \\

&&&& $>$200 & 0.045 & 4\\

&&&& $>$300 & 0.013 & 1 \\

\cline{4-7}

BPA3 & 101 & 6.23 & $>$200  & $>$ 100    & 0.006  & 1\\

&&&& $>$200 & 0.005 & $<~1$ \\

&&&& $>$300 & 0.004 & $<~1$ \\

\cline{4-7}
\hline 

& & & $>$100 & $>$100  & 0.305 &  30 \\

&&&& $>$200 & 0.138 & 14\\

&&&& $>$300 & 0.044 & 4 \\

\cline{4-7}

BPA4 & 193 & 2.47 & $>$200  & $>$ 100    & 0.032  & 3\\

&&&& $>$200 & 0.031 & 3 \\

&&&& $>$300 & 0.017 & 2 \\

%
%
%
%
%
\cline{4-7}
\hline 

& & & $>$100 & $>$100  & 0.113 &  11 \\

&&&& $>$200 & 0.075 & 7\\

&&&& $>$300 & 0.031 & 3 \\

\cline{4-7}

BPA5 & 295 & 0.54 & $>$200  & $>$ 100    & 0.032  & 3\\

&&&& $>$200 & 0.031 & 3 \\

&&&& $>$300 & 0.016 & 2 \\

\cline{4-7}

&  &  &$>$300  & $>$ 100  & 0.006  & 1 \\

&&&& $>$200 &  0.006 & 1 \\

&&&& $>$300  & 0.005 & $<~1$ \\

\cline{4-7}
\hline 

& & & $>$100 & $>$100  & 0.067 &  7 \\

&&&& $>$200 & 0.052 & 5\\

&&&& $>$300 & 0.027 & 3 \\

\cline{4-7}

BPA6 & 377 & 0.27 & $>$200  & $>$ 100    & 0.027  & 3\\

&&&& $>$200 & 0.027 & 3 \\

&&&& $>$300 & 0.016 & 2 \\

\cline{4-7}

&  &  &$>$300  & $>$ 100  & 0.007  & 1 \\

&&&& $>$200 &  0.007 & 1 \\

&&&& $>$300  & 0.006 & 1 \\

\cline{4-7}
\hline 

& & & $>$100 & $>$100  & 0.017 &  2 \\

&&&& $>$200 & 0.015 & 1\\

&&&& $>$300 & 0.011 & 1 \\

\cline{4-7}

BPA7 & 541 & 0.06 & $>$200  & $>$ 100    & 0.011  & 1\\

&&&& $>$200 & 0.010 & 1 \\

&&&& $>$300 & 0.008 & 1 \\

%
%
%
%
%
\cline{4-7}
\hline 
\hline
\end{tabular}
\end{center}
\caption {Signal events for few selected benchmark points (BPA2-BPA7, see Table \ref{tab:BP1}) with $\sqrt{s}$ = 14 TeV at the LHC for the luminosity $\mathcal{L} = 100~fb^{-1}$ after $\slashed{E}_T$, $H_T$ and $m_{\ell\ell}$ cuts.} 
\label{tab:Sig}
\end{table}  
\begin{table}[H]
\begin{center}
\begin{tabular}{|c|c|c|c|c|c|c|c|}
\hline
SM Backgrounds & $\sigma_{p~p \rightarrow SM}$ (fb) & $\slashed{E}_T$ (GeV)  & $H_T$ (GeV) & $\sigma^{\text{OSD}}$(fb) & $N^{\text{OSD}}_{\text{eff}}$ \\
\hline\hline 

& & $>$100 & $>$100  & 17.11 & 1711  \\

&&& $>$200 & 2.44 & 244\\

&&& $>$300 & $< 0.81$ & $<$ 1 \\

\cline{3-6}

$t~\bar {t}$ & $814.78 \times 10^3$ & $>$200  & $>$ 100    & $<0.81$  & $<$ 1 \\

&&& $>$200 & $<0.81$ & $<$ 1 \\

&&& $>$300 & $<0.81$ & $<1$ \\

\cline{3-6}

&  & $>$300  & $>$ 100  & $<$ 0.81  & $<$ 1 \\

&&& $>$200 &  $<0.81$ & $<$ 1 \\

&&& $>$300  & $<0.81$ & $<$ 1 \\

\cline{3-6}
\hline 


& & $>$100 & $>$100  & 20.51 & 2051  \\

&&& $>$200 & 10.01 & 1001 \\

&&& $>$300 & 2.00 & 200 \\

\cline{3-6}

$W^+ ~W^-$ & $100.06 \times 10^3$ & $>$200  & $>$ 100    & 2.00  & 200 \\

&&& $>$200 & 2.00 & 200 \\

&&& $>$300 & 0.50 & 50 \\

\cline{3-6}

&  & $>$300  & $>$ 100  & $<$ 0.50  & $<$ 1 \\

&&& $>$200 &  $<0.50$ & $<$ 1 \\

&&& $>$300  & $<0.50$ & $<$ 1 \\

\cline{3-6}
\hline 

& & $>$100 & $>$100  & 0.21 & 21  \\

&&& $>$200 & 0.14 & 14\\

&&& $>$300 & 0.07 & 7 \\

\cline{3-6}

$Z ~Z$ & $ 14.03 \times 10^3$ & $>$200  & $>$ 100    & $<0.07$  & $<$ 1 \\

&&& $>$200 & $<0.07$ & $<$ 1 \\

&&& $>$300 & $<0.07$ & $<1$ \\

\cline{3-6}

&  & $>$300  & $>$ 100  & $<$ 0.07  & $<$ 1 \\

&&& $>$200 &  $<0.07$ & $<$ 1 \\

&&& $>$300  & $<0.07$ & $<$ 1 \\

\cline{3-6}
\hline 

& & $>$100 & $>$100  & 0.17 & 17  \\

&&& $>$200 & 0.09 & 9\\

&&& $>$300 & 0.03 & 3 \\

\cline{3-6}

$W^+ ~W^-~Z$ & $0.16 \times 10^3$ & $>$200  & $>$ 100    & 0.04  & 4 \\

&&& $>$200 & 0.04 & 4 \\

&&& $>$300 & 0.02 & 2 \\

\cline{3-6}

&  & $>$300  & $>$ 100  & 0.01  & 1 \\

&&& $>$200 &  0.01 & 1 \\

&&& $>$300  & 0.01 &  1 \\

\cline{3-6}
\hline 

\hline
\end{tabular}
\end{center}
\caption {Dominant SM background contribution to $\ell^+\ell^{-} +({\slashed E}_T)$ signal events with $\sqrt{s}$ = 14 TeV at the LHC for  luminosity $\mathcal{L} = 100~fb^{-1}$ after $\slashed{E}_T$, $H_T$ and $m_{\ell\ell}$ cuts. The variation of effective number of final state background events with cut-flow are also tabulated.} 
\label{tab:background}
\end{table}  

Most important outcome of this analysis is summarised in Fig.~\ref{fig:signal-dist}, where the distribution of the signal events with respect to $\slashed{E}_T$, 
$m_{ \ell \ell}$ and $H_T$ are shown in top left, top right and bottom panel. We have chosen different $\Delta m$ (from the benchmark points as in Table \ref{tab:BP1}) 
upto as large as $\sim$500 GeV allowed by relic density and direct search for illustration. In top left figure, we see that with larger $\Delta m$, 
missing energy distribution becomes flatter and the peak shifts to a higher value. When this is contrasted with the same distributions from those of 
SM background contributions as pointed out in Fig.~\ref{fig:signal+backg-dist}, we see that the separation of signal events from those of the background becomes 
easier at high $\Delta m$. Therefore, for signal events with large $\Delta m$ can survive a large $\slashed{E}_T$ cut while reducing the SM background significantly. 
This should be contrasted with low $\Delta m$ ($\sim$ 50 GeV, BPA2 case), where the peak of missing energy falls within the same ballpark as those of SM 
backgrounds and therefore can not be separated. Therefore even if the signal cross-section is higher for such cases (as in the single component fermion DM case), 
the events are submerged into SM background. This feature is not very difficult to understand. With $\Delta m <m_W$, the $W$ decay is off-shell and $N^\pm$ momenta is 
shared amongst all the final state particles yielding a missing energy peak at lower value. For $\Delta m >m_W$, $W$ is produced on-shell and dominant momenta is 
carried by the dark matter ($N_1$)  as $m_{N_1}>m_W$.  The higher the $\Delta m$ is, the higher is the available momenta for DM. This therefore yields missing 
energy peak at larger values with larger mass splitting $\Delta m$. We also note that such distinction is also possible 
with $H_T$ distribution. Again, the larger the $\Delta m$, the larger will be the available momenta for the leptons as well. Therefore, large $H_T$ cut can also reduce 
SM background retaining signals particularly for benchmark points with higher $\Delta m$. On the other hand, invariant mass cut can effectively reduce SM background 
events coming from $ZZ$ and $WZ$ background, when a cut is applied within the $Z$ 
mass window where the peak of the distribution lies. Therefore, to eliminate SM background from the signal event, we further employ some combination of the following cuts:

\begin{itemize}
 \item $  m_{\ell \ell} < |m_z - 15|$ and  $m_{\ell \ell} > |m_z + 15|$,
 \item $H_T > 100,~200,~300$ GeV,
 \item $\slashed{E}_T > 100, ~200, ~300$ GeV.
\end{itemize}

\begin{figure}[htb!]
$$
 \includegraphics[height=5.0cm]{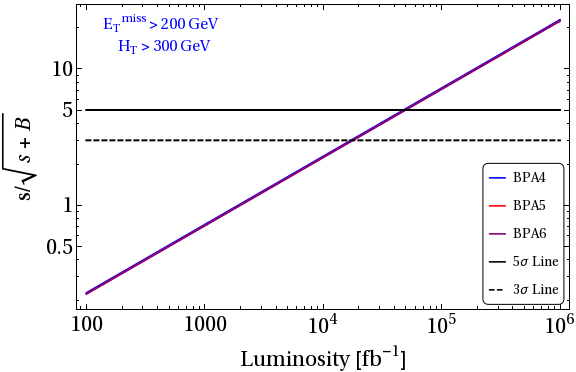}
 \includegraphics[height=5.0cm]{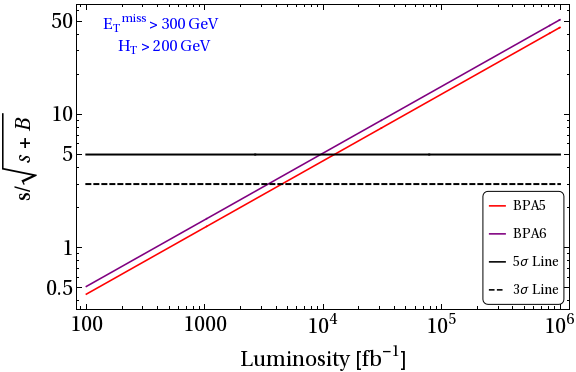}                             
$$
 \caption{Signal significance $\sigma=\frac{S}{\sqrt{S+B}}$ of OSD events for select few benchmark points (see Table \ref{tab:BP1}) at LHC with $E_{CM}=14$ TeV 
 as a function of integrated luminosity. Different combinations of ${\slashed{E_T}},~H_T$ cuts are chosen in left and right panel (mentioned in inset). 3$\sigma$ (black dashed) and 5$\sigma$ (black thick) lines are indicated as references.}
 \label{fig:sigbgst}
\end{figure}

Signal events with $\Delta m=\{50,101,193,295,377,541\}$ GeV corresponding to benchmark points BPA2, BPA4-BPA7 (as in Table \ref{tab:BP1}), are summarised in Table \ref{tab:Sig}, where the cut flow 
with different $H_T$ and $\slashed{E}_T$ are furnished. The final state event rates ($N_{eff}$) at a desired luminosity $\mathcal{L}$ is computed by:
 \bea
N_{eff} = \frac{\sigma_{p}~n}{N}\times\mathcal{L},
\eea
where $N$ is the simulated number of events and $n$ is the obtained final state events corresponding to production cross-section of $\sigma_p$. 
We see that although with larger $\Delta m$, the production cross-sections get diminished by the phase space suppression (as already pointed out in RHS of Fig.~\ref{fig:LHC-production}), 
the shift in the peak of the distribution compensates it to ensure the survival of more number of signal events for such cases. 
With $\Delta m=101$ GeV, the combination of $\slashed{E}_T>200$ GeV and $H_T>100$, leaves with a very few events to be observed. 
$\mathcal{L}=100~\rm{fb}^{-1}$ turns out to be rather low to see the signals from such events and we need higher luminosity. The main take however is to note that 
only those cases where $\Delta m$ is large, has a prospect of discovery by reducing SM background through effective cuts, while those with small $\Delta m$ as in the single 
component framework is almost hopeless.  The signal event rates can be contrasted with the SM background events with similar cut flow at 14 TeV at 
LHC as detailed in Table~\ref{tab:background}. We also note that the limitation in warranting any final state event with number of simulated points yield a limit on the effective background cross-section as indicated in 
the Table. We see that the dominant SM backgrounds can be tamed down significantly with a combination of $\slashed{E}_T$ and $H_T$ cut. 
The reach of the signal significance $\sigma=\frac{S}{\sqrt{S+B}}$ is plotted with integrated luminosity $\mathcal{L}$ for selected benchmark points with two different 
combinations of $\slashed{E}_T$ and $H_T$ cut in left and right panel of Fig.~\ref{fig:sigbgst}. It shows that 5$\sigma$ significance can be reached with luminosity as high as $\sim 10^4~fb^{-1}$.


We also note here that small $\Delta m$ along with small $\sin\theta$ predicts a delay in the decay of the charged fermion, yielding displaced vertex or stable charge 
track signature and serves as a characteristic signal for the fermion dark sector with singlet-doublet mixing, as has already been noted in ref. \cite{Bhattacharya:2017sml}. However, in 
that case, signal excess in dilepton channel can not be seen. On the contrary, with large $\Delta m$, when excess in opposite sign dilepton events can be seen, the decay of 
the charged fermion is quick and therefore no displaced vertex signature can be observed. Therefore the signal of singlet-doublet fermion DM in presence of a second lighter 
DM component has a complementarity to that of the same DM in a single component framework as far as collider search is concerned. The presence of a heavy scalar 
(as illustrated in Sec.~\ref{two_component_DM_SH}) doesn't of course change the fermion DM signal discussed here, but allows one to choose the scalar DM in a large mass range.

\section{Possible implications to Inflation and Reheating}
In this section, we comment briefly on the possibility of production of dark sector particles in the early Universe. It is usually assumed that the early Universe 
has gone through a period inflation driven by a scalar field, the so called inflaton. Subsequently the inflaton decays perturbatively 
(nonperturbatively) to bring back a thermal bath, the so called reheating (preheating) phase, with a temperature $T_R > 4$ MeV to pave 
a path for Big-Bang nucleosynthesis (BBN). See for a review~\cite{Martin:2013tda,Martin:2015dha}. During the reheating (preheating) phase all the 
elementary particles, including dark sector, are assumed to be produced. The process of reheating is quite model dependent and accordingly 
the temperature $T_R$ of thermal bath is set in a large range. 

It has been pointed out that during Inflation, the SM Higgs boson may develop a non-zero and large vev $h_I \sim H_I$~\cite{Freese:2017ace}, where $H_I$ is 
the Hubble scale at the end of inflation. Therefore, the particles which couple to the Higgs will also acquire very high mass during this period. This may result 
in a kinematic blocking of the inflaton decay if the mass of the inflaton is lighter that than the decay products induced by the non-zero Higgs vev\cite{Freese:2017ace}. 
The phenomena is involved and model dependent. We just provide a brief sketch of the main idea. The presence of the scalar DM which couples to Higgs in our scenario through Higgs portal, 
may add to the phenomena. A simple illustration of the above situation can be made by looking into the perturbative inflaton decay neglecting backreaction. 
If we assume a simple inflaton ($\phi$) potential given by $V_\phi=m_\phi^2 \phi^2/2$, the perturbative reheating temperature ($T_R$) is obtained through the 
solution of the following coupled Boltzmann Equations:
\bea
\nonumber \dot{\rho_\phi}+3H\rho_\phi &=& -\Gamma_\phi\rho_\phi, \\
 \dot{\rho_R}+3H\rho_R &=& \Gamma_\phi\rho_\phi,
\eea   
where $\rho_\phi$ is the density of the inflaton and $\rho_R$ is the density of radiation resulting from the decay of the inflaton with decay width $\Gamma_\phi$. 
$H$ is the Hubble constant with $H^2=\frac{8\pi}{3}(\rho_\phi+\rho_R)$. In presence of the DM, the inflaton also decays to DM in addition to SM particles and the total decay width is given by:
\bea
\Gamma_\phi=\Gamma_0\left(1-\frac{4m_f^2}{m_\phi^2}\right)^{3/2}\Theta(m_\phi^2-4m_f^2)+\Gamma_0\left(1-\frac{4m_s^2}{m_\phi^2}\right)^{3/2}\Theta(m_\phi^2-4m_s^2).
\label{eq:theta}
\eea 
In above equation, for simplicity, we just incorporate the decay to SM fermions ($f$) and to the scalar DM $S$. $\Gamma_0$ denotes the decay width at zero mass limit. The mass term 
for the SM fermion and DM are generated from Yukawa interactions ($y$ and $\lambda_{SH}$) followed by the large vev ($h_I$) that Higgs acquires during inflation and will be given by: 
\bea
m_f^2=\frac{1}{2} y^2 h_I^2; ~m_S^2=\frac{1}{2}\lambda_{SH}^2h_I^2.
\eea
The $\Theta$ function in Eq.\ref{eq:theta} denotes the phase space blocking. Depending on whichever is lighter between $m_f$ and $m_S$, the effective blocking condition for the 
inflaton decay (assuming $m_S<m_f$) reads:
\bea
\frac{h_I^2}{m_\phi^2} >\frac{1}{2\lambda_{SH}^2}.
\eea 
As a result the reheating temperature $T_R$ can drop significantly and can be even be less than the Higgs mass depending on the coupling and vev. The delay in reheating 
may alter the CMB spectrum in terms of the spectral index of density perturbation ($n_s$) and tensor-to-scalar ratio ($r$) or affect the heavy particle production.  
However, one should note here that the maximum temperature $T_{\rm max}$ during reheating can be much larger than reheating temperature $T_R$. During reheating, the temperature
rises to $T_{max}$ and then falls to $T_R$, see for example, ~\cite{Rangarajan:2008zb}: 
 \bea
 T_{max}=0.6 g_{*}^{-1/4}(\Gamma_\phi M_{Pl})^{1/4}M_I^{1/2},
 \eea
where $M_I=V_I^{1/4}$, $V_I$ depicts the energy density at $t_{osc}$. Depending on the model, maximum temperature can be as high as $T_{max}\sim 10^3 T_R$. 
As a result, the heavy particles (including dark sector particles $\chi_1$, $\chi_2$, $N$, $S$ {\it etc.}) 
in general can be produced during reheating phase itself. Once these particles are produced, irrespective of their initial number density, they can easily thermalise due 
to their coupling with the SM Higgs and other SM particles. For instance in our case $N$ is a doublet. So it can be easily thermalise due to 
its gauge coupling. On the other hand, $\chi_1$, and $S$ are singlet under the SM gauge group. However, these particles couple to 
the SM Higgs through (large) Yukawa interaction ($Y_1$ and $\lambda_{SH}$ respectively). Therefore, the dark sector particles in our case are no more in danger being over produced 
even if the kinematic blocking effects in a lower reheat temperature as discussed above.

\section{Summary}
 The dark sector of the universe is still a mystery to us. In this work, we have discussed a possible two component (WIMP-like) DM scenario with a vector like fermion 
(an admixture of a singlet and a doublet) and a scalar singlet stabilised by $\mathcal{Z}_2\times \mathcal{Z}^{'}_2$ symmetry. 
The proposed scenario crucially addresses the possibility of DM-DM interaction between fermion and scalar DM candidates through 
another heavy vectorlike fermion singlet which acts as mediator. We show that in absence of the mediator 
(which means the absence of $t$- channel heavy fermion mediated DM-DM interaction), 
both fermion and scalar DM components behave like two decoupled single component DMs. 
This is due to suppressed s-channel Higgs mediated interaction between the DM components. 
In such a situation, both of the sector turns out to fill up the corresponding under-abundant regions to add to the observed relic density. 
Although such a non interacting situation satisfy observed DM relic density, the direct search limit (XENON1T) 
rules-out most of the parameter space, particularly for the scalar DM to a very heavy mass $\gsim1$ TeV. For fermion DM, 
the necessity of co-annihilation contribution limits the mass difference with the charged doublet component to a small value ($\lsim 12$ GeV).  

However, in presence of a heavy fermion mediated $t$-channel DM-DM conversion, 
with moderate values of mediator mass $\sim$ 500 GeV, the freeze out and relic density of DM components get affected significantly. 
The change is observed mostly in the relic density of the heavier DM component, which has the liberty of annihilating to the lighter DM, 
unconstrained by direct search limit; while lighter DM component behaves mostly as in single component framework. 
So, by allowing DM-DM conversion in the interacting picture, we open up large parameter space allowed by both 
relic density and direct search bounds which otherwise yields over-abundance in non-interacting cases. 
For fermion DM (when it is heavier than scalar DM), large $\Delta m$ regions become allowed, but scalar DM is restricted to the Higgs resonance
region. In presence of a heavy scalar, which helps co-annihilating the scalar DM component, allow a larger mass range for scalar DM 
even when it is lighter than fermion DM. On the other hand, when scalar DM is heavier than fermion DM, DM-DM conversion allows 
the presence of smaller Higgs portal couplings, hiding the scalar DM from direct search to allow a larger mass range upto TeV and beyond. 

The work also demonstrates the importance of DM-DM conversion in seeing signals of a dark sector at LHC in relic density 
and direct search allowed parameter space. In the model, fermion dark sector is composed of a doublet and a singlet. 
Hence, the charged companions can be produced at LHC which yields hadronically quiet oppsite sign dilepton events 
plus missing energy through their decays to fermion DM. However, in a single component framework, 
relic density and direct search constraints restrict the fermion DM to have a small mass difference with the charged companion ($\Delta m$), 
which makes the signal submerged into SM background. On the contrary, in presence of a lighter DM component and an effective DM-DM conversion, 
$\Delta m$ can be large, which can segregate the signal from SM background by a combination of large missing 
energy and effective mass cuts as detailed in the analysis. The discovery limit of such a signal still might be delayed to an 
integrated luminosity $\sim 10^4~\rm{fb}^{-1}$.

\paragraph*{Acknowledgments\,:} 
The authors would like to acknowledge discussions with Dr. Nirakar Sahoo, Dr. Debaprasad Maity and Dr. Pankaj Saha. 
PG would like to thank Dr. Rashidul Islam and Basabendu Barman for fruitful discussions and MHRD, Government of India for research fellowship. 
SB would like to acknowledge the DST-INSPIRE research grant IFA13-PH-57 at IIT Guwahati. 

\appendix
\section{Single Component vector-like fermion DM}
\label{apnd:SVF}

The freeze-out of $N_1$ DM is controlled by the annihilation and co-annihilation channels as shown in Fig.~\ref{fd:an-coan}, \ref{co-ann-2}, \ref{co-ann-3}. 
This is mainly driven by gauge mediation and Higgs mediation apart from the $t$-channel heavy fermion $(N_2,N^\pm)$ mediation.   

\begin{figure}[htb!]
\begin{center}
    \begin{tikzpicture}[line width=0.5 pt, scale=0.9]
        \draw[solid] (-3,1.0)--(-1.0,1.0);
        \draw[solid] (-3,-1.0)--(-1.0,-1.0);
        \draw[solid] (-1.0,1.0)--(-1.0,-1.0);
        \draw[solid] (-1.0,1.0)--(1.0,1.0);
        \draw[solid] (-1.0,-1.0)--(1.0,-1.0);
        \node at (-3.4,1.0) {$\overline{N_i}$};
        \node at (-3.4,-1.0) {$N_j$};
        \node [right] at (-1.05,0.0) {$N_k$};
        \node at (1.8,1.0) {$Z/Z/h$};
        \node at (1.8,-1.0) {$h/Z/h$};
        \draw[solid] (5.0,1.0)--(6.5,0.0);
        \draw[solid] (5.0,-1.0)--(6.5,0.0);
        \draw[dashed] (6.5,0.0)--(8.5,0.0);
        \draw[solid] (8.5,0.0)--(10.0,1.0);
        \draw[solid] (8.5,0.0)--(10.0,-1.0);
        \node at (4.7,1.0) {$\overline{N_i}$};
        \node at (4.7,-1.0) {$N_j$};
        \node [above] at (7.4,0.05) {$h$};
        \node at (11.1,1.0) {$f/W^+/Z/h$};
        \node at (11.1,-1.0) {$\overline{f}/W^-/Z/h$};
     \end{tikzpicture}
 \end{center}
\begin{center}
    \begin{tikzpicture}[line width=0.5 pt, scale=0.9]
        \draw[solid] (-3,1.0)--(-1.0,1.0);
        \draw[solid] (-3,-1.0)--(-1.0,-1.0);
        \draw[solid](-1.0,1.0)--(-1.0,-1.0);
        \draw[solid] (-1.0,1.0)--(1.0,1.0);
        \draw[solid] (-1.0,-1.0)--(1.0,-1.0);
        \node at (-3.4,1.0) {$\overline{N_i}$};
        \node at (-3.4,-1.0) {$N_j$};
        \node [right] at (-1.05,0.0) {$N^-$};
        \node at (1.8,1.0) {$W^+$};
        \node at (1.8,-1.0) {$W^-$};
        \draw[solid] (5.0,1.0)--(6.5,0.0);
        \draw[solid] (5.0,-1.0)--(6.5,0.0);
        \draw[snake] (6.5,0.0)--(8.5,0.0);
        \draw[solid] (8.5,0.0)--(10.0,1.0);
        \draw[solid] (8.5,0.0)--(10.0,-1.0);
        \node at (4.7,1.0) {$\overline{N_i}$};
        \node at (4.7,-1.0) {$N_j$};
        \node [above] at (7.4,0.05) {$Z$};
        \node at (11.1,1.0) {$f/W^+/h$};
        \node at (11.1,-1.0) {$\overline{f}/W^-/Z$};
     \end{tikzpicture}
 \end{center}
\caption{Annihilation ($i=j$) and Co-annihilation ($i\neq j$) of fermion DM. Here $(i,j,k=1,2)$. }
\label{fd:an-coan}
 \end{figure}
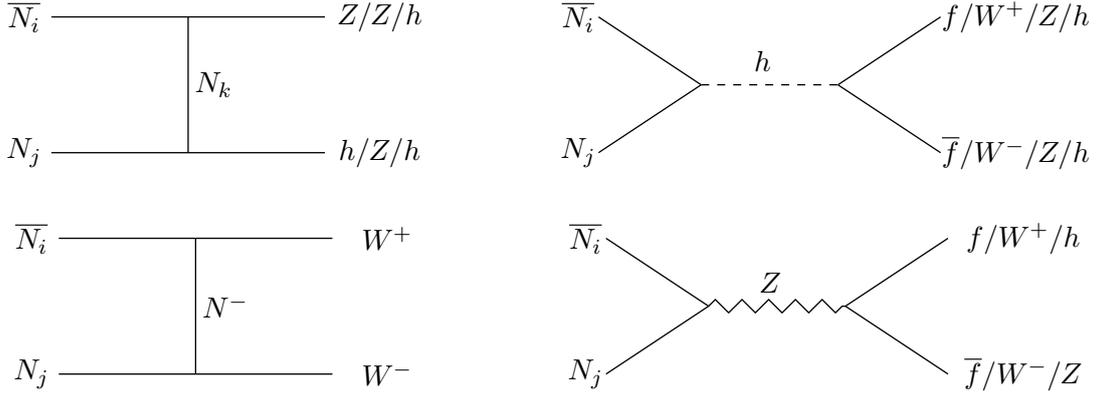
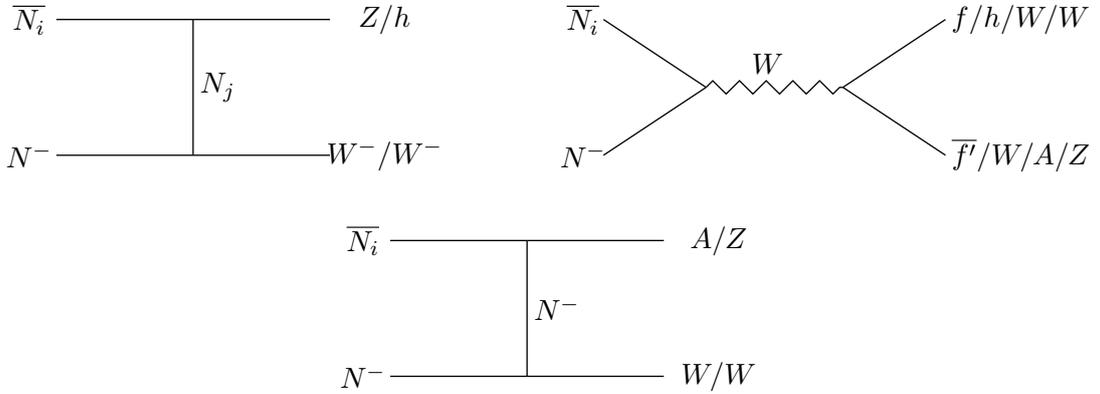
\begin{figure}[htb!]
\begin{center}
    \begin{tikzpicture}[line width=0.5 pt, scale=0.9]
        \draw[solid] (-3,1.0)--(-1.0,1.0);
        \draw[solid] (-3,-1.0)--(-1.0,-1.0);
        \draw[solid] (-1.0,1.0)--(-1.0,-1.0);
        \draw[solid] (-1.0,1.0)--(1.0,1.0);
        \draw[solid] (-1.0,-1.0)--(1.0,-1.0);
        \node at (-3.4,1.0) {$\overline{N_i}$};
        \node at (-3.4,-1.0) {$N^-$};
        \node [right] at (-1.05,0.0) {$N_j$};
        \node at (1.8,1.0) {$Z/h$};
        \node at (1.8,-1.0) {$W^-/W^-$};
        \draw[solid] (5.0,1.0)--(6.5,0.0);
        \draw[solid] (5.0,-1.0)--(6.5,0.0);
        \draw[snake] (6.5,0.0)--(8.5,0.0);
        \draw[solid] (8.5,0.0)--(10.0,1.0);
        \draw[solid] (8.5,0.0)--(10.0,-1.0);
        \node at (4.7,1.0) {$\overline{N_i}$};
        \node at (4.7,-1.0) {$N^-$};
        \node [above] at (7.4,0.05) {$W$};
        \node at (11.1,1.0) {$f/h/W/W$};
        \node at (11.1,-1.0) {$\overline{f^\prime}/W/A/Z$};
     \end{tikzpicture}
 \end{center}
\begin{center}
    \begin{tikzpicture}[line width=0.5 pt, scale=0.9]
        \draw[solid] (-3,1.0)--(-1.0,1.0);
        \draw[solid] (-3,-1.0)--(-1.0,-1.0);
        \draw[solid](-1.0,1.0)--(-1.0,-1.0);
        \draw[solid] (-1.0,1.0)--(1.0,1.0);
        \draw[solid] (-1.0,-1.0)--(1.0,-1.0);
        \node at (-3.4,1.0) {$\overline{N_i}$};
        \node at (-3.4,-1.0) {$N^-$};
        \node [right] at (-1.05,0.0) {$N^-$};
        \node at (1.8,1.0) {$A/Z$};
        \node at (1.8,-1.0) {$W/W$};
     \end{tikzpicture}
 \end{center}
\caption{ Co-annihilation process of $N_i ~(i=1,2)$ with the charge component $N^-$ to SM particles. }
\label{co-ann-2}
 \end{figure}
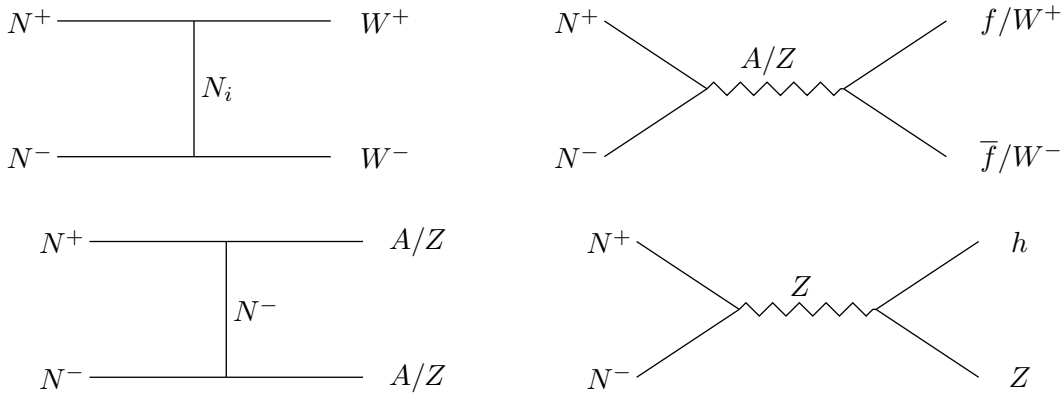
\begin{figure}[htb!]
\begin{center}
    \begin{tikzpicture}[line width=0.5 pt, scale=0.9]
        \draw[solid] (-3,1.0)--(-1.0,1.0);
        \draw[solid] (-3,-1.0)--(-1.0,-1.0);
        \draw[solid] (-1.0,1.0)--(-1.0,-1.0);
        \draw[solid] (-1.0,1.0)--(1.0,1.0);
        \draw[solid] (-1.0,-1.0)--(1.0,-1.0);
        \node at (-3.4,1.0) {$N^+$};
        \node at (-3.4,-1.0) {$N^-$};
        \node [right] at (-1.05,0.0) {$N_i$};
        \node at (1.8,1.0) {$W^+$};
        \node at (1.8,-1.0) {$W^-$};
        \draw[solid] (5.0,1.0)--(6.5,0.0);
        \draw[solid] (5.0,-1.0)--(6.5,0.0);
        \draw[snake] (6.5,0.0)--(8.5,0.0);
        \draw[solid] (8.5,0.0)--(10.0,1.0);
        \draw[solid] (8.5,0.0)--(10.0,-1.0);
        \node at (4.6,1.0) {$N^+$};
        \node at (4.6,-1.0) {$N^-$};
        \node [above] at (7.4,0.05) {$A/Z$};
        \node at (11.1,1.0) {$f/W^+$};
        \node at (11.1,-1.0) {$\overline{f}/W^-$};
     \end{tikzpicture}
 \end{center}
\begin{center}
    \begin{tikzpicture}[line width=0.5 pt, scale=0.9]
        \draw[solid] (-3,1.0)--(-1.0,1.0);
        \draw[solid] (-3,-1.0)--(-1.0,-1.0);
        \draw[solid](-1.0,1.0)--(-1.0,-1.0);
        \draw[solid] (-1.0,1.0)--(1.0,1.0);
        \draw[solid] (-1.0,-1.0)--(1.0,-1.0);
        \node at (-3.4,1.0) {$N^+$};
        \node at (-3.4,-1.0) {$N^-$};
        \node [right] at (-1.05,0.0) {$N^-$};
        \node at (1.8,1.0) {$A/Z$};
        \node at (1.8,-1.0) {$A/Z$};
        \draw[solid] (5.0,1.0)--(6.5,0.0);
        \draw[solid] (5.0,-1.0)--(6.5,0.0);
        \draw[snake] (6.5,0.0)--(8.5,0.0);
        \draw[solid] (8.5,0.0)--(10.0,1.0);
        \draw[solid] (8.5,0.0)--(10.0,-1.0);
        \node at (4.6,1.0) {$N^+$};
        \node at (4.6,-1.0) {$N^-$};
        \node [above] at (7.4,0.05) {$Z$};
        \node at (10.6,1.0) {$h$};
        \node at (10.6,-1.0) {$Z$};
     \end{tikzpicture}
 \end{center}
\caption{Co-Annihilation process of charged fermions $N^\pm$ to SM particles in final states . }
\label{co-ann-3}
 \end{figure}
 
Relic density of vector like fermion DM is then governed by the effective number changing cross-section following \cite{Griest:1990kh},
 
\bea
{\langle \sigma v\rangle}^{eff}_{N_1}&&= \frac{g_1^2}{g_{eff}^2} {\langle \sigma v \rangle}_{\overline{N_1}N_1}+\frac{2 g_1 g_2}{g_{eff}^2} {\langle \sigma v \rangle}_{\overline{N_1}N_2}\Big(1+\frac{\Delta m}{m_{N_1}}\Big)^\frac{3}{2} e^{-x \frac{\Delta m}{m_{N_1}}} \nonumber \\
&&+\frac{2 g_1 g_3}{g_{eff}^2} {\langle \sigma v \rangle}_{\overline{N_1}N^-}\Big(1+\frac{\Delta m}{m_{N_1}}\Big)^\frac{3}{2} e^{-x \frac{\Delta m}{m_{N_1}}} \nonumber \\
&& +\frac{2 g_2 g_3}{g_{eff}^2} {\langle \sigma v \rangle}_{N^+N_2}\Big(1+\frac{\Delta m}{m_{N_1}}\Big)^3 e^{- 2 x \frac{\Delta m}{m_{N_1}}} \nonumber \\
&& +\frac{g_2^2}{g_{eff}^2} {\langle \sigma v \rangle}_{\overline{N_2}N_2}\Big(1+\frac{\Delta m}{m_{N_1}}\Big)^3 e^{- 2 x \frac{\Delta m}{m_{N_1}}} \nonumber \\
&& +\frac{g_3^2}{g_{eff}^2} {\langle \sigma v \rangle}_{N^+N^-}\Big(1+\frac{\Delta m}{m_{N_1}}\Big)^3 e^{- 2 x \frac{\Delta m}{m_{N_1}}}. 
\label{eq:vf-ann}
\eea

In above equation, $g_{eff}$, defined as effective degrees of freedom, is given by
\bea
g_{eff}=g_1 + g_2 \Big(1+\frac{\Delta m}{m_{N_1}}\Big)^\frac{3}{2} e^{-x \frac{\Delta m}{m_{N_1}}} + g_3\Big(1+\frac{\Delta m}{m_{N_1}}\Big)^\frac{3}{2} e^{-x \frac{\Delta m}{m_{N_1}}} ,
\eea
where $g_1 ,~ g_2 \rm ~and~ g_3$ are the degrees of freedom of $N_1, ~N_2 \rm ~and~ N^-$ respectively and $x=x_f=\frac{m_{N_1}}{T_f}$, where $T_f$ is the freeze out temperature of $N_1$. Then relic density will be given by~\cite{Kolb:1990vq,Bhattacharya:2016ysw} :
\bea
\Omega h^2 =\frac{854.45 \times 10^{-13}}{\sqrt{106.7}} \frac{x_f}{{\langle \sigma v\rangle}^{eff}_{N_1}},
\eea
 assuming $x_f \sim 20$. 
 
 \begin{figure}[H]
$$
 \includegraphics[height=5.0cm]{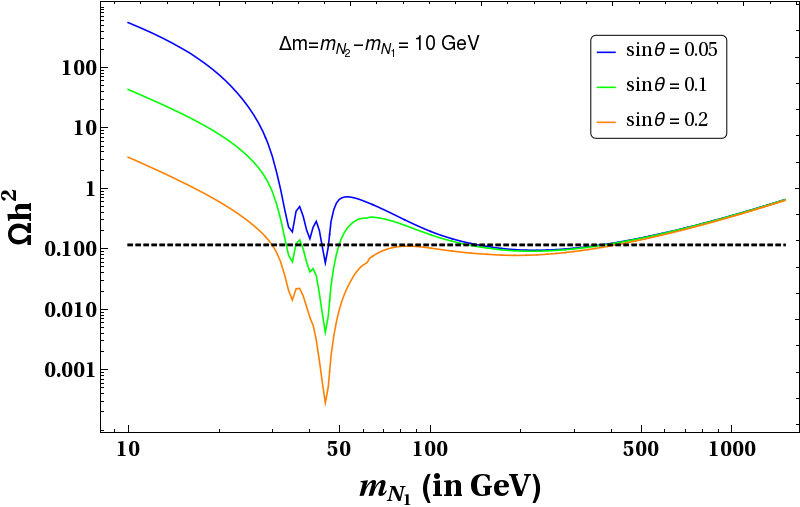} 
 \includegraphics[height=5.0cm]{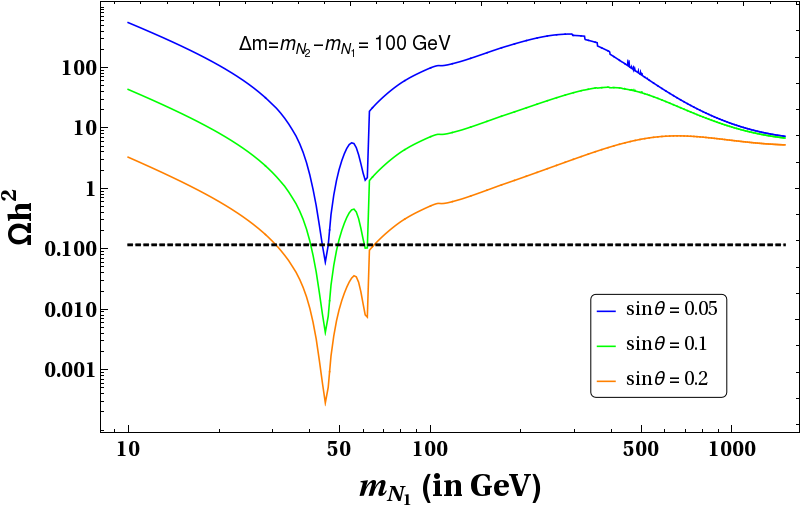}
$$
 \caption{Relic density of $N_1$ as a function of DM mass, $m_{N_1}$  with different mixing angle, $\sin\theta = 0.05$ (blue), $\sin\theta = 0.1$ (green) and $\sin\theta = 0.2$ (orange). Each plot corresponds to fixed $\Delta m$: $ = 10$ GeV (left), $ 100$ GeV (right). Black dashed line indicates  observed relic density $0.1133 \leq \Omega_{DM}h^2 \leq 0.1189$. }
 \label{fig:omega-f-m1}
\end{figure}

Variation of relic density of fermion DM is shown as a function of DM mass, for a fixed $\Delta m=10$ GeV (left panel of the Fig.~\ref{fig:omega-f-m1})~ and~$ 100$ GeV (right panel of the Fig.~\ref{fig:omega-f-m1}) and different choices of mixing angle,$\sin\theta$. We note that the annihilation cross-section is larger when we take larger values of $\sin\theta$, due to larger $SU(2)$ component, resulting smaller relic density. The resonance drop at $m_Z/2$ and at $m_h/2$ is observed due to $s$-channel $Z$ and $H$ mediated contributions. For $\Delta m = 100$ GeV, due to smaller co-annihilation contribution relic density increases compared to $\Delta m = 10$ GeV case.

\section{Higgs Invisible Decay Constraint}
\label{HiggsInv}

When masses of DMs are smaller than the Higgs mass i.e. $m_{DM} < m_h/2 $, then Higgs can decay to DM (invisible particles) and will contribute to invisible decay width. LHC data puts strong constraint on the invisible branching fraction of Higgs as $Br(h\rightarrow inv) < 0.24$ ~\cite{Tanabashi:2018oca}. This can be interpreted as follows:
\bea
Br(h \to  inv.) && < 0.24 \nonumber \\
\frac{\Gamma(h \rightarrow inv. )}{\Gamma(h \rightarrow SM) + \Gamma(h \rightarrow inv.  )} && < 0.24,  
\eea

where $\Gamma(h\to SM)=4.2$ MeV for SM Higgs (with mass $m_h=125.09~ {\rm GeV}$) is measured at LHC~\cite{Tanabashi:2018oca}. This then yields, 

\bea\label{eq:HinvA}
\Gamma(h \rightarrow inv.  ) < 1.32 ~\rm~ MeV~. 
\eea

In our two component DM scenario, the invisible decay may have two contributions if both $m_{N_1},~m_S < m_h/2$:
\bea \label{eq:HinvB}
\Gamma(h \rightarrow inv.  )= \Gamma(h \rightarrow \overline{N_1} N_1  ) +  \Gamma(h \rightarrow S~ S  ). \nonumber  \\ 
\eea

\begin{figure}[htb!]
$$
 \includegraphics[height=7.0cm]{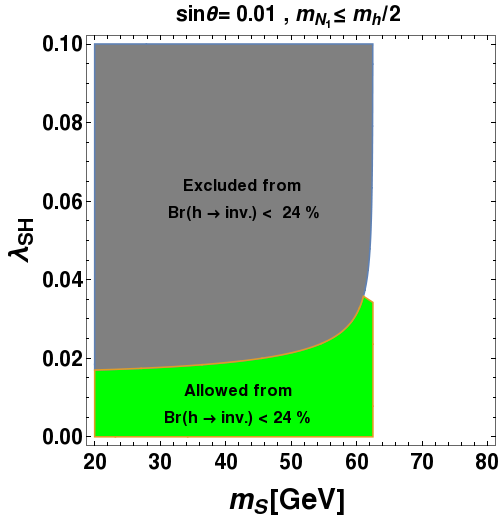} ~~~
 \includegraphics[height=7.0cm]{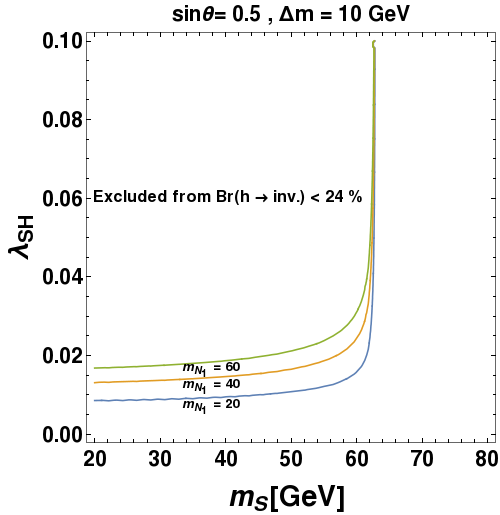} 
 $$
 $$
 \includegraphics[height=7.0cm]{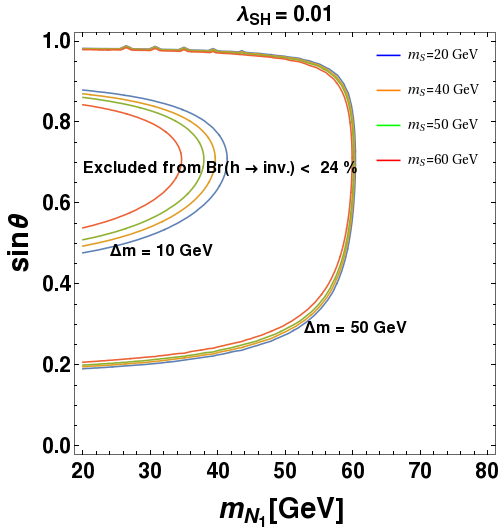} 
 $$
 \caption{Constraints on scalar and fermion DM from Higgs invisible branching ratio $Br(h \to  inv.) < 0.24 $~\cite{Tanabashi:2018oca} in $m_S -\lambda_{SH}$ plane (top panel) and $m_{N_1}-\sin\theta$ plane (bottom panel) keeping other parameter fixed (mentioned in the figure inset). }
\label{fig:Higgs-inv-S}
\end{figure}

The decay width of Higgs to $S ~\rm ~and ~N_1 $ can easily be calculated as:
\bea\label{eq:HinvC}
\Gamma_{h \to S ~S}= \frac{\lambda_{SH}^2 v^2}{32 \pi m_h^2}\sqrt{m_h^2-4m_{S}^2} ~~\Theta(m_h- 2 m_S),\nonumber \\
\Gamma_{h \to \overline{N_1} N_1}= \frac{1}{16 \pi} \Big(Y_1 \sin2\theta\Big)^2  m_h~~ \Big(1-\frac{4 m_{N_1}^2}{m_h^2}\Big)^\frac{3}{2} ~~\Theta(m_h-  2m_{N_1}) .
\eea

Invisible Higgs decay constraint from Eq.~\ref{eq:HinvA} together with Eq.~\ref{eq:HinvB} and Eq.~\ref{eq:HinvC} is shown in Fig.~\ref{fig:Higgs-inv-S}. 
 In top left panel of Fig.~\ref{fig:Higgs-inv-S}, the constraint is shown in $m_S-\lambda_{SH}$ plane. Here, the green region is allowed from Higgs invisible decay while grey region excluded for a fixed $\sin\theta=0.01$. The allowed (or excluded) region remains almost unchanged for any fermion DM mass ($m_{N_1} < m_h/2$) and $\Delta m$ for the small $\sin\theta$ due to negligible contribution of $\Gamma(h\rightarrow \overline{N_1} N_1)$. In the top right panel we consider larger mixing angle, $\sin\theta=0.5$. As the contribution the contribution of $\Gamma(h\rightarrow \overline{N_1} N_1)$  plays a important role to $\Gamma(h \rightarrow inv. )$. And therefore, choices of other parameters like $m_{N_1},\Delta m$ becomes relevant. The inner region of each contour in $m_S-\lambda_{SH}$ plane (top right panel) is excluded from  Higgs invisible decay constraint~\cite{Tanabashi:2018oca}. Note here however that such large $\sin\theta$($=0.5$) is disfavoured from direct search bounds~\cite{Akerib:2017kat,Aprile:2018dbl}. In the bottom panel we have shown excluded region in $m_{N_1}-\sin\theta$ plane keeping other parameters, $m_{S}, \lambda_{SH}$ and $\Delta m$ fixed. Similarly here the inner region of each contour line (which corresponds to different fixed values of $\Delta m$ and scalar DM mass ($m_S$), depicted in the figure) is excluded from Higgs invisible decay~\cite{Tanabashi:2018oca}. 

\section{Invisible Decay Constraint of $Z$}
\label{ZInv}

As the fermion DM has a doublet component in it, it has $Z$ mediated interaction. Hence, if fermion DM mass is below $m_Z/2$, then $Z$ can invisibly decay to dark particles. From current observation, invisible decay width of $Z$ is strongly constrained. The upper limit of invisible Z decay width is following ~\cite{Tanabashi:2018oca}:
\bea
\Gamma(Z \rightarrow inv.) \leq 499.0\pm 1.5 ~~ \rm ~MeV ,
\eea

where in our model,

\bea 
\Gamma(Z \rightarrow inv.  )&= &\Gamma(Z \rightarrow \overline{N_1} N_1  )  \nonumber  \\ 
&=& \frac{1}{48 \pi} \Big(\frac{g~ \sin^2\theta}{\cos\theta_W}\Big)^2 ~m_Z~~ \Big(1+\frac{2 m_{N_1}^2}{m_Z^2}\Big)\sqrt{1-\frac{4 m_{N_1}^2}{m_Z^2}}~~\Theta(m_Z-2 m_{N_1})~
\eea

Invisible decay of $Z$ mainly depend on mixing angle $\sin\theta$. Choice of small mixing angle, with $\sin\theta <0.1$ is preferable from direct search bound in which all fermion DM mass $m_{N_1}<m_Z/2$ is allowed from invisible decay width of $Z$ \cite{Bhattacharya:2016ysw}. We note that as the scalar DM component is a gauge singlet, it doesn't have a $Z$ mediated interaction and therefore no constraint from invisible $Z$ decay applies to it. 


\bibliographystyle{JHEP}
\bibliography{ref}


\end{document}